\documentclass[twocolumn,showpacs,preprintnumbers]{revtex4}

\usepackage{amsmath}
\usepackage{graphicx}
\usepackage{amsbsy}

\begin{document}

\begin{widetext}
\thispagestyle{empty}
\begin{Large}
\textbf{DEUTSCHES ELEKTRONEN-SYNCHROTRON}

\textbf{\large{in der HELMHOLTZ-GEMEINSCHAFT}\\}
\end{Large}

DESY 05-032

February 2005

\begin{eqnarray}
\nonumber &&\cr \nonumber && \cr \nonumber &&\cr
\end{eqnarray}
\begin{eqnarray}
\nonumber
\end{eqnarray}
\begin{center}
\begin{Large}
\textbf{Paraxial Green's functions in Synchrotron Radiation
theory}
\end{Large}
\begin{eqnarray}
\nonumber &&\cr \nonumber && \cr
\end{eqnarray}

\begin{large}
Gianluca Geloni, Evgeni Saldin, Evgeni Schneidmiller and Mikhail
Yurkov
\end{large}
\textsl{\\Deutsches Elektronen-Synchrotron DESY, Hamburg}
\begin{eqnarray}
\nonumber
\end{eqnarray}
\begin{eqnarray}
\nonumber
\end{eqnarray}
\begin{eqnarray}
\nonumber
\end{eqnarray}
\begin{eqnarray}
\nonumber
\end{eqnarray}
\begin{eqnarray}
\nonumber
\end{eqnarray}
\begin{eqnarray}
\nonumber
\end{eqnarray}
\begin{eqnarray}
\nonumber
\end{eqnarray}
\begin{eqnarray}
\nonumber
\end{eqnarray}
\begin{eqnarray}
\nonumber
\end{eqnarray}
\begin{eqnarray}
\nonumber
\end{eqnarray}
\begin{eqnarray}
\nonumber
\end{eqnarray}
ISSN 0418-9833
\begin{eqnarray}
\nonumber
\end{eqnarray}
\begin{large}
\textbf{NOTKESTRASSE 85 - 22607 HAMBURG}
\end{large}
\end{center}
\end{widetext}
\newpage


\title{Paraxial Green's functions in Synchrotron Radiation theory}

\author{Gianluca Geloni}
 \email{gianluca.aldo.geloni@desy.de}
\author{Evgeni Saldin}%
\author{Evgeni Schneidmiller}
\author{Mikhail Yurkov}
\affiliation{%
Deutsches Elektronen-Synchrotron, DESY \\ Notkestrasse 85, 22607
Hamburg,
\\ Germany}

\date{\today}

\begin{abstract}

This work contains a systematic treatment of single particle
Synchrotron Radiation and some application to realistic beams with
given cross section area, divergence and energy spread. Standard
theory relies on several approximations whose applicability limits
and accuracy are often forgotten. We begin remarking that on the
one hand, a paraxial approximation can always be applied without
loss of generality and with ultra relativistic accuracy. On the
other hand, dominance of the acceleration field over the velocity
part in the Lienard-Wiechert expressions is not always granted and
constitutes a separate assumption, whose applicability is
discussed. Treating Synchrotron Radiation in paraxial
approximation we derive the equation for the slow varying envelope
function of the Fourier components of the electric field vector.
Calculations of Synchrotron Radiation properties performed by
others showed that the phase of the Fourier components of the
electric field vector differs from the phase of a virtual point
source. In this paper we present a systematic, analytical
description of this phase shift, calculating amplitude and phase
of electric field from bending magnets, short magnets, two bending
magnet system separated by a straight section (edge radiation) and
undulator devices. We pay particular attention to region of
applicability and accuracy of approximations used. Finally, taking
advantage of results of analytical calculation presented in
reduced form we analyze various features of radiation from a
complex insertion device (set of two undulators with a focusing
triplet in between) accounting for the influence of energy spread
and electron beam emittance.
\end{abstract}

\maketitle

\section{\label{sec:intro} Introduction}

About sixty years have passed since the first, pioneering works on
Synchrotron Radiation have been published (see \cite{SCHW} and
references therein). The way scientists consider this phenomenon
has drastically changed during this period. At first, Synchrotron
Radiation was regarded only as a detrimental factor, a limitation
on the maximal particle energy attainable with accelerators (see,
for instance, \cite{IWAN}). Nowadays it is an outstanding research
tool allowing continuous advancement of both fundamental and
applied sciences and it is used worldwide by tens of thousands of
scientists from many different disciplines like physics,
chemistry, material sciences and structural biology.

The properties of Synchrotron Radiation can be derived by applying
the methods of classical electrodynamics to the motion of
relativistic electrons (or positrons) in magnetic structures like
bending magnets or undulators. The theory of Synchrotron Radiation
based on the works by Iwanenko, Pomeranchuk  and Schwinger has
been described in detail both in convenient summaries \cite{TERN}
as well as in textbooks \cite{JACK,WIED,WIE2,HOFM,HOF2,PINO,DUKE}.

On the one hand, these texts contain derivation of fundamental
equations describing spectral-angular properties of Synchrotron
Radiation in the far-zone for single electrons. A usual starting
point for quantitative descriptions of these properties is
provided by the formulas for the Lienard-Wiechert fields
\cite{JACK}.

On the other hand, in relation with the construction of beam lines
at Synchrotron Radiation facilities it is important to understand
the modifications to single particle treatment which is due to
realistic beams with given cross section area, divergence and
energy spread.

This raises several questions. First, discussing finite electron
beam emittance effects, an obvious remark is that, in the case of
second generation light sources, the emittance of the electron
beam was much larger than the radiation wavelength of interest. As
a result, both photons and electrons phase space had the same
hamiltonian structure, and properties of radiation could be
treated on the basis of geometrical optics. In recent years
though, more and more specialized and optimized Synchrotron
Radiation sources have started operation. During design and
optimization of these sources much emphasis has been directed at
reducing the transverse phase space area of the electrons. Among
the most exciting properties of today's third generation sources
is the tiny vertical emittance which is an order of magnitude
smaller than $1~ \AA$ (for design parameters of an up-to-date
source see, for instance, \cite{PETR}). In this case geometrical
optics cannot be applied any more and even basic properties like
angular distribution of intensity at fixed frequency, or spectral
intensity distribution at fixed observation angle should be
calculated applying full electromagnetic equations.

A second question arises considering the features of undulator
insertion devices. In older sources, typical length of insertion
devices was about $1 \div 2$ m, while the distance to the user
station was of order $20\div30$ m. In this situation the
asymptotic formulas for the far zone could be applied and the
question of their applicability region was of theoretical interest
only. Nowadays installation of $20$ meter-long undulators has been
planned at PETRA III (see \cite{PETR}) and distances between
source and observer of several tens of meters have been proposed.
A similar device (a $25$ m long segmented undulator with no
focusing elements between the segments) has been installed and
operates at SPring-8 (see \cite{KITS}). In this situation
applicability region of far-field formulas and correction for near
field effects are of actual practical importance, not to mention
the interest for far infrared edge radiation.

Another problem is related with the increasing complexity of
insertion devices. For instance at PETRA III (see again
\cite{PETR}), installation of two undulators segments separated by
a focusing triplet in between has been proposed to obtain small
vertical betatron functions. Computation of radiation
characteristics from these novel devices constitute a rather
challenging problem.

Furthermore, a question which is somewhat related to all the
previous ones is linked with the dramatic increase of brilliance
achieved in third generation light sources with respect to older
designs, which has triggered a number of new techniques and
experiments unthinkable before. Among the most exciting properties
of today third generation facilities is the high flux of coherent
x-rays provided. The availability of intense coherent x-ray beams
has fostered the development of new coherence based techniques
like fluctuation correlation dynamics
\cite{THUR,MOCH,TSUI,RIES,SEYD,SIKH}, phase imaging
\cite{SNIG,WILK,GURE,CLOE}, and coherent x-ray diffraction (CXD)
\cite{MIAO,ROBI,PITN,LETO}. In all these fields, the understanding
of the evolution of transverse coherence properties along the beam
line is of uttermost importance: in particular, both the beam size
and divergence should be taken into account in the evolution
model.

As one can see these questions address quite different physical
phenomena; yet they all belong to the field of Synchrotron
Radiation. This number of different issues stems from the fact
that practical applications make use of a very wide range of
radiation wavelengths, from $0.1~\AA$ to $100 ~\mu$m, which span
over seven orders of magnitude. It is no surprise that very
different problems arise when the wavelength of interest is tuned
to such different regions of the electromagnetic spectrum.
Considering, for instance,  bending magnet radiation, it is clear
that near field effects will be of practical importance for FIR
(Far Infra-Red) applications but hardly for X-ray radiation,
because of its much shorter radiation formation length. On the
other hand, the influence of the emittance on X-ray radiation will
be important because the magnitude of the emittance is comparable
with the radiation wavelength, but it will be negligible when it
comes to the characterization of FIR radiation properties, since
in this case the wavelength of interest is way larger than the
emittance.

As a result, all the matters mentioned above are of great
practical importance in some particular region of the spectrum,
although they are systematically neglected in Synchrotron
Radiation textbooks and reviews. Reaction of
synchrotron-radiation-users communities to this set of problems
was the development of computer codes capable to calculate
radiation properties from realistic setups starting from first
principles \cite{CHUB, TANA}. At first glance this settles all
issues since, at least from a practical viewpoint, users have the
possibility to calculate all the properties they need.

Yet, computer codes can calculate properties for a given set of
parameters, but can hardly improve physical understanding, which
is particularly important in the stage of planning experiments.
Understanding of correct approximations and their region of
applicability can simplify many tasks a lot, including practical
and non-trivial ones. The use of dimensional analysis in an
analytical framework is of uttermost importance to this goal. By
that, dimensionless quantities of physical interest can be easily
identified. In particular, dimensionless parameters much smaller
(or much larger) than unity always correspond to simplifications
of fundamental equations: most physical theories originate from
recognizing and taking advantage of such small (or large)
quantities.

In contrast, relying only on computational power to solve physical
problems, besides being inelegant, would automatically neglect the
presence of small (or large) parameters and simplifications
related to them. The lack, or poor understanding of approximations
involved in a certain theory will result, in its turn, in poor
understanding of the mechanisms involved in physical phenomena
which would undoubtedly result in gross mistakes. Moreover, just a
few codes are available, which can be modified to account for
particular situations (like coherence properties or radiation
characteristics from a complex insertion device for instance) by a
few experts only, while there are about twenty facilities each
with its own set of particular setups and situation to be
accounted for.

We propose to overcome all these difficulties by developing a
theory of Synchrotron Radiation which accounts for all
complexities analytically. Analytical approach will help to
understand physics of real Synchrotron Radiation sources, thus
filling the gap between idealized textbook analysis and actual
situations. We will make explicit use of small (or large)
parameters involved in the description of cases of practical
interest. In this way we will reduce problems as much as possible,
to a level where physical insight is easy, so that simple computer
codes, capable to describe the situation under study, can be
developed also by non-expert programmers. In this process we will
put particular care in the specification of the region of
applicability and of the accuracy of the approximations made,
which is usually neglected in analytical calculations.


In the next Section we discuss methods for computation of
Synchrotron Radiation from a single particle on a generic
trajectory. We start reviewing the two algorithms used today, the
first based on Lienard-Wiechert fields, the second on
Lienard-Wiechert potentials. We propose to start
calculations from the very beginning by solving paraxial Maxwell
equations for a given harmonic of the field. This is, from a
logical and educational viewpoint, the most natural way to perform
calculations; from a computational viewpoint it is also the
simplest way. The use of a paraxial approximation is justified by
the features of an ultra relativistic system characterized by a
Lorentz factor $\gamma^2 \gg 1$: radiation formation length is,
then, much longer than the wavelength, and the radiation is
distributed within a cone with opening angle much smaller than
unity. Maxwell equation for a given harmonic of the field
$\bar{\boldsymbol{E}}(\boldsymbol{r},\omega)$, which has elliptic characteristic,
can be then transformed into a parabolic equation. We solve this
equation with the help of an appropriate Green's function thus
coming, in Section \ref{sec:gene}, to a very generic expression
for $\bar{\boldsymbol{E}}$. We end the Section by addressing the obvious
question of the relation between our method, and the two algorithm
introduced at the beginning.

Subsequently we apply our expression to recover some well-known
and less well-known properties of Synchrotron Radiation from
bending magnets and undulators, in Section \ref{sec:bend} and
Section \ref{sec:undu} respectively. This is far from being a mere
repetition of already known results. In fact our method will prove
superior when it comes to physical understanding and determination
of applicability region and accuracy of approximations: this has,
of course, important application in estimation of practical
quantities, like field intensities, \textit{with their accuracy}.
We will also discuss in detail the methodological issues about
definition of fields and intensities from single devices. As an
example of application of our new understanding we will show, in
Section \ref{sec:appl}, how radiation characteristics from a
complex setup can be discussed in very simple terms. Finally, in
Section \ref{sec:conc} we come to conclusions.


\section{\label{sec:gene} Method}
\subsection{\label{sub:rev} Review of known methods}

There are two basic methods which are used to calculate
Synchrotron Radiation characteristics.

The first \cite{JACK,WIED,WIE2,HOFM,HOF2,PINO,DUKE} is based on
Lienard-Wiechert fields. It is the better known of the two and it
is used in a very widespread way in textbooks. Lienard-Wiechert
fields can be customary separated in a velocity and an
acceleration term. Usually, the acceleration part alone is
analyzed in Fourier components of the electric field vector:

\begin{eqnarray}
\bar{\boldsymbol{E}}(\boldsymbol{r_o},\omega) = -{e\over{c}}
\int_{-\infty}^{\infty} dt'{\boldsymbol{n}\times
\left[\left(\boldsymbol{n}-\boldsymbol{\beta}\right)\times \boldsymbol{\dot{\beta}}
\right]\over{|\boldsymbol{r_o}-\boldsymbol{r'}(t')|\left(1-\boldsymbol{n}\cdot{\beta}\right)^2}}
&&\cr \times
\exp\left\{i\omega\left(t'+{|\boldsymbol{r_o}-\boldsymbol{r'}(t')|\over{c}}\right)\right\}~,
\label{rev1}
\end{eqnarray}
where $(-e)$ is the electron charge, $\boldsymbol{n}(t')$ is the unit
vector from the particle to the observer position, $\boldsymbol{r'}(t')$
is the particle position, $\boldsymbol{r_o}$ is the observer position
$\boldsymbol{\beta}(t')$ and $\boldsymbol{\dot{\beta}}(t')$ are, respectively, the
particle velocity, $\boldsymbol{v}(t')$, normalized to the speed of light
$c$ and its derivative, all calculated at time $t'$. It is clear
that use of Eq. (\ref{rev1}) implies that the so-called velocity
part of the field is neglected in the computation.

The second method \cite{CHU2,CHU3,CHUB,TANA} is based on
Lienard-Wiechert potentials. As an alternative to the use of
Lienard-Wiechert fields as a starting point for computations based
on first principles, a few authors of scientific research papers
start with the Lienard-Wiechert potentials
$(\boldsymbol{A}(\boldsymbol{r_o},t),\phi(\boldsymbol{r_o},t))$. These can be decomposed
in their harmonics
$(\bar{\boldsymbol{A}}(\boldsymbol{r_o},\omega),\bar{\phi}(\boldsymbol{r_o},\omega))$
which can be subsequently used to calculate
$\bar{\boldsymbol{E}}(\boldsymbol{r_o},\omega)$ according to

\begin{eqnarray}
\bar{\boldsymbol{E}}(\boldsymbol{r_o},\omega) = -{i\omega e\over{c}}
\int_{-\infty}^{\infty} dt' &&\cr \times
\left[{\boldsymbol{\beta}-\boldsymbol{n}\over{|\boldsymbol{r_o}-\boldsymbol{r'}(t')|}}-{ic\over{\omega}}{\boldsymbol{n}\over{
|\boldsymbol{r_o}-\boldsymbol{r'}(t')|^2}}\right]&&\cr \times
\exp\left\{i\omega\left(t'+{|\boldsymbol{r_o}-\boldsymbol{r'}(t')|\over{c}}\right)\right\}~.
\label{rev2}
\end{eqnarray}
We failed to find textbooks using this second method for
educational purposes: although some authors (see \cite{DUKE} for
instance) start their derivation with harmonic analysis of
potentials, when it comes to actual calculation of Synchrotron
Radiation they go back to the first method based on
Lienard-Wiechert fields.

Questions arise about the relation between Eq. (\ref{rev1}) and
Eq. (\ref{rev2}). First, in contrast to Eq. (\ref{rev1}), Eq.
(\ref{rev2}) is exact, in the sense that the velocity field term
is not neglected. It is then natural to ask what are the
conditions for which the velocity field term in Eq. (\ref{rev1})
can be dropped, and with what accuracy this can be done.  In fact,
when approximated expressions like Eq. (\ref{rev1}) and its
specialization to particular magnetic systems (like undulators and
bending magnets) are found, special care should be taken in
specifying their region of applicability and accuracy; yet we
failed to find, through references
\cite{JACK,WIED,WIE2,HOFM,HOF2,PINO,DUKE} quantitative
specification of accuracy of results obtained under given
approximations. Second, Eq. (\ref{rev1}) and Eq. (\ref{rev2}) are
used both in the far and in the near field region and simplified
under the paraxial approximation: it is important to understand
relations between the paraxial approximation and the far or near
field assumptions as well as region of applicability and accuracy
of results. Third, Eq. (\ref{rev1}) is often used together with
integration by parts and neglecting the edge terms after
integration: in this case it is interesting to ask what are the
conditions which allow the edge terms to be dropped. Moreover, as
a remark to both methods, we should note that, in general, one
needs to know the entire history of the electron from $t'=-\infty$
to $t'=\infty$ since integration in Eq. (\ref{rev1}) and Eq.
(\ref{rev2}) is performed between these limits. This statement
should be interpreted, physically, depending on the situation
under study: integration should in fact be performed from and up
to times when the electron does not contribute to the field
anymore. This does not present any problem in the case, for
instance, of a circular motion because the particle trajectory is
limited in space. 
In practical situation though, one deals with radiation from a
beamline which is a composition of insertion devices, like
undulators and bending magnets, and straight sections. The
requirement of integration over the entire history of the electron
poses, then, the following methodological problem: one has to give
a meaning to the concept of \textit{radiation from a given
insertion device}, independently of the trajectory followed by the
particle before and after the device. This problem will be
discussed in more detail as we go through our paper. In general
one can break up the electron trajectory in several parts,
corresponding to physical devices like undulators, drifts and
bending magnets, thus calculating the integral in Eq. (\ref{rev1})
or Eq. (\ref{rev2}) along a finite time interval which corresponds
to a particular device. In doing this, it is assumed that the
observer is located far away enough from the device, so that the
velocity part of the field is negligible with respect to the
acceleration part. In this way, again, dropping velocity field
part in Eq. (\ref{rev1}) is justified and the results from Eq.
(\ref{rev1}) and Eq. (\ref{rev2}) coincide. It is understood that
different contributions from different devices must be accounted
for with the correct relative phase relation to get the total
$\bar{\boldsymbol{E}}(\boldsymbol{r},\omega)$ as a sum of the separate
contributions from single devices. In this way concepts like
\textit{radiation field from a undulator} or \textit{radiation
field from a bending magnet} have a well defined, practically
useful meaning as partial contribution to the total field at the
observation point. The concept of \textit{intensity} is, instead,
subtler, since it involves calculation of $\mid
\bar{\boldsymbol{E}}(\boldsymbol{r},\omega)\mid^2$. In this case then, also
interference terms between different devices should be accounted
for, in the most generic case. Only when these are not important
the total spectral intensity from a given beamline can be broken
up in the sum of separate terms from the single devices, and
\textit{radiation intensity from a undulator} or \textit{radiation
intensity from a bending magnet} are, then, well-defined.

Finally, Eq. (\ref{rev1}) is based on the integration of Maxwell
equation in time and in the subsequent harmonic analysis. This is
quite involved from a logical viewpoint: it is much simpler to
start with the equations for a given harmonic $\omega$ so to
obtain directly $\bar{\boldsymbol{E}}$ after integration of Maxwell
equations. A similar comment can be made regarding the way Eq.
(\ref{rev2}) was obtained: first the potentials were found, then
their Fourier transform was taken. In both cases one is bound to
solve the full Maxwell equation even though the ultra relativistic
nature of the systems considered in practice allows systematic
application of a much more convenient paraxial approximation. Note
that, usually, paraxial approximation is applied, but only at a
later stage, which often makes analytical derivations more
involved, and region of applicability and accuracy of results
difficult to specify.

\subsection{\label{sub:deri} Derivation using paraxial Green's function}

A system of electromagnetic sources can be conveniently described
by its charge density $\rho(\boldsymbol{r},t)$ and current density
$\boldsymbol{j}(\boldsymbol{r},t)$. In this paper we will be concerned about a
single electron so that, using the Dirac delta distribution, we
can write

\begin{equation}
\rho(\boldsymbol{r},t) = -e \delta(\boldsymbol{r}-\boldsymbol{r'}(t)) \label{charge}
\end{equation}
and

\begin{equation}
\boldsymbol{j}(\boldsymbol{r},t) = -{e} \boldsymbol{v}(t) \delta(\boldsymbol{r}-\boldsymbol{r'}(t))~,
\label{curr}
\end{equation}
where $\boldsymbol{r}'(t)$ and $\boldsymbol{v}(t)$ are, respectively, the position
and the velocity of the particle at a given time $t$ in a fixed
reference frame. We will assume that the particle trajectory is
given $a~priori$, and that Maxwell equations in vacuum allow a
good description of the source field. In this case the magnetic
field $\boldsymbol{B}$ can be easily recovered when the electric field
$\boldsymbol{E}$ is known simply remembering

\begin{equation}
\boldsymbol{B}(\boldsymbol{r_0},t) = {1\over{c}} \boldsymbol{n}\times\boldsymbol{E} \label{field}
\end{equation}
where $\boldsymbol{n}$ is the unit vector pointing from the retarded
position of the particle, $\boldsymbol{r'}(t')$ to the position $\boldsymbol{r_0}$
at time $t=t'+r'(t')/c$. Therefore in this paper we will
concentrate on the electric field $\boldsymbol{E}(\boldsymbol{r_0},t)$.

Trivial manipulations of Maxwell equations in $cgs$ system:

\begin{equation}
\boldsymbol{\nabla} \times \boldsymbol{E} = - {1\over{c}}{\partial
\boldsymbol{B}\over{\partial t}} \label{M1}
\end{equation}
and

\begin{equation}
{\boldsymbol{\nabla}} \times \boldsymbol{B} =  {4 \pi\over{c}} \boldsymbol{j} +
{1\over{c}} {\partial \boldsymbol{E}\over{\partial t}}\label{M2}
\end{equation}
give the well-known equation for the electric field:

\begin{equation}
c^2 \boldsymbol{\nabla}\times(\boldsymbol{\nabla}\times{\boldsymbol{E}}) = - {\partial^2
\boldsymbol{E}\over{\partial t^2}} - 4 \pi {\partial \boldsymbol{j}\over{\partial
t}}~. \label{elec}
\end{equation}
With the help of the identity

\begin{equation}
\boldsymbol{\nabla}\times(\boldsymbol{\nabla}\times{\boldsymbol{E}}) =
\boldsymbol{\nabla}(\boldsymbol{\nabla}\cdot{\boldsymbol{E}})-\nabla^2 \boldsymbol{E}\label{iden}
\end{equation}
and Poisson equation
\begin{equation}
\boldsymbol{\nabla}\cdot\boldsymbol{E} = 4 \pi \rho \label{pois}
\end{equation}
we obtain the inhomogeneous wave equation for $\boldsymbol{E}$

\begin{equation}
c^2 \nabla^2 \boldsymbol{E} - {\partial^2 \boldsymbol{E}\over{\partial t^2}} = 4
\pi c^2 \boldsymbol{\nabla} \rho + 4 \pi {\partial \boldsymbol{j}\over{\partial
t}}~. \label{diso}
\end{equation}
Eq. (\ref{diso}) is a partial differential equation of
hyperbolic type. We introduce the Fourier transform
$\bar{f}(\omega)$ of a quantity $f(t)$ as follows:

\begin{equation}
\bar{f}(\omega) = \int_{-\infty}^{\infty} dt f(t) e^{i \omega t}~,
\label{ftran}
\end{equation}
so that
\begin{equation}
f(t) = {1\over{2\pi}} \int_{-\infty}^{\infty} d\omega
\bar{f}(\omega) e^{-i \omega t}~.
 \label{fanti}
\end{equation}
Using the representation in Eq (\ref{fanti}) for the quantities in
Eq. (\ref{diso}) we get

\begin{equation}
c^2 \nabla^2 \bar{\boldsymbol{E}} + \omega^2 \bar{\boldsymbol{E}} = 4 \pi c^2
\boldsymbol{\nabla} \bar{\rho} - 4 \pi i \omega \bar{\boldsymbol{j}}~.
\label{trdiso}
\end{equation}
Eq. (\ref{trdiso}) is the well-known Helmholtz equation which has
elliptic characteristic. Upon introduction of a fixed cartesian
reference system $(x,y,z)$, it is always possible to define a
quantity $\widetilde{\boldsymbol{E}}$ such that

\begin{equation}
\bar{\boldsymbol{E}} = \widetilde{\boldsymbol{E}} e^{i \omega z/c}~.
\label{tildaE}
\end{equation}
However this definition is useful only in the case when
$\widetilde{\boldsymbol{E}}$ varies slowly along $z$ with respect to the
length $\lambda = 2 \pi c/\omega$: only in that case, in fact, Eq.
(\ref{tildaE}) is a useful factorization of $\bar{\boldsymbol{E}}$ as the
product of a fast and a slowly varying function of $z$. With the
help of Eq. (\ref{tildaE}) one can write Eq. (\ref{trdiso}) as

\begin{equation}
c^2 e^{i\omega z/c} \left( \nabla^2 +{2 i
\omega\over{c}}{\partial\over{\partial z}}\right)
\widetilde{\boldsymbol{E}} = 4 \pi c^2 \boldsymbol{\nabla} \bar{\rho} - 4 \pi i
\omega \bar{\boldsymbol{j}}~. \label{trdiso2}
\end{equation}
Let us now write the equation for the sources with the help of the
curvilinear abscissa $s$, simply defined as $s = v t$.  Here we
assume that $v = \mid \boldsymbol{v}(t)\mid$ is a constant. We will use
the general property

\begin{equation}
\delta[f(x)]={\sum_i {\delta(x-x_i)\over{\mid f'(x_i)\mid}}}~,
\label{diracd}
\end{equation}
where $x_i$ are the zeros of $f(x)$. Then, we will regard
$\delta(z-z'(t))$ as $\delta[f(t)]$, for any fixed value of $z$.
The only zero of $f(t)$ will be readily indicated with $t(z)$,
which is the value of $t$ for which $z-z'(t)=0$, while $\mid
f'(t(z))\mid=v_z(z(t))$. Therefore, with the help of Eq.
(\ref{diracd}) we can write

\begin{equation}
\delta(z-z'(t)) = {1\over{v_z(z)}} \delta(t-t(z))~.
\label{middledirac}
\end{equation}
Then, using $s(z)=vt(z)$, Eq. (\ref{charge}) and Eq. (\ref{curr})
give

\begin{equation}
\rho(\boldsymbol{r_\bot},z,t) = -{e\over{v_z(z)}}
\delta\left(\boldsymbol{r_\bot}-\boldsymbol{r'_\bot}(z)\right)
\delta\left({s(z)\over{v}}-t\right) \label{charge2}
\end{equation}
and

\begin{equation}
\boldsymbol{j}(\boldsymbol{r_\bot},z,t) = -{e\over{v_z(z)}} \boldsymbol{v}(z)
\delta\left(\boldsymbol{r_\bot}-\boldsymbol{r'_\bot}(z)\right)
\delta\left({s(z)\over{v}}-t\right) \label{curr2}
\end{equation}
so that

\begin{equation}
\bar{\rho}(\boldsymbol{r_\bot},z,\omega) = -{e\over{v_z(z)}}
\delta\left(\boldsymbol{r_\bot}-\boldsymbol{r'_\bot}(z)\right) e^{i \omega s(z)/v}
\label{charge2tr}
\end{equation}
and

\begin{equation}
\boldsymbol{\bar{j}}(\boldsymbol{r_\bot},z,\omega) = -{e\over{v_z(z)}} \boldsymbol{v}(z)
\delta\left(\boldsymbol{r_\bot}-\boldsymbol{r'_\bot}(z)\right) e^{i \omega
s(z)/v}~. \label{curr2tr}
\end{equation}
By substitution in Eq. (\ref{trdiso2}) we obtain

\begin{eqnarray}
\left({\nabla}^2 + {2 i \omega \over{c}} {\partial\over{\partial
z}}\right) \widetilde{\boldsymbol{E}} = {4 \pi e\over{v_z(z)}}
\exp\left\{{i \omega
\left({s(z)\over{v}}-{z\over{c}}\right)}\right\} && \cr \times
\left[{i\omega\over{c^2}}\boldsymbol{v}(z)
\delta\left(\boldsymbol{r_\bot}-\boldsymbol{r'_\bot}(z)\right)-\boldsymbol{\nabla}
\delta\left(\boldsymbol{r_\bot}-\boldsymbol{r'_\bot}(z)\right)
\right]~.\label{incipit}
\end{eqnarray}
Eq. (\ref{incipit}) is still fully general and may be solved in
any fixed reference system $(x,y,z)$ of choice with the help of an
appropriate Green's function. In general, $\widetilde{\boldsymbol{E}}$
will vary, in $z'$, on a characteristic length which is related
with $\exp\left\{{i \omega \left({s/{v}}-{z/{c}}\right)}\right\}$,
which enters in the Green's function solution of Eq.
(\ref{incipit}) as a factor in the integrand. As it will be
clearer after Paragraph \ref{sub:disc}, as we integrate along
$z'$, the factor $\omega(s(z')/v - z'/c)$ grows larger and larger
leading, eventually, to a highly oscillatory behavior of the
integrand which does not contribute anymore to the final
integration result. The value of $z'$ for which $\omega(s(z')/v -
z'/c) \sim 1$ can be considered as a measure of the value of $z'$
for which the integrand starts to display such oscillatory
behavior and it is naturally defined as the radiation formation
length $L_f$ of the system at frequency $\omega$. It is easy to
see by inspection that if $v$ is sensibly smaller than $c$ (but
still of order $c$), i.e. $v\sim c$ but $1/\gamma^2 \sim 1$, then
$L_f \sim \lambda$. On the contrary, when $v$ is very close to
$c$, i.e. $1/\gamma^2 \ll 1$, the terms $\omega s(z')/v$ and  $-
\omega z'/c$ tend to compensate so that $L_f \gg \lambda$. The
exact expression for $L_f$ depends on the particular magnetic
system, i.e. on $s(z')$, and on the frequency of interest
$\omega$.

Let us consider the assumption $1/\gamma^2 \ll 1$. We will hold
this assumption fulfilled throughout our paper. In general,
introduction of a small (or large) parameter in any theory brings
simplifications. In particular the ultra relativistic
approximation has two consequences: first, as just remarked, the
radiation formation length is much larger than $\lambda$ and,
second, as it will be discussed in Paragraph \ref{sub:disc},
radiation is emitted in a narrow cone of opening angle much
smaller than unity. Accounting for these features of the
radiation, and considering an electron moving on a given
trajectory, there will always be, in practical cases of interest,
some privileged reference system $(x,y,z)$ in which Eq.
(\ref{incipit}) is simplified in a paraxial form, due to the ultra
relativistic assumption, for some set of observer points.

\begin{figure}
\begin{center}
\includegraphics*[width=90mm]{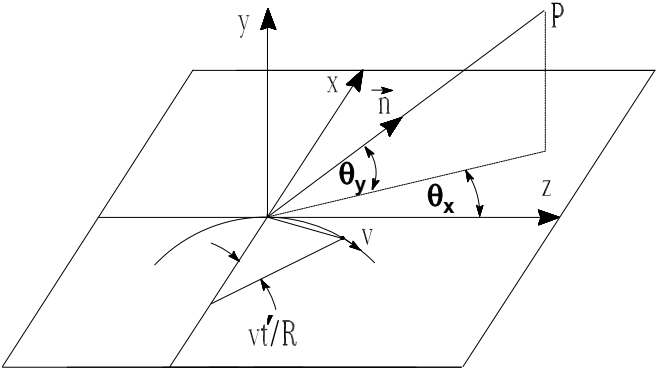}
\caption{\label{srgeo} Geometry and reference system for
Synchrotron Radiation problems in paraxial approximation.}
\end{center}
\end{figure}
\begin{figure}
\begin{center}
\includegraphics*[width=90mm]{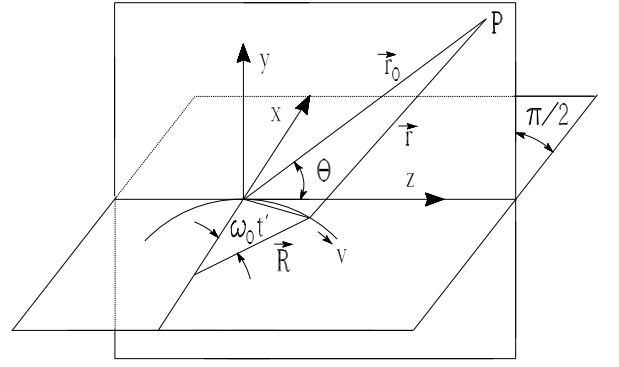}
\caption{\label{srstand} Standard reference frame for Synchrotron
Radiation computations.}
\end{center}
\end{figure}

To apply such paraxial approximation we should first assume ultra
relativistic motion and identify a main observer $P$. Then we can
define a frame with the $z$ axis oriented as the tangent from some
trajectory point $T$, to $P$. As it will be discussed in Paragraph
\ref{sub:disc}, using this frame we can describe radiation at $P$
from all points of the trajectory with velocity vector forming
with the $z$ axis an angle much smaller than unity. In other
words, for all these points $\gamma_z^2 \gg 1$: this condition is
necessary and sufficient for the applicability of the paraxial
approximation. This is not a restriction, in practice, because it
still keeps open the possibility of angles much larger or much
smaller than $1/{\gamma}$ (bending magnet or undulator case). The
situation is described with a particular example for the case of a
circular motion, in Fig. \ref{srgeo}. Note that the reference
frame used to describe radiation at $P$ can be also used for
observers in the neighborhood of $P$, provided that these form an
angle much smaller than unity with respect to the velocity at $T$.
Since we presented, as particular example, the case of a circular
motion, it is worth to underline the difference between the frame
depicted in Fig. \ref{srgeo} and the standard frame used for
Synchrotron Radiation computations in Fig. \ref{srstand}: using
the frame in Fig. \ref{srstand} we will never be able to describe
the field at to observer locations displaced along the $x$ axis.
This is required, for instance, if one needs to calculate the
autocorrelation function of the field at those two points. On the
contrary, such a description is allowed with the choice of a frame
like in Fig. \ref{srgeo}.

%
Since the radiation formation length is much longer than the
wavelength, $\widetilde{\boldsymbol{E}}$ does not vary much along $z$ on
the scale of $\lambda$, that is $\mid
\partial_z \widetilde{E}_{x,y}\mid \ll \omega/c \mid
\widetilde{E}_{x,y}\mid$. Therefore, the second order derivative
with respect to $z$ in the $\nabla^2$ operator on the left hand
side of Eq. (\ref{incipit}) is negligible with respect to the
first order derivative. This means that we can apply a paraxial
approximation which considerably simplifies Eq. (\ref{incipit}) as

\begin{eqnarray}
\left({\nabla_\bot}^2 + {2 i \omega \over{c}}
{\partial\over{\partial z}}\right) \widetilde{\boldsymbol{E}}_\bot = {4
\pi e\over{c}} \exp\left\{{i \omega
\left({s(z)\over{v}}-{z\over{c}}\right)}\right\} && \cr \times
\left[{i\omega\over{c^2}}\boldsymbol{v_\bot}(z)
\delta\left(\boldsymbol{r_\bot}-\boldsymbol{r'_\bot}(z)\right)-\boldsymbol{\nabla_\bot}
\delta\left(\boldsymbol{r_\bot}-\boldsymbol{r'_\bot}(z)\right)
\right]~,\label{incipit2}
\end{eqnarray}
where we consider transverse components of $\widetilde{\boldsymbol{E}}$
only and we substituted $v_z(z)$ with $c$, having used the fact
that $1/\gamma_z^2 \ll 1$. Eq. (\ref{incipit2}) is Maxwell's
equation in paraxial approximation. Note that this approximation
transformed Eq. (\ref{incipit}) which is an elliptic partial
differential equation, into Eq. (\ref{incipit2}), which is of
parabolic type.

The Green's function for Eq. (\ref{incipit2}), namely the solution
corresponding to the unit point source, satisfies the equation:

\begin{eqnarray}
\left({\nabla_\bot}^2 + {2 i \omega \over{c}}
{\partial\over{\partial z}}\right) G(z_o-z;\boldsymbol{r_{\bot
o}}-\boldsymbol{r_\bot}) &&\cr =  \delta(\boldsymbol{r_{\bot
o}}-\boldsymbol{r_\bot})\delta(z_o-z)~, \label{greeneq}
\end{eqnarray}
and, in an unbounded region, can be written explicitly as

\begin{eqnarray}
G(z_o-z';\boldsymbol{r_{\bot o}}-\boldsymbol{r'_\bot}) = -{1\over{4\pi (z_o-z')}}
&&\cr \times \exp\left\{ i\omega{\mid \boldsymbol{r_{\bot o}}
-\boldsymbol{r'_\bot}\mid^2\over{2c (z_o-z')}}\right\} \label{green}~.
\end{eqnarray}
%
As usual we will denote source coordinates with \textit{primes}
and observer coordinates with the index \textit{o}. With the aid
of the Green's function $G$, the solution of Eq. (\ref{incipit2})
can be represented as

\begin{eqnarray}
\widetilde{\boldsymbol{E}}_\bot(z_o, \boldsymbol{r_{\bot o}},\omega)= -{e\over{c}}
\int_{-\infty}^{\infty} dz' {1\over{z_o-z'}} \int d \boldsymbol{r'_{\bot}}
&&\cr \times \left[{i\omega\over{c^2}}\boldsymbol{v_\bot}(z')
\delta\left(\boldsymbol{r'_{\bot}}-\boldsymbol{r'_\bot}(z')\right)
-\boldsymbol{\nabla'_\bot} \delta\left(\boldsymbol{r'_{\bot}
}-\boldsymbol{r'_\bot}(z')\right)\right] &&\cr \times
\exp\left\{i\omega\left[{\mid \boldsymbol{{r_{\bot o}}}-\boldsymbol{r'_\bot}
\mid^2\over{2c (z_o-z')}}+
\left({s(z')\over{v}}-{z'\over{c}}\right)\right] \right\} ~,&& \cr
\label{blob}
\end{eqnarray}
where $\boldsymbol{\nabla'_\bot}$ represents the gradient operator with
respect to the source point. The integration over transverse
coordinates can be carried out leading to the final result:

\begin{eqnarray}
\widetilde{\boldsymbol{E}}_\bot(z_o, \boldsymbol{r_{\bot o}},\omega)= -{i \omega
e\over{c^2}} \int_{-\infty}^{\infty} dz' {e^{i
\Phi_T}\over{z_o-z'}}  \left[\left({v_x(z')\over{c}}
\right.\right.&& \cr\left.\left.
-{x_o-x'(z')\over{z_o-z'}}\right)\boldsymbol{\hat{x}}
+\left({v_y(z')\over{c}}-{y_o-y'(z')\over{z_o-z'}}\right)\boldsymbol{\hat{y}}\right]
~, \label{generalfin}
\end{eqnarray}
where the total phase $\Phi_T$ is given by

\begin{equation}
\Phi_T = \omega \left[{s(z')\over{v}}-{z'\over{c}}\right]+ \omega
\left[
{\left(x_o-x'(z')\right)^2+\left(y_o-y'(z')\right)^2\over{2c
(z_o-z')}}\right] \label{totph}
\end{equation}
Eq. (\ref{generalfin}) can be used in all generality to
characterize radiation from an electron moving on any trajectory
as long as the ultra relativistic approximation is satisfied, and
a reference system suitable for paraxial approximation exists,
which is always the case in situation of practical interest. Note
that both $\boldsymbol{\hat{x}}$ and $\boldsymbol{\hat{y}}$ polarizations terms in
Eq. (\ref{generalfin}) are a sum of two parts which can be traced
back to current and charge densities: in fact, the first part,
proportional to the particle velocity in the $x$ or $y$ direction,
follows from the transverse current density $\boldsymbol{j_{\bot}}$. The
second instead, follows from the gradient of the charge density
$\boldsymbol{\nabla} \rho$. Our result makes a consistent use of paraxial
approximation and of harmonic analysis which brings simplicity and
power to the method. In the following Sections we will show how
Eq. (\ref{generalfin}) can be used to characterize bending magnet
radiation and undulator radiation.

\subsection{\label{sub:disc} Discussion}

Following the previous derivation very natural questions arise
which regard the relation between our expression, Eq.
(\ref{generalfin}), with Eq. (\ref{rev1}) and Eq. (\ref{rev2}) as
well as the applicability region and the accuracy of our Eq.
(\ref{generalfin}).

In order to investigate these subjects we go back to the most
general Eq. (\ref{trdiso}) and we note that it can be solved by
direct application of the Green's function for the Helmholtz
equation:

\begin{eqnarray}
G(\boldsymbol{r_{o}}-\boldsymbol{r'}) = -{1\over{4\pi
\left|\boldsymbol{r_{o}}-\boldsymbol{r'}\right| }} \exp\left\{
i{\omega\over{c}}{\mid \boldsymbol{r_{o}} -\boldsymbol{r'}\mid}\right\}
\label{greenhyp}~.
\end{eqnarray}
Integrating by parts the term in $\boldsymbol{\nabla}\bar{\rho}$ we have

\begin{eqnarray}
\boldsymbol{\bar{E}} = -\int d\boldsymbol{r'} \left[{i\omega \over{c
\left|\boldsymbol{r_{o}}-\boldsymbol{r'}\right| }} \left(\bar{\rho} \boldsymbol{n} -
{\boldsymbol{\bar{j}}\over{c}}\right)+{\bar{\rho}
\boldsymbol{n}\over{\left|\boldsymbol{r_{o}}-\boldsymbol{r'}\right|^2}}\right]&&\cr\times
\exp\left\{ i{\omega\over{c}}{\mid \boldsymbol{r_{o}}
-\boldsymbol{r'}\mid}\right\} \label{hypE}~.
\end{eqnarray}
Use of explicit expressions for $\bar{\rho}$, Eq.
(\ref{charge2tr}), and $\boldsymbol{\bar{j}}$, Eq. (\ref{curr2tr}), leads
straightforwardly to

\begin{eqnarray}
\bar{\boldsymbol{E}}(\boldsymbol{r_o},\omega) = -{i\omega e\over{c}}
\int_{-\infty}^{\infty} dz' &&\cr \times
{1\over{v_z(z')}}\left[{\boldsymbol{\beta}-\boldsymbol{n}\over{|\boldsymbol{r_o}-\boldsymbol{r'}(z')|}}-{ic\over{\omega}}{\boldsymbol{n}\over{
|\boldsymbol{r_o}-\boldsymbol{r'}(z')|^2}}\right]&&\cr \times
\exp\left\{i\omega\left({s(z')\over{v}}+
{|\boldsymbol{r_o}-\boldsymbol{r'}(z')|\over{c}}\right)\right\}~. \label{rev2zeta}
\end{eqnarray}
Finally, remembering $z'=v_z t'$ we obtain Eq.
(\ref{rev2}). 
This result simply confirms that Eq. (\ref{rev2}) is derived under
most general conditions; the derivation that we proposed from
direct use of fields  is much less involved, from a logical
viewpoint, with respect to the original, which makes use of
potentials (see \cite{CHU2,CHU3}) and it may be interesting for
educational purposes.


It is interesting to note that the only approximation applied to
Eq. (\ref{trdiso}) in order to obtain, finally, Eq.
(\ref{generalfin}) was simply the paraxial approximation. We can
easily see that if we apply paraxial approximation to Eq.
(\ref{rev2}) we get back Eq. (\ref{generalfin}). Let us show how
this is possible.

The choice of $\omega$ depends on our interest. Then, our ultra
relativistic approximation implies a formation length $L_f \gg
\lambda$. Moreover in practical cases, $|\boldsymbol{r_o}-\boldsymbol{r'}|$ will
always be at least of order $L_f$, so that $|\boldsymbol{r_o}-\boldsymbol{r'}| \gg
c/\omega$ and, with an accuracy $\lambda/(2\pi L_f)$, Eq.
(\ref{rev2}) can be simplified as

\begin{eqnarray}
\bar{\boldsymbol{E}}(\boldsymbol{r_o},\omega) = -{i\omega e\over{c}}
\int_{-\infty}^{\infty}
dt'{\boldsymbol{\beta}-\boldsymbol{n}\over{|\boldsymbol{r_o}-\boldsymbol{r'}(t')|}} &&\cr \times
\exp\left\{i\omega\left(t'+{|\boldsymbol{r_o}-\boldsymbol{r'}(t')|\over{c}}\right)\right\}~
\label{rev3}
\end{eqnarray}
or, using again $z'=v_z t'$, $v_z \simeq c$ and also $s = v t'$

\begin{eqnarray}
\bar{\boldsymbol{E}}(\boldsymbol{r_o},\omega) = -{i\omega e\over{c^2}}
\int_{-\infty}^{\infty}
dz'{\boldsymbol{\beta}-\boldsymbol{n}\over{|\boldsymbol{r_o}-\boldsymbol{r'}(z')|}} &&\cr \times
\exp\left\{i\omega\left({s(z')\over{v}}+
{|\boldsymbol{r_o}-\boldsymbol{r'}(z')|\over{c}}\right)\right\}~. \label{rev4}
\end{eqnarray}
Remembering that $\boldsymbol{\bar{E}_\bot}=\boldsymbol{\widetilde{E}_\bot}
\exp\{i\omega z_o/c\}$ we write the transverse field components as

\begin{eqnarray}
{\boldsymbol{\widetilde{E}_\bot}}(\boldsymbol{r_o},\omega) = -{i\omega
e\over{c^2}} \int_{-\infty}^{\infty}
dz'{\boldsymbol{\beta_\bot}-\boldsymbol{n_\bot}\over{|\boldsymbol{r_o}-\boldsymbol{r'}(z')|}}
&&\cr \times
\exp\left\{i\omega\left[\left({s(z')\over{v}}-{z'\over{c}}\right)+
\right.\right.&&\cr\left.\left.
\left({|\boldsymbol{r_o}-\boldsymbol{r'}(z')|\over{c}}-{z_o-z'\over{c}}\right)\right]\right\}~.
\label{rev444}
\end{eqnarray}
In any situation, it is \textit{mathematically} correct
to expand $|\boldsymbol{r_o}-\boldsymbol{r'}|$ as an \textit{infinite sum} of
terms

\begin{equation}
|\boldsymbol{r_o}-\boldsymbol{r'}| \simeq (z_o-z') + {|\boldsymbol{r_{\bot
o}}-\boldsymbol{r'_\bot}|^2\over{2(z_o-z')}}+...\label{parexpr}
\end{equation}
Truncation of the series in Eq. (\ref{parexpr}) and subsequent
simplification of the phase in the integrand of Eq. (\ref{rev444})
is a delicate business.

The first term of the expansion Eq. (\ref{parexpr}) naturally
cancels the term $-i\omega(z_o-z')/c$ in the phase of Eq.
(\ref{rev444}) so that one is left with phase contributions due to
higher order terms from the expansion Eq. (\ref{parexpr}) and with
$i\omega(s(z')/v - z'/c)$. This last contribution depends on
$s(z')$ and can be specified only when the magnetic system is
specified. As has already been said in Paragraph \ref{sub:deri},
as we integrate along $z'$, $\omega(s(z')/v - z'/c)$ grows larger
and larger leading, eventually, to a highly oscillatory behavior
of the integrand which does not contribute anymore to the final
integration result. The value of $z'$ for which $\omega(s(z')/v -
z'/c) \sim 1$ can be considered as a measure of the value of $z'$
for which the integrand starts to display such oscillatory
behavior and it is naturally defined as the radiation formation
length $L_f$ of the system at frequency $\omega$.


Another length dictated by the physics of the problem is the
natural size of the system. We will refer to it as the
characteristic length of the system and we will indicate it with
$L_{ch}$; for instance, in the case of a circular motion, $L_{ch}$
is simply the circle radius.


Relation between $L_{ch}$ and $L_f$ depends on the system. In the
ultra relativistic approximation we can say that $L_{ch}$ can be
either much larger or comparable to $L_f$, but in any case never
smaller, since integration of the Green's function is performed
along a path length comparable with $L_{ch}$. Moreover, in the
ultra relativistic approximation the two terms $\omega s(z')/v$
and - $\omega z'/c$ nearly compensate leading to a formation
length much longer than $\lambda$ so that $L_f \gg \lambda$, and
therefore also $L_{ch} \gg \lambda$. It should be stressed that,
although the previous properties are very general, the relation
between $L_f$, $L_{ch}$ and $\lambda$ depends on the particular
physical situation under study and can be better specified only in
relation with that situation: for instance, in the case of ultra
relativistic motion on a circular trajectory  $L_f = [\lambda
R^2/(2\pi)]^{1/3}$. Also, note that if the ultra relativistic
approximation cannot be applied anymore, then even the general
scaling laws between $L_{ch}$, $L_f$ and $\lambda$ change. For
instance, when if $v\sim c$ but $\gamma \sim 1$, one has $L_{ch}
\sim L_f \sim \lambda$.

The expansion Eq. (\ref{parexpr}) makes sense only if
$\omega(s(z')/v - z'/c)$ is smaller or comparable to unity since
the two terms $\omega s(z')/v$ and $- \omega z'/c$ can be grouped
together in a useful way. Once a certain wavelength of interest is
fixed, the condition $\omega(s(z')/v - z'/c) \sim 1$ determines
the formation length $L_f$, which can be used in the second term
of Eq. (\ref{parexpr}). By imposing that also this second term is
not larger than unity one obtains 
a condition on the observation points of interest:

\begin{equation}
{|\boldsymbol{r_{\bot o}}-\boldsymbol{r'_\bot}|^2\over{2(z_o-z')^2}}\lesssim
{\lambda\over{2\pi L_f}}\ll 1 ~,\label{sectermcondb}
\end{equation}
where we assumed $z_o-z' \gtrsim L_f$. The third term in the
expansion can be then neglected, together with second and next
terms in the expansion of the denominator of the integrand in Eq.
(\ref{rev444}), with an accuracy ${\lambda/(2\pi L_f)}$. Note
that, whatever the magnitude of $\boldsymbol{r'_\bot}$ across the
beamline, it follows from Eq. (\ref{sectermcondb}) that the square
of the opening angle of radiation $r_{\bot o}^2/z_o^2$, is much
smaller than unity. This justify our attention to the transverse
components of $\boldsymbol{\widetilde{E}}$ only.

If we substitute the first two terms of Eq. (\ref{parexpr}) in the
phase of the integrand of Eq. (\ref{rev444}), and the first term
in its denominator we find

\begin{eqnarray}
\boldsymbol{\widetilde{E}_\bot}(\boldsymbol{r_o},\omega) = -{i\omega e\over{c^2}}
\int_{-\infty}^{\infty}
dz'{\boldsymbol{\beta_\bot}-\boldsymbol{n_\bot}\over{z_o-z'}} &&\cr \times
\exp\left\{i\omega\left({s\over{v}}-{z'\over{c}} + {|\boldsymbol{r_{\bot
o}} -\boldsymbol{r'_\bot}|^2\over{2c(z_o-z')}}\right)\right\}~.
\label{rev5}
\end{eqnarray}
that is just Eq. (\ref{generalfin}).

We demonstrated that application of paraxial approximation to Eq.
(\ref{rev2}) gives us back Eq. (\ref{generalfin}). This fact is,
however, quite obvious. In our derivation of Eq.
(\ref{generalfin}) we applied paraxial approximation to Maxwell
equations from the very beginning, while here we simply apply the
same approximation \textit{after} solving Maxwell equations in
their more generic form. For consistency reasons we had to obtain
the same result.

The previous discussion tells that our approach is a
specialization of the more general approach described by Eq.
(\ref{rev2}). The novelty in this, stems from the fact that
Synchrotron Radiation, by definition, applies to radiation by
ultrarelativistic particles in magnetic structures; in Paragraph
\ref{sub:deri} we demonstrated that paraxial approximation can
\textit{always} be applied to Maxwell equations in the case of
ultrarelativistic particles. Eq. (\ref{rev2}) can \textit{always}
be reduced to Eq. (\ref{generalfin}) in the treatment of
Synchrotron Radiation. In other words, there is no point in
starting from Eq. (\ref{rev2}) when treating Synchrotron Radiation
problems: one may start, directly, with the simpler Eq.
(\ref{generalfin}) without loss of generality.

Compared to Eq. (\ref{rev2}), Eq. (\ref{generalfin}) is easier to
apply for analytical investigations, whose importance we already
stressed in the introduction, because the paraxial approximation
has been applied from the very beginning, so that the
approximations made, their region of applicability and their
accuracy could be discussed independently of the magnetic system
selected.

The relation between our approach and the method in Eq.
(\ref{rev1}) simplifies then, to the relation between Eq.
(\ref{rev2}) and Eq. (\ref{rev1}). It is not straightforward to
show this relation because Eq. (\ref{rev2}) does not coincide with
Eq. (\ref{rev1}) once the term in $| \boldsymbol{r_{o}} -\boldsymbol{r'}|^{-2}$ is
dropped. However, it is straightforward to show that Eq.
(\ref{rev2}) is completely equivalent to the expression for the
Fourier transform of the  Lienard-Wiechert fields
\textit{including} the velocity field part. This can be shown in
full generality, regardless of the paraxial approximation. Eq.
(\ref{rev2}) can be written as

\begin{eqnarray}
\bar{\boldsymbol{E}}(\boldsymbol{r_o},\omega) = -{ e} \int_{-\infty}^{\infty}
dt'{\boldsymbol{n}\over{|\boldsymbol{r_o}-\boldsymbol{r'}(t')|^2}} &&\cr \times \exp
\left\{i\omega\left(t'+{|\boldsymbol{r_o}-\boldsymbol{r'}(t')|\over{c}}\right)\right\}
&&\cr- {e\over{c}} \int_{-\infty}^{\infty}
dt'{\boldsymbol{\beta}-\boldsymbol{n}\over{({1-\boldsymbol{n}\cdot\boldsymbol{\beta}})|\boldsymbol{r_o}-\boldsymbol{r'}(t')|}}
&&\cr \times {d\over{dt'}}\exp
\left\{i\omega\left(t'+{|\boldsymbol{r_o}-\boldsymbol{r'}(t')|\over{c}}\right)\right\}~,
\label{revtrasf}
\end{eqnarray}
where we have used relation

\begin{eqnarray}
{1\over{c}}{d\over{dt'}}|\boldsymbol{r_o}-\boldsymbol{r'}(t')| =
-\boldsymbol{n}\cdot\boldsymbol{\beta}~. \label{usefulrel}
\end{eqnarray}
Eq. (\ref{revtrasf}) can be integrated by parts. When the edge
terms can be dropped (we will discuss this assumption below) one
obtains

\begin{eqnarray}
\bar{\boldsymbol{E}}(\boldsymbol{r_o},\omega) = -{ e} \int_{-\infty}^{\infty}
dt'{\boldsymbol{n}\over{|\boldsymbol{r_o}-\boldsymbol{r'}(t')|^2}} &&\cr \times \exp
\left\{i\omega\left(t'+{|\boldsymbol{r_o}-\boldsymbol{r'}(t')|\over{c}}\right)\right\}
&&\cr+ {e\over{c}} \int_{-\infty}^{\infty} dt'{d\over{dt'}}\left[
{\boldsymbol{\beta}-\boldsymbol{n}\over{({1-\boldsymbol{n}\cdot\boldsymbol{\beta}})|\boldsymbol{r_o}-\boldsymbol{r'}(t')|}}\right]
&&\cr \times \exp
\left\{i\omega\left(t'+{|\boldsymbol{r_o}-\boldsymbol{r'}(t')|\over{c}}\right)\right\}
\label{revtrasfbiss}
\end{eqnarray}

With the help of Eq. (\ref{usefulrel}) and

\begin{eqnarray}
{d\boldsymbol{n}\over{dt'}} = {c\over{|\boldsymbol{r_o}-\boldsymbol{r'}(t')|}}
\left[-\boldsymbol{\beta}+\boldsymbol{n}\left(\boldsymbol{n}\cdot\boldsymbol{\beta}\right)\right]
\label{usefulrel0}
\end{eqnarray}
one obtains, from Eq. (\ref{revtrasfbiss}) the Fourier transform
of the Lienard-Wiechert fields:

\begin{eqnarray}
\bar{\boldsymbol{E}}(\boldsymbol{r_o},\omega) = -e\int_{-\infty}^{\infty} dt'
\left[{\boldsymbol{n}-\boldsymbol{\beta}\over{\gamma^2 (1-\boldsymbol{n}\cdot\boldsymbol{\beta})^2
|\boldsymbol{r_o}-\boldsymbol{r'}|^2}}\right] && \cr \times \exp
\left\{i\omega\left(t'+{|\boldsymbol{r_o}-\boldsymbol{r'}(t')|\over{c}}\right)\right\}
&&\cr-{e
\over{c}}\int_{-\infty}^{\infty}dt'\left[{1\over{(1-\boldsymbol{n}\cdot\boldsymbol{\beta})^2}}
{\boldsymbol{n}\times[(\boldsymbol{n}-\boldsymbol{\beta})\times
\boldsymbol{\dot{\beta}}]\over{|\boldsymbol{r_o}-\boldsymbol{r'}|}}  \right] && \cr \times
\exp
\left\{i\omega\left(t'+{|\boldsymbol{r_o}-\boldsymbol{r'}(t')|\over{c}}\right)\right\}~.
\label{LW}
\end{eqnarray}
The only assumption made going from Eq. (\ref{revtrasf}) to Eq.
(\ref{LW}) is that the edge term in the integration by parts is
simply zero. This assumption can be justified by means of physical
arguments in the most general situation accounting for the fact
that the integral in $dt'$ has to be performed over the entire
history of the particle and that at $t'=-\infty$ and $t'=+\infty$
the electron does not contribute to the field anymore. It is
obvious that the same line of reasoning can be followed starting
from Eq. (\ref{LW}) and going back to Eq. (\ref{revtrasf}): in
general, edge terms can be dropped.

\begin{figure}
\begin{center}
\includegraphics*[width=90mm]{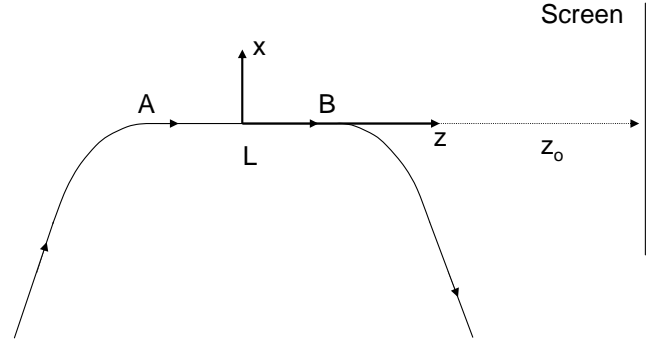}
\caption{\label{edgeuno} A system constituted by two bending
magnets connected by a straight section of length $L$. Radiation
is collected at a distance $z_o$ from the downstream magnet. }
\end{center}
\end{figure}
The previous statement is very general and \textit{per se} trivial
but it should be interpreted from a physical viewpoint depending
on the situation.

For instance, consider the case of a system like the one sketched
in Fig. \ref{edgeuno} (which will be treated extensively in
Paragraph \ref{sec:edge}).  Fig. \ref{edgeuno} shows two bending
magnets separated by a straight line. Before and after the
straight lines are two semi-infinite straight sections. A question
arises about what we may take as $t'=-\infty$ and $t'=+\infty$ in
this particular situation. Intuitively the magnets act like
switches: the first magnet switches radiation on, the second
switches it off. The statement that edge terms can be dropped,
combined with the paraxial approximation tell that we can take
$t'=-\infty$ and $t'=+\infty$ as the particle is well inside the
first and the second bend, respectively, and neglect other parts
of the trajectory.

Note that an expression alternative to Eq. (\ref{rev2}) for the
Fourier transform of the electric fields can be found in
\cite{LUCC}. After starting with Eq. (\ref{LW}), the authors of
\cite{LUCC} organized integration by part in a different way
compared with what has been done in Eq. (\ref{revtrasfbiss}).
First they found that

\begin{eqnarray}
{\boldsymbol{n}\times[(\boldsymbol{n}-\boldsymbol{\beta})\times\boldsymbol{\dot{\beta}}]
\over{|\boldsymbol{r_o}-\boldsymbol{r'}|(1-\boldsymbol{n}\cdot\boldsymbol{\beta})^2}} =
{1\over{|\boldsymbol{r_o}-\boldsymbol{r'}|}} {d\over{dt'}}
\left[{\boldsymbol{n}\times(\boldsymbol{n}\times\boldsymbol{\beta})\over{(1-\boldsymbol{n}\cdot\boldsymbol{\beta})}}\right]
&& \cr
-\left[{\boldsymbol{\dot{n}}(\boldsymbol{n}\cdot\boldsymbol{\beta})+\boldsymbol{n}(\boldsymbol{\dot{n}}\cdot\boldsymbol{\beta})
-
\boldsymbol{\dot{n}}(\boldsymbol{n}\cdot\boldsymbol{\beta})^2-\boldsymbol{\beta}(\boldsymbol{\dot{n}}\cdot\boldsymbol{\beta})
\over{|\boldsymbol{r_o}-\boldsymbol{r'}|(1-\boldsymbol{n}\cdot\boldsymbol{\beta})^2}}\right]~.
\label{luccioparti}
\end{eqnarray}
Note that Eq. (\ref{luccioparti}) accounts for the fact that
$\boldsymbol{n} = (\boldsymbol{r_o}-\boldsymbol{r'}(t'))/|\boldsymbol{r_o}-\boldsymbol{r'}(t')|$ is not a
constant in time. Second, using Eq. (\ref{luccioparti}) in the
integration by parts they got:

\begin{eqnarray}
\bar{\boldsymbol{E}}(\boldsymbol{r_o},\omega) = -{i \omega e\over{c}}
\int_{-\infty}^{\infty} dt'&&\cr \times
\left[-{\boldsymbol{n}\times(\boldsymbol{n}\times{\boldsymbol{\beta}})\over{|\boldsymbol{r_o}-\boldsymbol{r'}(t')|}}
+ {i {c}\over{\omega}}
{\boldsymbol{\beta}-\boldsymbol{n}-2\boldsymbol{n}(\boldsymbol{n}\cdot\boldsymbol{\beta})\over{|\boldsymbol{r_o}-\boldsymbol{r'}(t')|^2}}\right]
&&\cr \times
\exp\left\{i\omega\left(t'+{|\boldsymbol{r_o}-\boldsymbol{r'}(t')|\over{c}}\right)\right\}
~ \label{revluccio}
\end{eqnarray}
where, similarly as before,  the edge terms have been dropped. Eq.
(\ref{revluccio}) includes an integrand which is completely
different with respect to Eq. (\ref{rev2}). This is no mistake.
Both Eq. (\ref{rev2}) and (\ref{revluccio}) are correct: using
integration by parts in different ways simply gives different
integrands which anyway, after integration, yield the same value
for $\bar{\boldsymbol{E}}(\boldsymbol{r_o},\omega)$. It is interesting to note
that, both in Eq. (\ref{rev2}) and Eq. (\ref{revluccio}), terms in
the first and second powers of $|\boldsymbol{r_o}-\boldsymbol{r'}(t')|^{-1}$
cannot be interpreted as is usually done in time domain, as
acceleration and velocity terms respectively nor they constitute,
mathematically, Fourier pairs with
$\bar{\boldsymbol{E}}(\boldsymbol{r_o},\omega)$. In fact, a different organization
of the integration by parts leads to different results: such terms
have, here, no separate physical meaning. We have just shown that
this situation is obtained starting from the Lienard-Wiechert
field, applying integration by parts in two different ways and
dropping the edge terms.

Going back to the fact that edge terms can be dropped, we can make
now some remark which is particularly interesting from a
methodological viewpoint. Edge terms are important only when
calculation of radiation properties from a given system is
performed over a part of it, that is integration is not taken from
$t'=-\infty$ to $t'=+\infty$ but only on a part of the trajectory
arbitrarily chosen. However, summing up all non-negligible
contributions is equivalent to integrate the system from
$t'=-\infty$ to $t'=+\infty$. From this viewpoint edge radiation
can be calculated without accounting for edge terms, which are
artificial. As we will show in Paragraph \ref{sec:edge}, the
low-frequency component of the field generated by a particle
moving as in Fig. \ref{edgeuno}, usually referred to as edge
radiation, actually arises  from the straight line between the
bends when one integrates the system over all the particle
trajectory.

As said before, on the one hand  the term in $| \boldsymbol{r_{o}}
-\boldsymbol{r'}|^{-2}$ in Eq. (\ref{rev2}) can be dropped in paraxial
approximation but, on the other hand, Eq. (\ref{rev2}) does not
coincide with Eq. (\ref{rev1}) once the velocity term is dropped.
In other words, the term in $|\boldsymbol{r_o}-\boldsymbol{r'}(t')|^{-1}$ of Eq.
(\ref{LW}) is not the term in $|\boldsymbol{r_o}-\boldsymbol{r'}(t')|^{-1}$ of Eq.
(\ref{rev2}). Even in paraxial approximation, the velocity term in
Eq. (\ref{LW}), in general, cannot be dropped. In fact, in order
to do so, it is required that the ratio between the modulus of the
velocity field and the modulus of the acceleration field is much
smaller than unity:

\begin{equation}
{c \mid \boldsymbol{n}-\boldsymbol{\beta} \mid \over{\gamma^2 |\boldsymbol{r_o}-\boldsymbol{r'}|
\mid \boldsymbol{n}\times[(\boldsymbol{n}-\boldsymbol{\beta})\times \boldsymbol{\dot{\beta}}] \mid
}}\ll 1 \label{conveldrop}
\end{equation}
Then, in paraxial approximation we have

\begin{equation}
\boldsymbol{n}\times[(\boldsymbol{n}-\boldsymbol{\beta})\times \boldsymbol{\dot{\beta}}] = -
\boldsymbol{n}\times[\boldsymbol{\dot{\beta}}\times(\boldsymbol{n}-\boldsymbol{\beta})] \simeq
[\boldsymbol{\dot{\beta}}\cdot(\boldsymbol{n}-\boldsymbol{\beta})]\boldsymbol{n}
\label{paraxdoubleprod}
\end{equation}
Substitution in condition (\ref{conveldrop}) gives

\begin{equation}
{\gamma^2 |\boldsymbol{r_o}-\boldsymbol{r'}|\over{c}}\left(
{\boldsymbol{n}-\boldsymbol{\beta}\over{\mid\boldsymbol{n}-\boldsymbol{\beta}\mid }}\cdot
\boldsymbol{\dot{\beta}}\right)\gg 1 ~.\label{finaldropping}
\end{equation}
Condition (\ref{finaldropping}) depends on the system under
investigation, but in general is not automatically satisfied. For
instance, in the case of circular motion $\boldsymbol{\dot{\beta}}/c \sim
1/R$, $R$ being the circle radius. Then we have

\begin{equation}
{\gamma^2 |\boldsymbol{r_o}-\boldsymbol{r'}|\over{c}}\left(
{\boldsymbol{n}-\boldsymbol{\beta}\over{\mid\boldsymbol{n}-\boldsymbol{\beta}\mid }}\cdot
\boldsymbol{\dot{\beta}}\right)\le {\gamma^2 r_o \theta_{o}\over{R}}
~.\label{finaldropping2}
\end{equation}
where $\theta_{o}$ is the angle between $\boldsymbol{n}$ and
$\boldsymbol{\dot{\beta}}$. For instance, it is well known that the
radiation formation length at the critical wavelength $\lambda_c
\sim R/\gamma^3$ is simply $L_f \sim R/\gamma$ and that $\theta_o
\sim 1/\gamma$; then, in the case $r_o \sim L_f$, we have

\begin{equation}
{\gamma^2 r_o \theta_{o}\over{R}} \sim 1~.\label{finaldropping3}
\end{equation}
As a result, when $r_o \sim L_f$ and $\lambda \sim \lambda_c$
condition (\ref{finaldropping}) is not satisfied, although the
paraxial approximation is enforced. This counterexample shows that
paraxial approximation, alone, is not sufficient to guarantee that
the velocity field in Eq. (\ref{LW}) can be dropped.

The region of parameter space for which the velocity field can be
neglected  is usually referred to as the far field zone. In order
to be in the far field zone the paraxial approximation is
necessarily enforced.

We take the condition that $\boldsymbol{n}$ is constant as a definition of
the far field zone. When this is the case, the classical result
(see for instance \cite{JACK})

\begin{equation}
{\boldsymbol{n}\times[(\boldsymbol{n}-\boldsymbol{\beta})\times{\boldsymbol{\dot{\beta}}}]
\over{(1-\boldsymbol{n}\cdot\boldsymbol{\beta})|\boldsymbol{r_o}-\boldsymbol{r'}|}} ={1\over{r_o}}
{d\over{dt'}}\left[
{\boldsymbol{n}\times(\boldsymbol{n}\times{\boldsymbol{\beta}})\over{1-\boldsymbol{n}\cdot\boldsymbol{\beta}}}\right]
\label{classdiff}
\end{equation}
can be used to perform integration by parts. Starting from Eq.
(\ref{LW}) and assuming, as explained before, that the edge terms
can be dropped, we arrive at the widely used result presented in
textbooks (see again \cite{JACK})

\begin{eqnarray}
\bar{\boldsymbol{E}}(\boldsymbol{r_o},\omega) = -{i\omega e\over{c r_o}}
\int_{-\infty}^{\infty}
dt'\left[{\boldsymbol{n}\times(\boldsymbol{n}\times{\boldsymbol{\beta}})}\right]&& \cr
\times \exp \left\{i\omega\left(t'-{1\over{c}}{\boldsymbol{n}\cdot
\boldsymbol{r'}}\right)\right\} ~. \label{revwied}
\end{eqnarray}
Now, with ultra relativistic accuracy $1-\boldsymbol{n}\cdot\boldsymbol{\beta}$
one has $\boldsymbol{n}\times(\boldsymbol{n}\times{\boldsymbol{\beta}}) \simeq \boldsymbol{\beta}
- \boldsymbol{n}$: using this relation we see straightforwardly that, in
the far field zone, Eq. (\ref{revwied})  is equivalent to  Eq.
(\ref{rev3}). 
To conclude, in this discussion we have
seen that paraxial approximation can be applied in the calculation
of Synchrotron Radiation properties for any set of parameters.
Moreover, in general, we should account for both velocity and
acceleration term in the Lienard-Wiechert expression: however, in
the case $\boldsymbol{n}$ can be taken as a constant (far zone
approximation), the velocity term can always be dropped.

\section{\label{sec:bend} Bending magnet radiation}

\subsection{\label{sub:circle} Circular motion}

Consider a particle moving along a circular trajectory and an
observer as sketched in Fig. \ref{srgeo}. The radiation from a
particle moving with velocity $v$ along a circle of radius $R$ is
observed from a point $P$. $R$ is the only characteristic length
of the system so that we can impose naturally that $L_{ch}$ is
$R$. When $\theta_y$ and $\theta_x$, defined in Fig. \ref{srgeo},
are much smaller than unity the reference system $(x,y,z)$
indicated in Fig. \ref{srgeo} is a consistent with the conditions
for paraxial treatment explained in the previous Section and Eq.
(\ref{generalfin}) can be used to characterize the radiation at
$P$. The motion along the curvilinear abscissa $s$ can be
described as

\begin{equation}
\boldsymbol{r'_\bot}(s) = -R\left(1-\cos(s/R)\right) \boldsymbol{\hat{x}}
\label{trmot}
\end{equation}
and

\begin{equation}
z'(s) = R \sin(s/r) \label{zmot}
\end{equation}
where $s = vt'$ and the expansion in Eq. (\ref{trmot}) is
justified, once again, in the framework of the paraxial
approximation.

Since the integral in Eq. (\ref{generalfin}) is performed along
$z'$ we should invert $z(s)$ in Eq. (\ref{zmot}) and find the
explicit dependence $s(z')$:

\begin{equation}
s(z') = R \arcsin(z'/r) \simeq z' + {z'^3\over{6R^2}} \label{sz}
\end{equation}
so that

\begin{equation}
\boldsymbol{r'_\bot}(z') = - {z'^2\over{2 R}} \boldsymbol{\hat{x}}\label{rpdis}
\end{equation}

 Substituting Eq. (\ref{sz}), Eq.(\ref{zmot}) and Eq.
(\ref{trmot}) in Eq. (\ref{generalfin}) we can write

\begin{eqnarray}
\widetilde{\boldsymbol{E}}_\bot(z_o, \boldsymbol{r_{\bot o}},\omega)= {i \omega
e\over{c}} \int_{-\infty}^{\infty} dz' {e^{i \Phi_T}\over{z_o-z'}}
&&\cr \times\left[\left({v
z'\over{Rc}}+{x_o+z'^2/(2R)\over{z_o-z'}}\right)\boldsymbol{\hat{x}}
+\left({y_o\over{z_o-z'}}\right)\boldsymbol{\hat{y}}\right]~,&&\cr
\label{srone}
\end{eqnarray}
where $\Phi_T$ is given by

\begin{equation}
\Phi_T = \omega\left({z'\over{2\gamma^2c}}+{z'^3\over{6R^2v}}+
{\left[x_o+ z'^2/(2R)\right]^2+y_o^2 \over{2c(z_o-z')}}
\right)~.\label{phh}
\end{equation}
At this stage our expression is still very general and valid for
any observation distance $z_o$. We keep up to the third order in
$z'$ in the expression for $s(z')$ in the phase, since the term
$z'/({2\gamma^2c})$ includes the small parameter $1/\gamma^2$.

Note that, although integration is performed from $-\infty$ to
$\infty$, the only part of the trajectory contributing to the
integral is of order of the radiation formation length $L_f =
[\lambda R^2/(2\pi)]^{1/3}$. At the critical wavelength
$R/\gamma^3$ that is simply $\sim R/\gamma$. Physically, this is
included in our paraxial approximation. Mathematically, it is
reflected in the fact that $\Phi_T$ in Eq. (\ref{phh}) exhibits
more and more rapid oscillations as $z'$ becomes larger and larger
due to non linear terms: in particular note that, as $z' \sim
L_f$, the term in $z'^3$ in Eq. (\ref{phh}) is of order unity. If
$z_o \geq L_f$ we can expand all expressions in $(z_o-z')^{-1}$
around $z_o$.   We will be considering the so-called $far~field$
radiation limit so that, expanding $(z_o-z')^{-1}$ we will retain
up to the third order in $z'$ in the expression for the phase and
up to the first order in $z'$ in the rest of the integrand. We
introduce angles $\theta_x=x_o/z_o$ and $\theta_y=y_o/z_o$ as in
Fig. \ref{srgeo} (with the restriction $\theta_x \ll 1$ and
$\theta_y \ll 1$). Then, accounting for $v\simeq c$, Eq.
(\ref{srone}) and Eq. (\ref{phh}) read respectively

\begin{eqnarray}
{\widetilde{\boldsymbol{E}}}= {i \omega e\over{c^2 z_o}}
\int_{-\infty}^{\infty} dz' {e^{i  \Phi_T}} \left({
z'+R\theta_x\over{R}}\boldsymbol{\hat{x}}
+\theta_y\boldsymbol{\hat{y}}\right)~\label{srtwo}
\end{eqnarray}
and

\begin{eqnarray}
\Phi_T = \omega \left[
\left({\theta_x^2+\theta_y^2\over{2c}}z_o\right)
+\left({1\over{2\gamma^2c}} +
{\theta_x^2+\theta_y^2\over{2c}}\right)z' \right.  &&\cr \left. +
\left({\theta_x\over{2Rc}}\right)z'^2 +
\left(1\over{6R^2c}\right)z'^3\right]~.\label{phh2}
\end{eqnarray}
One can easily reorganize the terms in Eq. (\ref{phh2}) to obtain

\begin{eqnarray}
\Phi_T = \omega\left[
\left({\theta_x^2+\theta_y^2\over{2c}}z_o\right)-{R\theta_x\over{2c}}\left({1\over{\gamma^2}}
+{\theta_x^2\over{3}} +\theta_y^2\right) \right.&&\cr \left.
+\left({{1\over{\gamma^2}}+\theta_y^2}\right){\left(z'+R\theta_x\right)\over{2c}}
+ {\left(z'+R\theta_x\right)^3\over{6 R^2 c
}}\right]~.\label{phh2b}
\end{eqnarray}
Finally, redefinition of $z'$ as $z'+R\theta_x$ gives the final
result

\begin{eqnarray}
{\widetilde{\boldsymbol{E}}}= {i \omega e\over{c^2 z_o}} e^{i\Phi_s}
e^{i\Phi_o} \int_{-\infty}^{\infty} dz'
\left({z'\over{R}}\boldsymbol{\hat{x}}+\theta_y\boldsymbol{\hat{y}}\right) &&\cr
\times
\exp\left\{{i\omega\left[{z'\over{2\gamma^2c}}\left(1+\gamma^2\theta_y^2\right)
+{z'^3\over{6R^2c}}\right]}\right\}~,\label{srtwob}
\end{eqnarray}
where
\begin{equation}
\Phi_s ={\omega z_o\over{2c}}\left(\theta_x^2+\theta_y^2
\right)\label{phis}
\end{equation}
and

\begin{equation}
\Phi_o = -{\omega R \theta_x\over{2c}}\left( {1\over{\gamma^2}}
+{\theta_x^2\over{3}} +\theta_y^2 \right)~.\label{phio}
\end{equation}
Note that the linear term in $(\theta_x^2+\theta_y^2) z'$ in the
phase of Eq. (\ref{srtwo}) defines the maximal angle of interest.
In fact, when $z' \sim L_f$, this term is of order unity if
$\theta_x^2+\theta_y^2  \sim [\lambda/(2\pi R)]^{2/3}$; this is
simply the ratio of $L_f^2$ to $L_{ch}^2$. As the observation
angle grows more the linear term becomes larger and larger
resulting in an oscillatory behavior of the entire integral in Eq.
(\ref{srtwo}) and thus giving no net contribution to the field. A
similar mechanism has been discussed in relation with the
radiation formation length. Now that we have found an upper limit
to the observation angle, $L_f/L_{ch}$, it is interesting to
discuss the accuracy of our expression. We have

\begin{equation}
{|\boldsymbol{r_{\bot o}}-\boldsymbol{r'}_\bot|^2\over{2(z_o-z')^2}} \sim
\theta_x^2+\theta_y^2 - \theta_x {L_f^2\over{Rz_0}} +
{L_f^4\over{4R^2z_o^2}} \sim {L_f^2\over{R^2}} \ll 1 ~.
\label{parexprconcircle}
\end{equation}
In other words,  the accuracy of our calculations is given by the
square of the maximal observation angle.

To compare Eq. (\ref{srtwob}) with results in literature (for
instance \cite{WIED,HOFM}) we should remember that the latter are
obtained using a reference system like the one in Fig.
\ref{srstand}. Therefore, in order to perform a comparison, we
should let $\theta_x = 0$, so that $\Phi_o=0$. Then, since results
are often obtained in terms of integration along the retarded time
$t'$ we should use $z'\simeq v t'$. Finally we obtain

\begin{eqnarray}
{\widetilde{\boldsymbol{E}}}= {i \omega e\over{c z_o}} e^{i\Phi_s}
\int_{-\infty}^{\infty} dt' \left({c t'\over{R}}
\boldsymbol{\hat{x}}+\theta_y\boldsymbol{\hat{y}}\right) &&\cr \times
\exp\left\{{i\omega\left[{t'\over{2\gamma^2}}\left(1+\gamma^2\theta_y^2\right)
+{t'^3c^2\over{6 R^2}}\right]}\right\}~.\label{srcomp}
\end{eqnarray}
Every Synchrotron Radiation textbook (see, for instance
\cite{WIE2}) shows that Eq. (\ref{srcomp}) can be written as

\begin{eqnarray}
{\widetilde{\boldsymbol{E}}}= -{\sqrt{3} e\over{c z_o}} {2 R \omega\over{3
\gamma^3 c}}  e^{i\Phi_s} \gamma(1+\gamma^2\theta_y^2) &&\cr
\times\left[K_{2/3}(\xi) \boldsymbol{\hat{x}} - i {\gamma\theta_y
K_{1/3}(\xi) \over{(1+\gamma^2\theta_y^2)^{1/2}}}\right]
.\label{srcompbess}
\end{eqnarray}
Here $K_{2/3}$ and $K_{1/3}$ are the modified Bessel functions of
second kind of fractional order $2/3$ and $1/3$ respectively,
while

\begin{equation}
\xi={R \omega\over{3 \gamma^3 c}}
(1+\gamma^2\theta_y^2)^{3/2}~.\label{csidefin}
\end{equation}
Note that, depending on the definition of Fourier transform Eq.
(\ref{ftran}), textbook versions of Eq. (\ref{srcomp}) or Eq.
(\ref{srcompbess}) may differ by a factor $1/\sqrt{2\pi}$ and (or)
by the operation of complex conjugation; moreover textbooks often
neglect the exact phase factor for the field, since they are
usually more interested in the calculation of the intensity, which
involves the square modulus of Eq. (\ref{srcomp}).

The result in Eq. (\ref{srcompbess}) is far from being trivial.
Bessel functions are real, therefore the $\boldsymbol{\hat{x}}$ component
in Eq. (\ref{srcompbess}) is real, while the $\boldsymbol{\hat{y}}$
component is purely imaginary. In other words, the $\boldsymbol{\hat{x}}$
component of the integral in Eq. (\ref{srcomp}) is purely
imaginary, while the $\boldsymbol{\hat{y}}$ component is real. This is not
immediately obvious, but it can be easily seen by inspection,
accounting for the fact that the exponential function in the
integrand can be written as $\exp(i ~\cdot~) = \cos(\cdot) + i
\sin(\cdot)$; then, for parity reasons, the term in
$\boldsymbol{\hat{x}}$, being odd, couples with the sine function, thus
giving an imaginary result while the term in $\boldsymbol{\hat{y}}$, being
even, couples with the cosine function, giving a real result.

From the previous remarks it follows that the only non-trivial
phase factor is specified by the surviving exponential argument
$\Phi_s=\omega z_o \theta_y^2/(2c)$, which is usually neglected in
literature. This simply represents, in our paraxial approximation,
the phase difference between the point $(0,y_o,z_o)$ and the point
$(0,0,z_o)$. Physically, in the particular reference frame of Fig.
\ref{srstand}, which we have chosen by setting $\theta_x=0$, the
electric field is represented by a spherical wave propagating
outwards from the origin of the coordinate system.

\begin{figure}
\begin{center}
\includegraphics*[width=90mm]{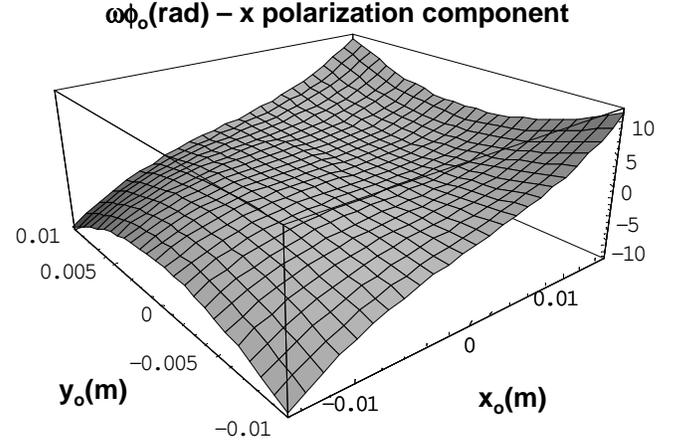}
\caption{\label{chubarx} Phase correction for the horizontal
polarization component of the field; case discussed in
\cite{CHUB}.}
\end{center}
\end{figure}
It is interesting now to investigate the meaning of the phase
$\Phi_o$, which is nonzero for nonzero values of $\theta_x$. This
term has no equivalent in the usual treatment of Synchrotron
Radiation, because displacement along the $x$ axis cannot be
considered in the reference system sketched in Fig. \ref{srstand}.
In the usual treatment, an $x_o$ displacement is treated
redefining the reference frame so that $x_o=0$. This redefinition
has simply the effect of shifting the phase of the field expressed
in the old reference frame of a quantity $i \Phi_o$.

\begin{figure}
\begin{center}
\includegraphics*[width=90mm]{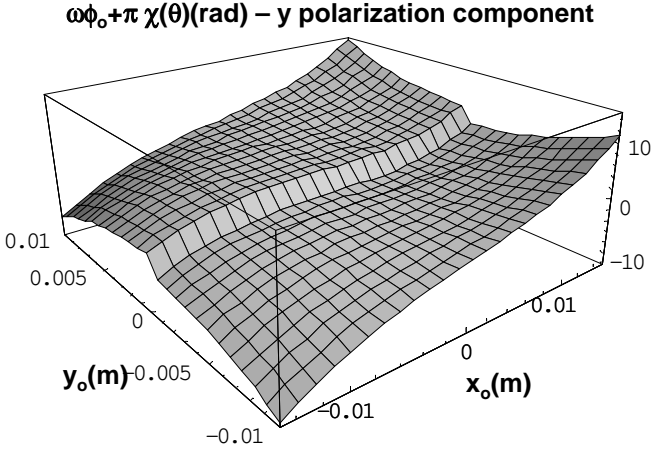}
\caption{\label{chubary} Phase correction for the vertical
polarization component of the field; case discussed in
\cite{CHUB}.}
\end{center}
\end{figure}

There is a considerable amount of literature dealing with the
phase of the Fourier components of the electric field vector in
Synchrotron Radiation (see, for instance \cite{CHUB}). In
particular, \cite{CHUB} reports the results of a thorough
simulation of single particle effects. It is obviously found that
the wavefront of a single particle is not spherical; a
correspondent phase shift (with respect to the spherical wave) for
the field of a single particle was computed and analytical
estimations within $1\%$ of the numerical result were presented
which are in complete agreement with Eq. (\ref{phio}). For the
sake of completeness and comparison we present, in Fig.
\ref{chubarx} and Fig. \ref{chubary} the phase shifts calculated
for the horizontal and vertical polarization components,
respectively by means of our Eq. (\ref{phio}) for the same example
considered in \cite{CHUB}: bending magnet emission from a $2.5$
GeV particle, constant magnetic field in bending magnet $1.56$ T,
photon energy 40 eV and distance from tangential source point to
optical component $5$ m. Note that these parameters correspond to
the far zone case where $z_o \gg R/\gamma$. The phase for the
$y$-polarization component, in Fig. \ref{chubary}, is simply the
phase for the $x$-polarization component (that is Fig.
\ref{chubarx}) added to $\pi H(\theta_y)$, where $H(\theta_y) = 1$
for $\theta_y \ge 0$ and $H(\theta_y) = 0$ for $\theta_y < 0$:
this last terms simply accounts for the fact that the $y$
component of Eq. (\ref{srcomp}) is odd.

%
%
%

\subsection{\label{sub:circledef} Circular motion with offset and deflection}

Electrons following a circular motion are, of course, an
approximation. Electron beams have always some small angular
spread and offset with respect to the nominal trajectory. Any beam
with a small geometrical emittance can be thought, in agreement
with paraxial treatment, as a composition of perfectly collimated
beams with different deflection angles with respect to the orbital
plane of the nominal trajectory. This representation is useful
when one is interested in calculating, for instance, the influence
of angular spread and offset on the electric field intensity: one
can simply compute the contribution for each collimated beam and
sum up the results. This method allows to simplify calculations of
more complicated quantities like the field autocorrelation
function, which is of uttermost importance in the characterization
of the statistical properties of a light source. We have already
discussed the advantage of our method regarding calculations of
the field autocorrelation function in connection with the choice
of a fixed reference frame as in Fig. \ref{srgeo}. Let us now
discuss how it can be used to calculate $\widetilde{\boldsymbol{E}}$ from
a single particle with a given angular deflection with respect to
the orbital plane of a nominal electron. Such an expression was
first calculated, starting from the Lienard-Wiechert fields, in
\cite{TAKA}.

The meaning of horizontal and vertical deflection angles $\eta_x$
and $\eta_y$ is clear once we specify the particle velocity

\begin{eqnarray}
\boldsymbol{v}(s) = v\left[\sin\left({s\over{R}}+\eta_x\right)\cos(\eta_y)
\boldsymbol{\hat{x}} + \sin(\eta_y) \boldsymbol{\hat{y}} \right.&&\cr\left.+
\cos\left({s\over{R}}+\eta_x\right)\cos(\eta_y) \boldsymbol{\hat{z}}
\right] \label{veloangle}
\end{eqnarray}
so that the trajectory can be expressed as a function of the
curvilinear abscissa $s$ as

\begin{eqnarray}
\boldsymbol{r'} = \left[l_x +
R\cos\left({s\over{R}}+\eta_x\right)\cos(\eta_y)
\right.&&\cr\left. - R\cos(\eta_x)\cos(\eta_y) \right]
\boldsymbol{\hat{x}} + \left[l_y+s \sin(\eta_y)\right]  \boldsymbol{\hat{y}}
&&\cr + \left[+ R\sin\left({s\over{R}}+\eta_x\right)\cos(\eta_y) -
R\sin(\eta_x)\sin(\eta_y) \right] \boldsymbol{\hat{z}} &&\cr
\label{trajangle}
\end{eqnarray}
Here we have introduced, also, an arbitrary offset $(l_x,l_y,0)$
in the trajectory. Using Eq. (\ref{trajangle}) an approximated
expression for $s(z')$ can be found:

\begin{equation}
s(z') = z'+ {z'^3\over{6 R^2}}+{z'^2 \eta_x\over{2 R}} +{z'
\eta_x^2\over{2}}+{z' \eta_y^2\over{2}} \label{szangle}
\end{equation}
so that

\begin{equation}
\boldsymbol{v_\bot}(z') =   \left(- {v z'\over{R}}+ v \eta_x  \right)
\boldsymbol{\hat{x}} + \left(v \eta_y  \right)
\boldsymbol{\hat{y}}~\label{vapprangle}
\end{equation}
and

\begin{equation}
\boldsymbol{r'_\bot}(z') = \left(- {z'^2\over{2R}}+ \eta_x z' + l_x\right)
\boldsymbol{\hat{x}} + \left(\eta_y z' +l_y \right)
\boldsymbol{\hat{y}}~.\label{trmot2}
\end{equation}
We will consider $\eta_x \ll 1$ and $\eta_y \ll 1$ in agreement
with the fact that we deal with electron \textit{beams} with small
angular deflection and not with more general kind of plasmas.

Substituting, as in the previous Paragraph,  Eq.
(\ref{szangle}), Eq. (\ref{trmot2}) and Eq. (\ref{vapprangle}) in
Eq. (\ref{generalfin}) we can write

\begin{eqnarray}
\widetilde{\boldsymbol{E}}_\bot(z_o, \boldsymbol{r_{\bot o}},\omega)= {i \omega
e\over{c}} \int_{-\infty}^{\infty} dz' {e^{i \Phi_T}\over{z_o-z'}}
&&\cr \times\left[\left({ z'\over{R}}-{\eta_x}+{x_o-l_x-\eta_x z'
+z'^2/(2R)\over{z_o-z'}}\right)\boldsymbol{\hat{x}} \right. && \cr \left.
+\left(-{\eta_y}+{y_o-l_y-\eta_y z'
\over{z_o-z'}}\right)\boldsymbol{\hat{y}}\right]~,&&\cr \label{srone2}
\end{eqnarray}
where $\Phi_T$ is given by

\begin{eqnarray}
\Phi_T = \omega \left(
{z'\over{2\gamma^2c}}+{z'^3\over{6R^2v}}+{z'^2 \eta_x\over{2 Rv}}
+{z' \eta_x^2\over{2v}}+{z' \eta_y^2\over{2v}} \right.&&\cr\left.
+{\left[x_o-l_x-\eta_x z' +z'^2/(2R)\right]^2
+\left[y_o-l_y-\eta_y z'\right]^2 \over{2c(z_o-z')}}\right)
~.\label{phh22}
\end{eqnarray}
At this stage, as in the previous Paragraph, our expression is
still very general and valid for any observation distance $z_o$,
and again as we did before, we have retained up to the third order
in $z'$ in the expression for $s(z')$ in the phase, since the term
$z'/({2\gamma^2c})$ includes the small parameter $1/\gamma^2$ and
up to the first order in $z'$ in the rest of the integrand. We
will now consider the \textit{far field} radiation limit which, as
the reader remembers, allows expansion of all expressions in
$(z_o-z')^{-1}$ around $z_o$: as before we will retain up to the
third order in $z'$ in the expression for the phase and up to the
first order in $z'$ in the rest of the integrand. From Eq.
(\ref{srone2}), Eq. (\ref{phh22}) and Eq. (\ref{trmot2}) it is
evident that the offsets $l_x$ and $l_y$ are always subtracted
from $x_o$ and $y_o$ respectively: shifting the particle
trajectory on the vertical plane is equivalent to a shift of the
observer in the opposite direction. With this in mind, in analogy
with Fig. \ref{srgeo}, we introduce angles
$\bar{\theta}_x=\theta_x-l_x/z_o$ and
$\bar{\theta}_y=\theta_y-l_y/z_o$ (with the restriction
$\bar{\theta}_x \ll 1$ and $\bar{\theta}_y \ll 1$). Then,
accounting for $\eta_x \ll 1$ and $\eta_y\ll 1$, Eq.
(\ref{srone2}) and Eq. (\ref{phh22}) can be written down
respectively, as follows:

\begin{eqnarray}
{\widetilde{\boldsymbol{E}}}= {i \omega e\over{c^2 z_o}}
\int_{-\infty}^{\infty} dz' {e^{i \Phi_T}}&&\cr \times \left({
z'+R(\bar{\theta}_x-\eta_x)\over{R}}\boldsymbol{\hat{x}}
+{(\bar{\theta}_y-\eta_y)}\boldsymbol{\hat{y}}\right)~\label{srtwoang}
\end{eqnarray}
and

\begin{eqnarray}
\Phi_T =
\left({\bar{\theta}_x^2+\bar{\theta}_y^2\over{2c}}z_o\right)
+{1\over{2c}}\left({1\over{\gamma^2}} +
\left(\bar{\theta}_x-\eta_x\right)^2 \right.&&\cr\left. +
\left(\bar{\theta}_y-\eta_y\right)^2\right)z' +
\left({\bar{\theta}_x\over{2Rc}}\right)z'^2 +
\left(1\over{6R^2c}\right)z'^3~.\label{phh2ang}
\end{eqnarray}
One can easily reorganize the terms in Eq. (\ref{phh2ang}) to
obtain

\begin{eqnarray}
\Phi_T =
\left({\bar{\theta}_x^2+\bar{\theta}_y^2\over{2c}}z_o\right)-
{R(\bar{\theta}_x-\eta_x)\over{2c}} && \cr \times
\left({1\over{\gamma^2}} +(\bar{\theta}_y-\eta_y)^2
+{(\bar{\theta}_x-\eta_x)^2\over{3}}\right) && \cr
+\left({{1\over{\gamma^2}}+(\bar{\theta}_y-\eta_y)^2}\right)
{\left(z'+R(\bar{\theta}_x-\eta_x)\right)\over{2c}} &&\cr +
{\left(z'+R (\bar{\theta}_x-\eta_x)\right)^3\over{6 R^2 c
}}~.\label{phh2angfin}
\end{eqnarray}
Redefinition of $z'$ as $z'+R(\bar{\theta}_x-\eta_x)$ gives the
result

\begin{eqnarray}
{\widetilde{\boldsymbol{E}}}= {i \omega e\over{c^2 z_o}} e^{i \Phi_s} e^{i
\Phi_o} \int_{-\infty}^{\infty} dz'
\left({z'\over{R}}\boldsymbol{\hat{x}}+(\bar{\theta}_y-\eta_y)\boldsymbol{\hat{y}}\right)
&&\cr \times
\exp\left\{{i\omega\left[{z'\over{2\gamma^2c}}\left(1+\gamma^2
(\bar{\theta}_y-\eta_y)^2\right)
+{z'^3\over{6R^2c}}\right]}\right\}~,\label{srtwoang2}
\end{eqnarray}
where
\begin{equation}
\Phi_s = {\omega z_o
\over{2c}}\left(\bar{\theta}_x^2+\bar{\theta}_y^2
\right)\label{phisang}
\end{equation}
and

\begin{equation}
\Phi_o = - {\omega R(\bar{\theta}_x-\eta_x)\over{2c}}
\left({1\over{\gamma^2}} +(\bar{\theta}_y-\eta_y)^2
+{(\bar{\theta}_x-\eta_x)^2\over{3}}\right)~.\label{phioang}
\end{equation}
Except for the phase term $\Phi_s$, Eq. (\ref{srtwoang2}) can be
obtained from Eq. (\ref{srtwob}) simply by substituting $\theta_x$
with $\bar{\theta}_x-\eta_x$ and $\theta_y$ with
$\bar{\theta}_y-\eta_y$: besides a common phase factor then, we
can say that including a deflection angle has the same effect of
shifting the observer position of the same angle. Still, we should
remember that $\bar{\theta}_x=\theta_x-l_x/z_o$ and
$\bar{\theta}_y=\theta_y-l_y/z_o$.

In the limit for $l_{x,y}/z_o \ll \theta_F$, with  $\theta_F =
[\lambda/(2\pi R)]^{1/3}$, one can simplify further Eq.
(\ref{srtwoang2}). Note how introduction of small parameters
allows increasing specialization and simplification of the theory:
for instance, in this Paragraph we started with an expressions for
the fields obtained in the limit for $1/\gamma_z^2 \gg 1$, that
could be cast into simpler form in the limit $L_f/z_o \ll 1$ and
now further specialized assuming $l_{x,y}/z_o \ll \theta_F$
leading to:

\begin{eqnarray}
{\widetilde{\boldsymbol{E}}}= {i \omega e\over{c^2 z_o}} e^{i \Phi_s} e^{i
\Phi_o} \int_{-\infty}^{\infty} dz'
\left({z'\over{R}}\boldsymbol{\hat{x}}+\left(\theta_y-\eta_y
\right)\boldsymbol{\hat{y}}\right) &&\cr \times
\exp\left\{{i\omega\left[{z'\over{2\gamma^2c}}\left(1+\gamma^2
\left(\theta_y-\eta_y\right)^2\right)
+{z'^3\over{6R^2c}}\right]}\right\}~,\label{srtwoang2bis}
\end{eqnarray}
where
\begin{equation}
\Phi_s = {\omega z_o \over{2c}}\left(\theta_x^2+\theta_y^2
\right)\label{phisangbis}
\end{equation}
and

\begin{eqnarray}
\Phi_o \simeq - {\omega R({\theta_x}-\eta_x)\over{2c}}
\left({1\over{\gamma^2}} +(\theta_y-\eta_y)^2 \right.&&\cr\left.
+{(\theta_x-\eta_x)^2\over{3}}\right)-{\omega\over{c}}(l_x
\theta_x+l_y\theta_y) ~.\label{phioangbis}
\end{eqnarray}
Eq. (\ref{srtwoang2}) is an extremely useful tool, because it
describes the radiation from an electron with offset and
deflection as in an electron beam with finite emittance, including
the correct phase factor for the field.  Starting from Eq.
(\ref{srtwoang2}) then, it is possible to calculate the field
correlation function, and to provide a study of transverse
coherence properties of the radiation from a given electron beam
by means of analytical techniques.

As has already been remarked, coherence is a special and important
property of Synchrotron Radiation sources. In the far zone, the
spatial coherence at different observer angles
$\theta_{x,y}^{(1)}$ and $\theta_{x,y}^{(2)}$ is characterized by
the autocorrelation function (see \cite{GOOD, MAND} and
\cite{ATTW} for applications to Synchrotron Radiation science):

\begin{eqnarray}
\Gamma_{ij}\left(\theta_{x,y}^{(1)},\theta_{x,y}^{(2)},\omega\right)
&&\cr= \left\langle
\widetilde{E}_i^*\left(\theta_{x,y}^{(1)},\eta_{x,y},l_{x,y}\right)
\widetilde{E}_j\left(\theta_{x,y}^{(2)},\eta_{x,y},l_{x,y}\right)
\right\rangle \label{autocorf}
\end{eqnarray}
where $\widetilde{E}_i$ is the $i$-th Fourier component of the
electric field given in Eq. (\ref{srtwoang2}), and brackets
$\langle ... \rangle$ denote an ensemble average with respect to
electron parameters. Note that a given electron is correlated just
with itself, that is why Eq. (\ref{autocorf}) only includes the
electric field from a single electron and not from two different
particles. The ensemble average can be replaced by integration
over the electron phase-space density. The intensity distribution
can be obtained directly from Eq. (\ref{autocorf}) by letting
$\theta_{x,y}^{(1)}=\theta_{x,y}^{(2)}$.

Of course a numerical code can always be developed, either
starting from Eq. (\ref{generalfin}) or just from the
Lienard-Wiechert fields, which calculates the field correlation
function in a generic case, but such a code would not help in
physical understanding of the situation. On the contrary, Eq.
(\ref{srtwoang2}) includes all relevant information about an
electron in a realistic beam (i.e. with offset and deflection)
and, being an analytically manageable expression, constitutes the
first step towards the characterization of transverse coherence
properties from bending magnet radiation.

\subsection{\label{sec:edge} Edge radiation}

In this Paragraph we will stress the importance of the knowledge
of the \textit{entire} trajectory followed by the electron as we
study the effect of a change in longitudinal velocity due to the
passage of an electron in a magnetic system. This results in
collimated emission of radiation in the low photon energy range, a
mechanism analogous to transition radiation, which is well-known
in literature under the name of edge radiation
\cite{CHUB,BOSC,PROY}.

We restrict ourselves to the system depicted in Fig.
\ref{edgeuno}, which shows two bending magnets separated by a
straight line of length $L$. Radiation is detected at a screen
positioned at distance $z_o$ from the downstream magnet. We will
require that the bending magnets deflect the electron trajectory
of an angle much larger than $[\lambda/(2\pi R)]^{1/3}$, $\lambda$
being the wavelength of interest: in this way the straight lines
before the upstream bend and after the downstream bend do not
contribute to the field detected at the screen position.
Intuitively, the magnets act like switches: the first magnet
switches the radiation on, the second switches it off. The
trajectory can be therefore split in three parts: the two bends
and the straight section of length $L$ between them. With the help
of Eq. (\ref{generalfin}) we write the contribution from the
straight line as

\begin{equation}
\widetilde{\boldsymbol{E}}_{\boldsymbol{AB}}={i \omega e\over{c^2 z_o}}
\int_{A}^{B} dz' e^{i\Phi_T} \left(\theta_x \boldsymbol{\hat{x}}+\theta_y
\boldsymbol{\hat{y}}\right) \label{ABcontr}
\end{equation}
where we assumed $z_o \gg L$. The previous assumption is not
always verified in cases of practical interest. Here, however, we
are only concerned with an example of application of our method.
$\Phi_T$ in Eq. (\ref{ABcontr}) is given by

\begin{equation}
\Phi_T = \omega \left[ {\theta_x^2+\theta_y^2\over{2c}} z_o +
{z'\over{2c}}\left({1\over{\gamma^2}}+\theta_x^2+\theta_y^2\right)\right]~.
\label{phiab}
\end{equation}
The characteristic length related to the straight section is
obviously $L_{ch} = L$. In general, the formation length of
radiation at wavelength $\lambda$ is given imposing $\omega z'/(2
c \gamma^2) \sim 1$, which gives $L_f \sim \gamma^2 \lambda$.
However, if we are interested in low frequencies such that $\omega
L/(2 \gamma^2 c) \ll 1$, we can simply consider $L_f \sim L$ and
neglect the term in $1/\gamma^2$ in Eq. (\ref{phiab}). Trivial
calculations show that

\begin{equation}
\widetilde{\boldsymbol{E}}_{\boldsymbol{AB}}=-{i \omega e L\over{c^2
z_o}}e^{i\Phi_s} \left(\theta_x \boldsymbol{\hat{x}}+\theta_y
\boldsymbol{\hat{y}}\right)
 {\sin\left[\omega L (\theta_x^2+\theta_y^2) /(4c)
\right]\over{[\omega L (\theta_x^2+\theta_y^2) /(4c)]}}~,
\label{lineab}
\end{equation}
where $\Phi_s$ has the usual meaning (compare, for instance, with
Eq. (\ref{phis})). The magnitude of the contribution from the
straight section AB can be estimated noting that the sinc function
in Eq. (\ref{lineab}) drops rapidly as $\theta_x^2+\theta_y^2$
reach the value $4 c/(\omega L) = 2 \lambda/(\pi L)$. This means
that a maximal observation angle of interest related to the
straight line can be found in the transverse direction:

\begin{equation}
\theta_{x,y}^2 \sim {\lambda\over{2\pi L}} ~.\label{maxinterline}
\end{equation}
At this angle, the formation length for the straight line
radiation at wavelength $\lambda$ is simply equal to the straight
section length, $L$, as it can be seen from Eq. (\ref{phiab}).
Then, the magnitude of $\widetilde{\boldsymbol{E}}_{\boldsymbol{AB}}$ is of order

\begin{equation}
\mid \widetilde{\boldsymbol{E}}_{\boldsymbol{AB}} \mid \sim {2 e \over{c z_o}}
\left({\omega L\over{c}}\right)^{1/2}~. \label{EABord}
\end{equation}
On the other hand, the magnitude of the contributions
from the bends can be estimated as

\begin{eqnarray}
{\widetilde{\boldsymbol{E}}}_{\boldsymbol{b}} \sim -{ \omega e\over{c^2 z_o}}
\int_{0}^{\infty} dz'
\left({z'\over{R}}\boldsymbol{\hat{x}}+\theta_y\boldsymbol{\hat{y}}\right) &&\cr
\times
\exp\left\{{i\omega\left[{z'\over{2\gamma^2c}}\left(1+\gamma^2\theta_y^2\right)
+{z'^3\over{6R^2c}}\right]}\right\}~.
 \label{bendcon}
\end{eqnarray}
In Eq. (\ref{bendcon}) we have redefined the origin of our
reference system, as it is clear from the integration limits;
however, here we are interested in the magnitude and not in the
phase of the field, so that we can use Eq. (\ref{bendcon}) without
further discussion on the correct phase. The linear term in $z'$
in the exponential function in Eq. (\ref{bendcon}) is of order
unity for angles $\theta_y$ such that

\begin{equation}
{\omega L_f\over{2 c}} \theta_y^2 \sim 1~.
\label{conditionangleblob}
\end{equation}
Substituting the formation length for the bends $L_{fB} \sim
[\lambda R^2/(2\pi)]^{1/3}$ we obtain a condition for the maximal
observation angle of interest related to the bending magnets

\begin{equation}
\theta_y^2 \sim \left({\lambda\over{2\pi R}}\right)^{2/3}~.
\label{conditionangleblob2}
\end{equation}
The previous condition on $\theta_y^2$ found for the bends has to
be compared with the condition on $\theta_y^2$ previously found
for the straight line, $\theta_y^2 \sim \lambda/(2\pi L)$. The
ratio between the latter and the former is equal to the ratio of
the formation length for the straight line to the formation length
for the bends respectively, that is $L/L_{fB}$. Assuming $L/L_{fB}
\gg 1$, that is always verified in practice, we obtain that the
maximal observation angle of interest related the straight line is
much smaller than the maximal observation angle of interest
related to the bend. As a result the entire linear term in the
phase of Eq. (\ref{bendcon}) can be neglected thus leading to the
following simplified expression for
${\widetilde{\boldsymbol{E}}}_{\boldsymbol{b}}$:


\begin{eqnarray}
{\widetilde{\boldsymbol{E}}}_{\boldsymbol{b}} \sim -{ \omega e\over{c^2 z_o}}
\int_{0}^{\infty} dz'
\left({z'\over{R}}\boldsymbol{\hat{x}}+\theta_y\boldsymbol{\hat{y}}\right)
\exp\left\{{i\omega z'^3\over{6R^2c}}\right\} ~.&&\cr
 \label{bendcon2}
\end{eqnarray}
An estimate of Eq. (\ref{bendcon2}) is

\begin{eqnarray}
{\widetilde{\boldsymbol{E}}}_{\boldsymbol{b}} \sim -{ 2 e\over{c z_o}}
{\omega\over{2 c}} \left({L_f^2\over{2 R}}\boldsymbol{\hat{x}}+\theta_y
L_f \boldsymbol{\hat{y}}\right) &&\cr \sim -{ 2 e\over{c z_o}}
\left[{1\over{2}} \left({\omega R
\over{c}}\right)^{1/3}\boldsymbol{\hat{x}}+{\left({\omega R
\over{c}}\right)^{2/3}} \left({c \over{\omega L }}\right)^{1/2}
\boldsymbol{\hat{y}}\right] ~.
 \label{bendcon3}
\end{eqnarray}
\begin{figure}
\begin{center}
\includegraphics*[width=90mm]{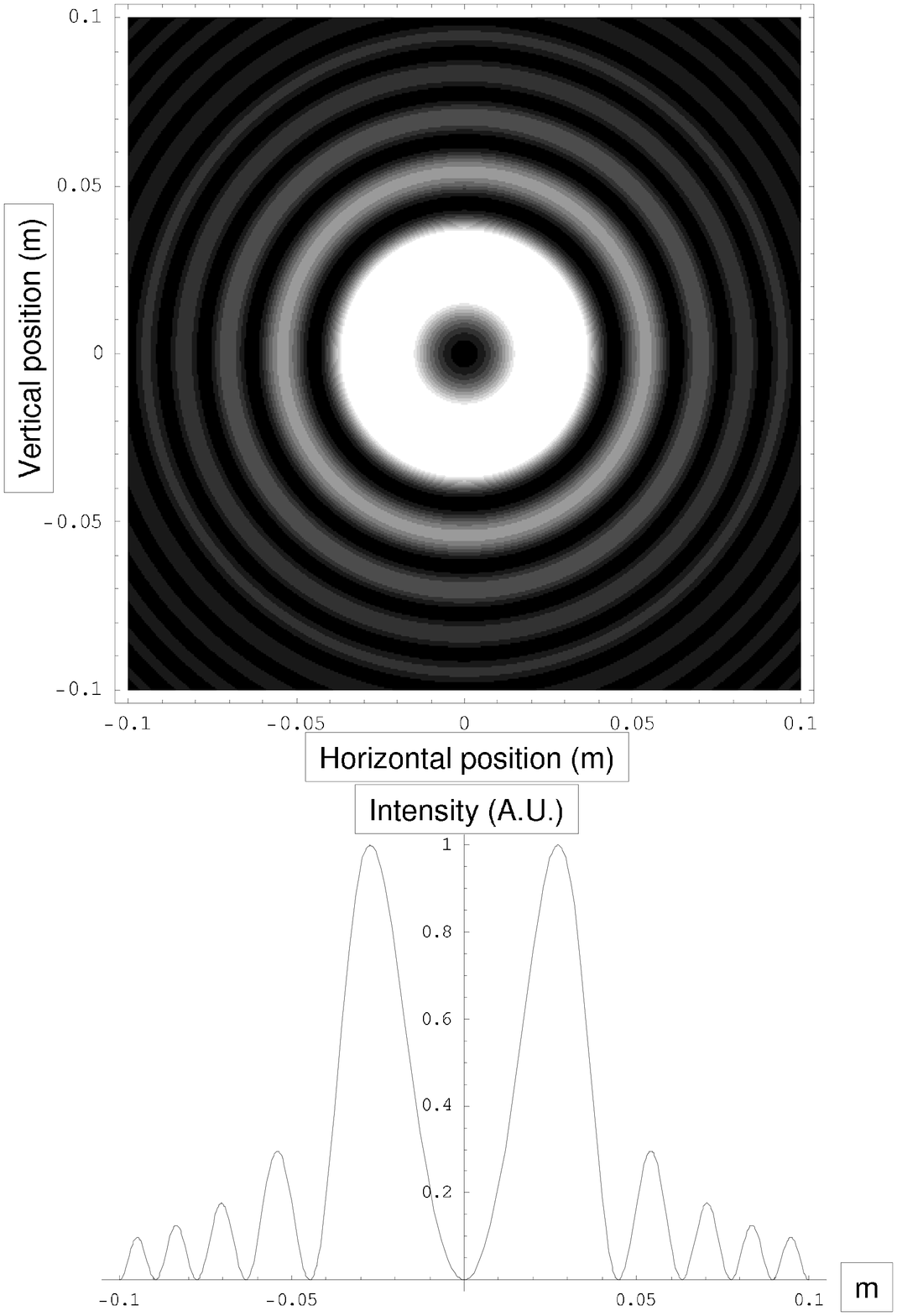}
\caption{\label{edge2} Intensity distribution of radiation from
the setup in Fig. \ref{edgeuno} at $\lambda = 10\mu$m and for
$L=1$ m. Radiation is collected at $z_o=10$ m from the downstream
magnet. A contour plot (upper figure) and a horizontal cut by the
median plane (lower figure) are presented. }
\end{center}
\end{figure}
The ratio between the magnitude of the bend and the straight
section contribution is then
\begin{eqnarray}
{{\widetilde{\boldsymbol{E}}}_{\boldsymbol{b}}\over{\mid
{\widetilde{\boldsymbol{E}}}_{\boldsymbol{AB}}\mid}} \sim -{1\over{2}}
\left({\omega R \over{c}}\right)^{1/3}  \left({c\over{\omega L
}}\right)^{1/2}\boldsymbol{\hat{x}}&&\cr -{\left({\omega R
\over{c}}\right)^{2/3}} \left({c \over{\omega L
}}\right)\boldsymbol{\hat{y}} = -{1\over{2}}\sqrt{{L_f\over{L}}}
\boldsymbol{\hat{x}} -{L_f\over{L}}\boldsymbol{\hat{y}} ~.
 \label{bendcon3bis}
\end{eqnarray}
Therefore, under the already accepted condition $L \gg L_f$ one
can neglect the contribution from the bending magnet. The only
remaining contribution is given by Eq. (\ref{lineab}), which
represents an expression for the field at the detector position.
We plot the intensity corresponding to the field in Eq.
(\ref{lineab}) in Fig. \ref{edge2} for the case $\lambda = 10
~\mu$m, $L = 4$ m and $z_o$ = 20 m.

Eq. (\ref{lineab}) is the same reported in \cite{PROY,BOS2}. It is
interesting to remark that in conventional treatments of
Synchrotron Radiation, edge radiation has its origin in the edge
term arising in the integration by parts of Eq. (\ref{rev1}); in
other words it is found from the acceleration part of the
Lienard-Wiechert field, which is present only in the magnets. On
the other hand, from our method, edge radiation arises as the
contribution from the straight section between the magnets. This
seems paradoxical, but one has to remember that there is no
physical meaning in the calculation of the field harmonic content
over \textit{a part} of the trajectory alone. It does not make any
sense, for instance, to calculate the field intensity from the
edge of a single bend, because there will be either a second bend,
as in Fig. \ref{edgeuno}, other structures downstream of the
second bend which must be taken into account in the calculation of
the total field: the sum of all these contributions will give
interference terms when the intensity is calculated. Then,
accepting the viewpoint that only the knowledge of the
\textit{entire} trajectory of the particle brings physical sense
to field calculations, there is no real contradiction: in our
method edge radiation appears from the straight section.  In the
usual approach, instead, it appears from the edge terms in the
integration by part of the acceleration field. Yet these terms,
alone, have no physical meaning. This solves our paradox.

\subsection{\label{sub:short} Short magnet radiation}

Our method can be used to compute radiation characteristics from a
short magnet, characterized by a bending angle $2 \psi \ll
1/\gamma$. This device has the quite interesting characteristic
that the Fourier transform of the electric field in the far field
limit has non-zero value for $\omega \rightarrow 0$ and that the
critical frequency does not depend on the magnet radius but only
on the magnet length. These features can be explained in a simple
way. For circular motion the far field from a single particle has
the property:

\begin{equation}
\int_{-\infty}^{\infty} \boldsymbol{E} dt= 0~. \label{zerointe}
\end{equation}
\begin{figure}[tb]
\begin{center}
\includegraphics*[width=90mm]{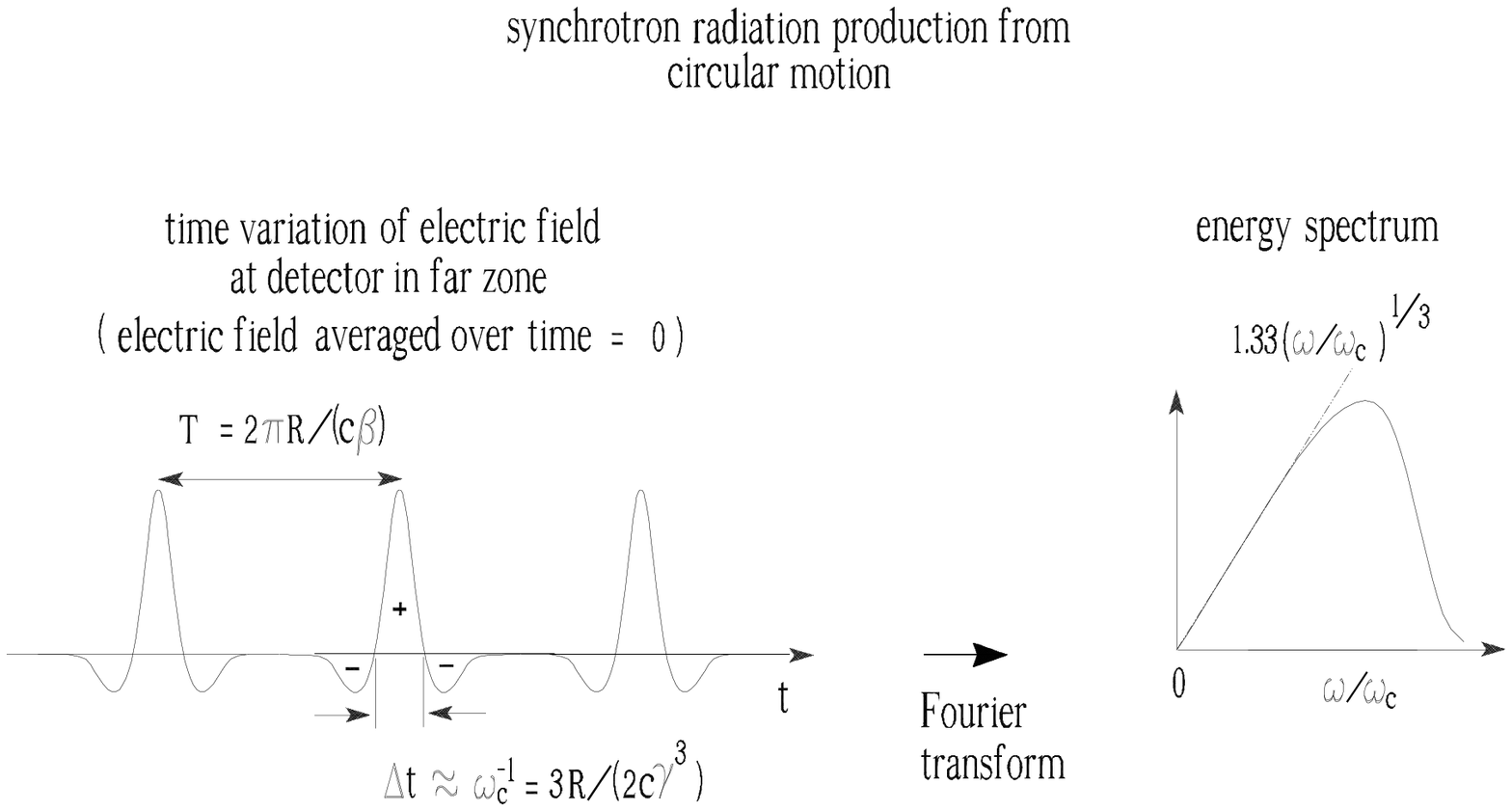}
\end{center}
\caption{ Radiation field from an electron moving along a circle
in the time and in the frequency domain} \label{fig:dsm2}
\end{figure}

\begin{figure}[tb]
\begin{center}
\includegraphics*[width=90mm]{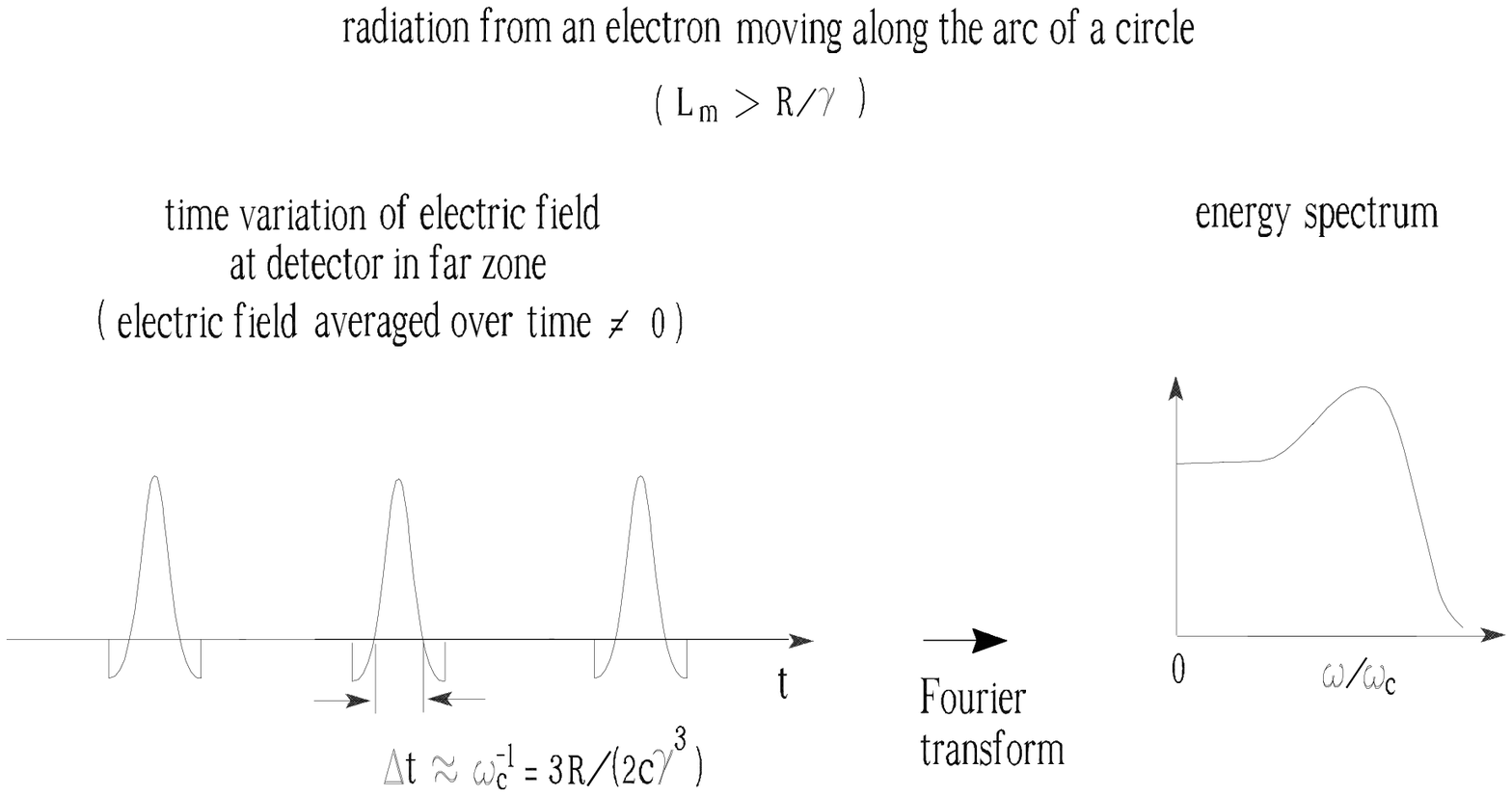}
\end{center}
\caption{ Radiation field from an electron moving along an arc of
a circle in the time and in the frequency domain} \label{fig:dsm3}
\end{figure}

\begin{figure}[tb]
\begin{center}
\includegraphics*[width=90mm]{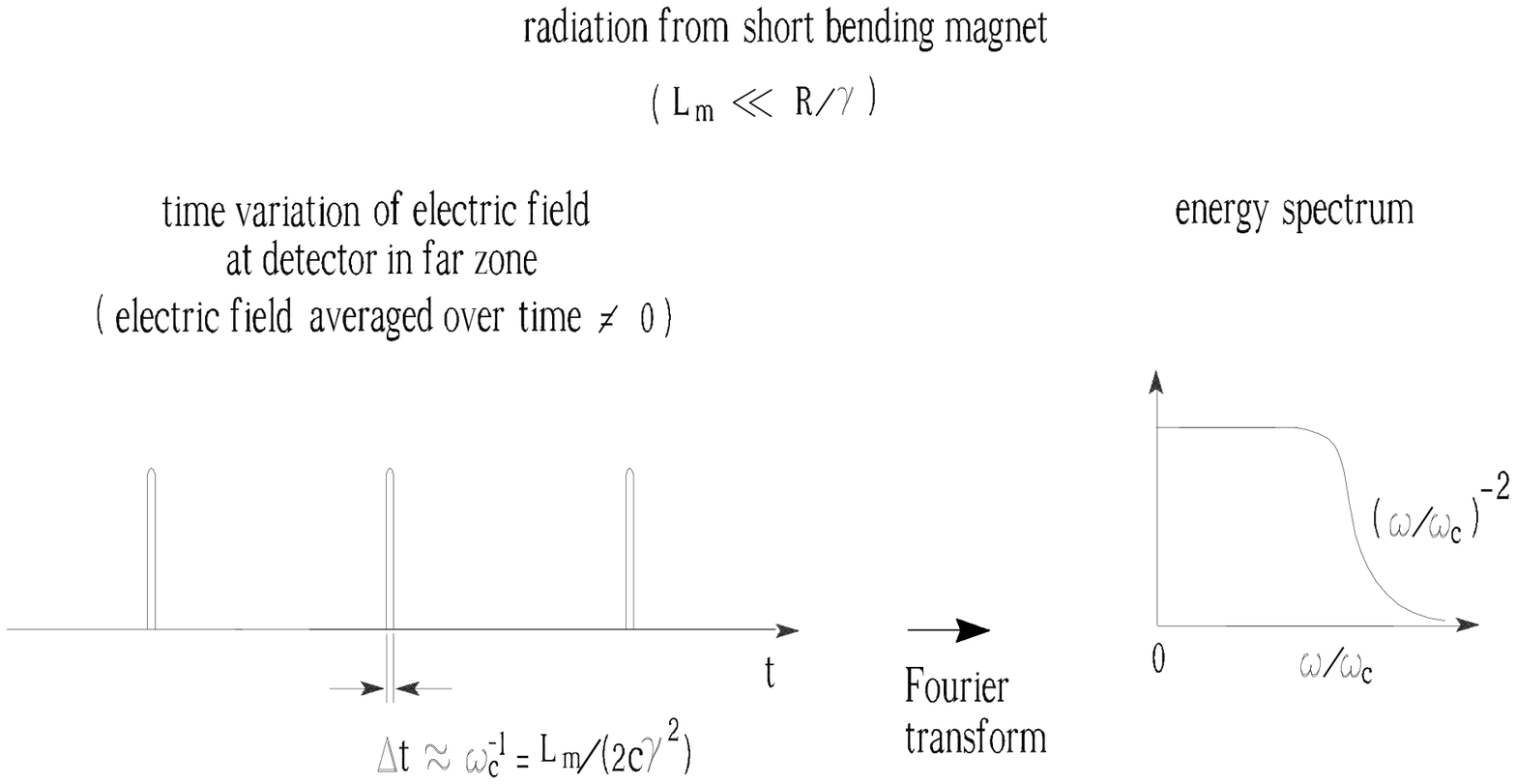}
\end{center}
\caption{ Radiation field from an electron moving along a short
bending magnet in the time and in the frequency domain}
\label{fig:dsm4}
\end{figure}
When the electron moves in arc of circles with different angular
extensions $2 \psi$ though, the time-average of the electric field
is nonzero; then, its Fourier transform has a nonzero component
for $\omega \rightarrow 0$.

The first studies on long wavelength asymptotic can be found in
\cite{BAGR}. For a review on the subject it may be interesting to
consult \cite{HOFM}; for a discussion dealing also with properties
of CSR (Coherent Synchrotron Radiation) see \cite{OURS}. In Fig.
\ref{fig:dsm2}, Fig. \ref{fig:dsm3} and Fig. \ref{fig:dsm4} we
plot the energy spectrum of the radiation in the case of an
electron moving on circular motion, dipole magnet and short bend,
respectively. All figures refer to the $\boldsymbol{\hat{x}}$ polarization
component. As Eq. (\ref{zerointe}) does not hold we find a nonzero
component of the energy spectrum at $\omega \rightarrow 0$. As the
magnet extension becomes smaller than $1/\gamma$ the critical
frequency depends only on the magnet length $L_m = 2 R \psi$
according to $\omega_c \simeq 2c \gamma^2/ L_m$.

The short magnet is a device of particular interest also because
its radiation characteristics are related to the methodological
issue introduced in Section \ref{sec:intro}. From the very
beginning, our method relies on harmonic analysis of the electric
field, so that the $entire$ trajectory of an electron is
considered known. This gives no particular problem in the case of
a closed trajectory, as discussed in the previous Subsections, but
a question arises, now that a short magnet is considered, about
the trajectory of our particle outside the short magnet. In
general, the field at the observation position depends strongly on
the trajectory followed by the electron before and after the
magnet. So one should clarify what is the meaning of the term
$short~magnet~radiation$ by itself, when it is not specified what
follows and what precedes the bend.

Computing the field from the magnet alone as it is usually done
has a meaning, but only in a particular sense: one can always
break up the beamline in several parts, calculate contributions
separately and finally add them up, accounting for the proper
relative phase, to get the total field at the observer position.
Then the short magnet radiation is simply a part of the total
field, a contribution to be added up to something else, which
depends, case by case, on what precedes and follows the magnet.

In conclusion, when we talk about radiation from a short magnet,
we mean the contribution to the total field at the observer
position $P$ due to the short magnet alone: this makes sense as
long as we consider it only a part of the total field, calculated
separately just for computational convenience.

With this in mind we calculate the short magnet field from Eq.
(\ref{srtwo}) limiting the integration to the magnet extent and
using $2 \psi \ll 1/\gamma$. This introduces a further small
parameter in our system which leads to extra simplification of
already known formulas derived for the case of a bending magnet
with arbitrary extension. From Eq. (\ref{srtwo}) we obtain

\begin{eqnarray}
{\widetilde{\boldsymbol{E}}}= {i \omega e\over{c^2 z_o}} e^{i\Phi_s}
\int_{-R\psi}^{R\psi} dz'
\left({z'+R\theta_x\over{R}}\boldsymbol{\hat{x}}+\theta_y\boldsymbol{\hat{y}}\right)
&&\cr \times \exp\left\{{i\omega
{z'\over{2\gamma^2c}}\left[1+\gamma^2(\theta_x^2+
\theta_y^2)\right]}\right\} &&\cr \times \exp\left\{i\omega
{\theta_x z'^2\over{2Rc}}\right\} \exp\left\{i \omega {z'^3\over{6
R^2 c }}\right\}~,\label{srsh}
\end{eqnarray}
where $\Phi_s$ is defined as in Eq. (\ref{phis}). In Eq.
(\ref{srsh}) we intentionally separated phase factors
$\exp\{i\omega {\theta_x z'^2/{2Rc}}\}$ and $\exp\{i \omega
z'^3/(6 R^2 c)\}$ because we intend to neglect the second and
expand the first around unity. For our system the characteristic
length is obviously $L_{ch} = R\psi$. In general, the formation
length of the radiation is fixed by the first phase factor, $L_f
\sim \gamma^2 \lambda$, but we will be interested in studying the
long wavelength region such that $L_f \sim R\psi$, that is the
maximum attainable value, since it is always $L_f \lesssim
L_{ch}$. Then, the first phase factor imposes, simultaneously,
$\omega R\psi/(2\gamma^2c) \lesssim 1$ and $\omega \theta_x^2
R\psi/c \lesssim 1$. If on the one hand, from the first of these
conditions $\omega_{max} \simeq 2\gamma^2 c/(R\psi)$, from the
second condition follows that $\theta_x^2 \lesssim 1/(2\gamma^2)$
and from the second phase factor one has $\omega {\theta_x
(R\psi)^2/(2Rc)\sim \gamma \psi \ll 1}$. On the other hand if,
from the first of these conditions, $\omega \ll 2\gamma^2
c/(R\psi)$, then from the second condition follows that
$\theta_{x}^2 \simeq c/(\omega R \psi)\gg 1/(2\gamma^2)$ and from
the second phase factor one has $\omega \theta_{x} (R\psi)^2/(2Rc)
\simeq (\gamma\psi)/(\gamma\theta_x) \ll 1$. It follows from these
considerations that we can perform the expansion of $\exp\{i\omega
{\theta_x z'^2/{2Rc}}\}$ up to the second term with an accuracy
$(\gamma\psi)^2$ or better and neglect $\exp\{i \omega z'^3/(6 R^2
c)\}$ with the same accuracy for any choice of $\omega$. Then we
can write

\begin{eqnarray}
{\widetilde{\boldsymbol{E}}}= {i \omega e\over{c^2 z_o}} e^{i\Phi_s}
\int_{-R\psi}^{R\psi} dz' \left({z'+R
\theta_x\over{R}}\boldsymbol{\hat{x}}+\theta_y\boldsymbol{\hat{y}}\right) &&\cr
\times \left(1+i\omega {\theta_x z'^2\over{2Rc}}\right)
\exp\left\{{i\omega
{z'\over{2\gamma^2c}}\left[1+\gamma^2(\theta_x^2+\theta_y^2)\right]}
\right\}~,\label{srsh2}
\end{eqnarray}
Performing calculations and dropping negligible terms we find
\begin{eqnarray}
{\widetilde{\boldsymbol{E}}}=-{e\over{z_o R c}} e^{i\Phi_s}
{2\gamma^2\over{[1+\gamma^2(\theta_x^2+\theta_y^2)]^2}}
\Big[\Big(1+\gamma^2(\theta_x^2+\theta_y^2)
&&\cr-2\gamma^2\theta_x^2\Big)\boldsymbol{\hat{x}}-\Big(2\gamma^2
\theta_x\theta_y\Big)\boldsymbol{\hat{y}}\Big]
\bar{\chi}_{[-R\psi,R\psi]}(\alpha)~, \label{srsh3}
\end{eqnarray}
where

\begin{equation}
\alpha = {\omega\over{2 c \gamma^2}}
\left[1+\gamma^2(\theta_x^2+\theta_y^2)\right]~ \label{alpha}
\end{equation}
and
\begin{eqnarray}
{\bar{\chi}}_{[-R\psi,R\psi]}(\alpha) = \int_{-R\psi}^{R\psi} dz'
\exp\left\{i \alpha z'\right\} ~.\label{chiftransf}
\end{eqnarray}
If we define a function $\chi_{[-R\psi,R\psi]}$ which assumes
unitary value on the interval $[-R\psi,R\psi]$ and zero value
elsewhere, then $\chi_{[-R\psi,R\psi]}$ simply represents the
magnetic field shape at the magnet position, and
$\bar{\chi}_{[-R\psi,R\psi]}$ is its Fourier transform with
respect to $\alpha$.

Suppose now the magnetic field has a generic shape $B(z')$.
Keeping the same accuracy as before, we can generalize Eq.
(\ref{srsh3}) substituting relations

\begin{equation}
x'(z') = -{e\over{\gamma m_e}} \int_{0}^{z'} dz'' \int_{0}^{z''}
dz''' B(z''') ~\label{xaltro}
\end{equation}
and

\begin{equation}
v_x(z') = -{e\over{\gamma m_e}} \int_{0}^{z'} dz'' B(z'')~,
~\label{vxaltro}
\end{equation}
where $m_e$ is the electron rest mass, in place of $x'=-z'^2/(2R)$
and $v_x/c = -z'/R$ directly in Eq. (\ref{srsh}). Performing
approximations as before and integrating by parts two time the
term in $x'$ and one time the term in $v_x$ we find that Eq.
(\ref{srsh3}) can be easily generalized.

Results in literature are often given in cylindrical coordinates
$(\phi,\theta,z)$ where $\sin\phi=y_o/\sqrt{x_o^2+y_o^2}$ and
$\theta=\theta_x^2+\theta_y^2$. The same choice can be found, for
instance, \cite{HOFM}. Making use of this system we obtain:

\begin{eqnarray}
{\widetilde{\boldsymbol{E}}}={2r_e \gamma\over{z_o}} e^{i\Phi_s}
{1\over{(1+\gamma^2\theta^2)^2}}
\left[\left(1-\gamma^2\theta^2\cos(2\phi)\right)\boldsymbol{\hat{x}}\right.
&&\cr\left.-\left(\gamma^2
\theta^2\sin(2\phi)\right)\boldsymbol{\hat{y}}\right]
\bar{B}\left({1+\gamma^2\theta^2\over{2c \gamma^2}}\omega\right)~
\label{srshcomp}
\end{eqnarray}
where $r_e$ is the classical radius of the electron

\begin{equation}
r_e = {e^2\over{m_e c^2}} \label{classicrad}
\end{equation}
and

\begin{eqnarray}
{\bar{B}}(\alpha) = \int_{-\infty}^{\infty} dz' B(z') \exp\left\{i
\alpha z'\right\} ~.\label{Bftransf}
\end{eqnarray}
As has been said, the field from a short magnet makes sense only
as a part of the total field from a given beamline. In this
Paragraph we have calculated, for such contribution, the
expression in Eq. (\ref{srshcomp}) which can also be found in
\cite{HOFM}. Our aim, here, is not to review a well-known result,
but to cast new light in its physical interpretation.

In order to do so we need first a digression. Our Eq.
(\ref{generalfin}) is based on the more general Eq. (\ref{rev2}).
In Paragraph \ref{sub:disc} we have shown that Eq. (\ref{rev2}) is
equivalent to the Fourier transform of the Lienard-Wiechert fields
Eq. (\ref{LW}) provided that the edge term in the integration by
parts going from Eq. (\ref{LW}) to Eq. (\ref{rev2}) (or viceversa)
can be dropped. On the one hand we showed that this can always be
done since the integral in Eq. (\ref{LW}) and Eq. (\ref{rev2})
should be performed over the entire electron history, and only a
finite part of the trajectory contributes, practically, to the
electric field at the observer position. On the other hand, one is
free to break up the beamline in parts and sum up partial
contributions to the total field using Eq. (\ref{LW}) followed by
integration by parts for each segment. In this case one must
retain edge terms to obtain the same result: as an example of this
fact, using our method, we have shown in Paragraph \ref{sec:edge}
that the edge radiation from the system depicted in Fig.
\ref{edgeuno} arises as the contribution from the straight section
between the magnets, while conventional calculations indicate its
origins in the edge term from the integration by parts of Eq.
(\ref{LW}) in the far field limit.

Now coming back to the subject of this Paragraph, typical
derivations of Eq. (\ref{srshcomp}) (see \cite{HOFM, BAGR, BORD})
start from Eq. (\ref{rev1}) and include the complete expression
for the acceleration field. Then, following conventional
derivations, it is not clear wether edge terms play an important
role in Eq. (\ref{srshcomp}) or not. Yet, we were able to obtain
Eq. (\ref{srshcomp}) without starting with Eq. (\ref{rev1}):  in
fact we began with Eq. (\ref{srsh}) which is Eq. (\ref{srtwo})
with different integration limits which is, in its turn, a
reduction of Eq. (\ref{generalfin}) that, finally, is a
simplification of Eq. (\ref{rev2}). Since our result coincide with
the one in literature it follows that edge radiation from the
presence of the short magnet is completely ignorable.

Of course, one should still account for contributions from the
other parts of the beamline: for instance, if the short magnet
were installed in the middle of the straight section in Fig.
\ref{edgeuno} one should consider the edge radiation contribution
as in Paragraph \ref{sec:edge}.

However, according to our reasoning we can conclude that one can
safely drop the contribution of edge terms due to the presence of
the short magnet. On the contrary, beginning with Eq. (\ref{rev1})
as is done in usual treatments, one is not able to separate the
two physical phenomena and, indeed, may be easily misunderstand
and misinterpret results concluding, erroneously, that short
magnet radiation cannot be derived without edge contributions.
Here we have seen that a critical study of the theoretical status
of well-known formulas, can sometimes yield surprises.

\section{\label{sec:undu} Radiation from insertion devices}

\subsection{\label{sec:plu} Standard undulator}

For a review of up to date knowledge on undulator radiation one
may be interested in consulting references \cite{WIED} to
\cite{DUKE}. An experimental characterization of radiation from
insertion devices at third generation light sources can be found
in \cite{PTRI}. Eq. (\ref{generalfin}) can be used to derive the
expression for $\bar{E}(\boldsymbol{r},\omega)$ in the case of an
undulator as well.

\begin{figure}
\begin{center}
\includegraphics*[width=90mm]{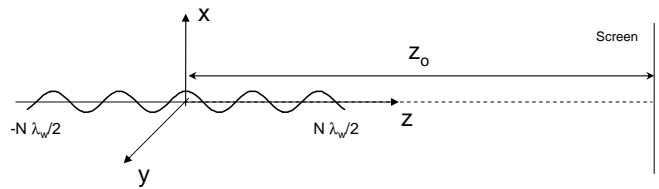}
\caption{\label{undugeo} Geometry for undulator radiation. }
\end{center}
\end{figure}
Again, as before, we remark that, with the term \textit{undulator
field} we actually mean a part of the total field seen by an
observer and that one should account, in general, for the entire
motion of the particle. The situation is sketched in Fig.
\ref{undugeo}, where a planar undulator consisting of $N_w$
periods is sketched. We will follow \cite{WIED} in deriving
well-known relations. For the electron transverse velocity we
assume

\begin{equation}
\boldsymbol{v_\bot}(z') = - {c K\over{\gamma}} \sin{\left(k_w z'\right)}
\boldsymbol{x}~. \label{vuz}
\end{equation}
Here $K$ is the deflection parameter, $k_w=2\pi/\lambda_w$, where
$\lambda_w$ is the undulator period, so that the undulator length
is $L_w = N_w \lambda_w$. The undulator length $L_w$ is also,
naturally, the characteristic length of the system, so that
$L_{ch}=L_w$. The transverse position of the electron is therefore

\begin{equation}
\boldsymbol{r'_\bot}(z') = {K\over{\gamma k_w}} \cos{\left(k_w z'\right)}
\boldsymbol{x}~. \label{erz}
\end{equation}
An expression for the curvilinear abscissa $s$ as a function of
$z$ is given by

\begin{equation}
s(z') =  {{\beta}\over{\beta_{av}}}z' - {K^2\over{8\gamma^2 k_w}}
\sin\left(2 k_w z'\right)~.\label{zsss}
\end{equation}
where $\beta_{av}$ is the time-averaged velocity along the $z$
direction, that can be expressed as:

\begin{equation}
\beta_{av} = \beta \left(1-{K^2\over{4\gamma^2}}\right)
\label{bbar}
\end{equation}
\begin{figure}
\begin{center}
\includegraphics*[width=90mm]{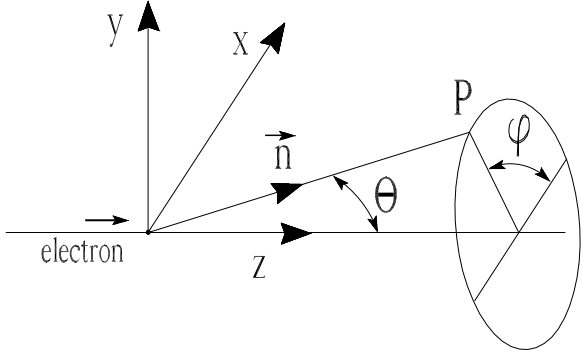}
\caption{\label{angle} Radiation geometry for the undulator field.
}
\end{center}
\end{figure}
We can now substitute Eq. (\ref{erz}) and Eq. (\ref{vuz}) in our
Eq. (\ref{generalfin}). Such a substitution leads to a general
expression, valid for any observer distance $z_o$.  It is possible
to obtain, similarly to many Synchrotron Radiation textbooks, a
simplified expression valid in the limit for large values of
$z_o$. Since we are interested in the contribution of the
undulator device to the total field at the observer position, we
will integrate Eq. (\ref{generalfin}) only along the undulator.
Then all terms in $(z_o-z')^{-1}$ in the phase factor of Eq.
(\ref{generalfin}) can be expanded around $z_o$. In the limit $z_o
\gg N_w \lambda_w$,  we can usually retain first order terms in
$z'$: later on, in Paragraph \ref{sec:near}, we will discuss the
applicability region of this approximation and we will see that
there are particular regions of parameters which do not allow
retention of first order terms alone. Dropping negligible terms
and defining $\phi$ and $\theta$ according to Fig. \ref{angle} (or
remembering the analogous definition given in Paragraph
\ref{sub:short}) we obtain

\begin{eqnarray}
\widetilde{\boldsymbol{E}}_\bot(z_o, \boldsymbol{r_{\bot o}},\omega)= {i \omega
e\over{c^2 z_o}} \int_{-N_w\lambda_w/2}^{N_w\lambda_w/2} dz' {e^{i
\Phi_T}} &&\cr \times\left[\left({K\over{\gamma}} \sin\left(k_w
z'\right)+\theta\cos\left(\phi\right)\right)\boldsymbol{\hat{x}}
+\theta\sin\left(\phi\right)\boldsymbol{\hat{y}}\right]~.&&\cr
\label{undurad}
\end{eqnarray}
Here

\begin{eqnarray}
\Phi_T = {\omega\over{\omega_1}} \left[k_w z' -
{K\theta\cos(\phi)\over{\gamma}}{\omega_1\over{k_w c}}\cos(k_w z')
\right.&&\cr \left.- {K^2\over{8\gamma^2}} {\omega_1\over{k_w c}}
\sin(2 k_w z') \right] +\Phi_s~,\label{phitundu}
\end{eqnarray}
where $\Phi_s = \omega \theta^2 z_o/(2c)$, as usual, and
$\omega_1$ is defined as

\begin{equation}
\omega_1^{-1} = {1\over{2 k_w c
\gamma^2}}\left(1+{K^2\over{2}}+\gamma^2\theta^2\right)
~.\label{om1}
\end{equation}
Note that as $\omega = \omega_1$, the formation length of our
system is simply given by $L_f = 1/k_w$ as it is seen immediately
from Eq. (\ref{phitundu}) imposing the first term in phase to be
of order unity. Yet it should be stressed here that the formation
length $L_f = 1/k_w$ is obtained in the most general case, without
accounting for special conditions which are very often chosen for
undulator operation. In fact,  when a large number of undulator
periods is selected an extra large-parameter is introduced in the
system yielding simplifications within the theory; then, as it
will be clear after reading Paragraph \ref{sec:res}, if the so
called resonance condition is met, the actual formation length of
the system becomes $L_f \sim L_w$.

The choice of the integration limits in Eq. (\ref{undurad})
express explicitly the fact that the reference system has its
origin in the center of the undulator as in Fig. \ref{undugeo}. As
said before, Eq. (\ref{undurad}) gives only a partial contribution
to the total field, which must be summed up with terms arising
from structures preceding and following the undulator. In
\cite{WIED}, for instance, no particular attention is given to
this problem: the calculation of undulator radiation distribution
is performed considering the acceleration term of the
Lienard-Wiechert fields alone, integrating by parts, and dropping
the edge term.

\begin{figure}
\begin{center}
\includegraphics*[width=90mm]{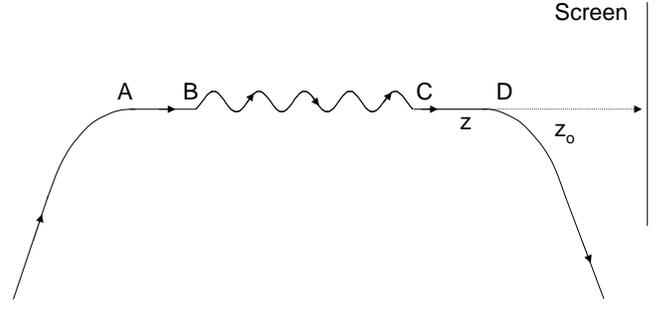}
\caption{\label{todo} Undulator followed and preceded by bending
magnets and straight sections. Information about the particle
trajectory before and after the undulator is necessary to
calculate the radiation spectrum.}
\end{center}
\end{figure}
It is indeed very useful, at this point, to comment some passage
of \cite{WIED}. At page 412 one can read: "No assumptions on the
magnetic field parameters have been made to derive the radiation
spectrum in the form of equation (7.87) which we use to calculate
the radiation spectrum from a wiggler magnet". Eq. (7.87) in
\cite{WIED} is the radiation spectrum from a particle moving on a
generic trajectory calculated, indeed, starting from the
Lienard-Wiechert field expression for the acceleration field,
integrating by parts and neglecting the edge terms; it obviously
appears in the form of an integration with limits from $-\infty$
to $\infty$: this is just our Eq. (\ref{revwied}) that was derived
exactly as Eq. (7.87) in \cite{WIED}. Dropping the velocity field
in the Lienard-Wiechert expression is done under the sufficient
(but not necessary, in general!) condition that $\boldsymbol{n}$ can be
considered constant. This is regarded as the far field
approximation. Then, neglecting of edge terms can be done under
physical assumptions discussed much earlier in Paragraph
\ref{sub:disc}. Again, the integral in $dt'$ has to be performed
over the \textit{entire history of the particle};  the physical
meaning of $t'=-\infty$ and $t'=+\infty$ is that at these times,
the electron does not contribute to the field anymore. From this
viewpoint, the geometry specified in Fig. \ref{undugeo} is no more
sufficient. One has to provide information about what precedes and
what follows the undulator; an example is depicted in Fig.
\ref{todo}. In the case of the scheme in Fig. \ref{todo} for
instance, $t'=-\infty$ and $t'=+\infty$ refer to moments when the
particle is well inside the first and the second bend,
respectively, and do not contribute anymore to the resulting
radiation spectrum. In order to drop edge terms in the particular
case of the trajectory in Fig. \ref{todo}, one has to consider
contributions from the first bend, then from $A$ to $B$ along the
straight line, from $B$ to $C$ inside the undulator, from $C$ to
$D$ along the second straight line and, finally, from the second
bending magnet.

Instead of following this prescription, the author of \cite{WIED}
substitutes characteristic quantities for a trajectory on an
infinitely long undulator in Eq. (7.87), and then he simply sets
the undulator strength parameter $K$ to zero outside of a given
temporal interval, which is simply the physical time that an
electron takes to pass through the undulator. This is equivalent
to consider two infinite straight lines before and after the end
of the undulator. Then, although results  are correct in some
particular case (namely under resonant approximation, which we
discuss in Paragraph \ref{sec:res}) the reader is left with the
unsolvable task of understanding what is the physical meaning of
two infinite straight lines and, especially, what is the
contribution of these infinite parts of the trajectory to the
total electric field at the observer position. Again, if one
wishes to drop the edge terms, one should start considering the
entire trajectory of the electron, not a part of it alone.

In the case depicted in Fig. \ref{todo} for instance, we will have
different interfering contributions from the straight lines before
and after the undulator and we would end up with both edge
radiation and transition undulator radiation. Again, in the most
general case, Eq. (\ref{undurad}) refers to a part of the electric
field seen by the observer and it should be added to other
contributions corresponding to the entire history of the particle.
Only in particular situations like the one discussed in the
following Paragraph \ref{sec:res} calculation of radiation
properties from Eq. (\ref{undurad}) alone makes sense: in
Paragraph \ref{sec:res} we will discuss the resonant approximation
which can be used in the case of a large number of undulator
periods, $N_w \gg 1$. Again, the presence of an extra large (or
small) parameter yields to further simplifications of the theory.
However it should be noted that, in practice, $N_w$ ranges from a
few units, in the case of Far Infrared insertion devices to about
one thousand, for X-ray FELs, so that the applicability of the
resonant approximation strongly depends on the situation under
study.

\subsection{\label{sec:res} Resonant
approximation}

In general, it does not make sense to calculate the intensity
distribution from Eq. (\ref{undurad}) alone, without extra
interfering terms. Yet, we can find \textit{particular situations}
for which the contribution from Eq. (\ref{undurad}) is dominant
with respect to others. In this case, and only in this case, Eq.
(\ref{undurad}), alone, is endowed with physical meaning.

The method proposed in \cite{ALFE} to calculate the integral in
Eq. (\ref{undurad}) is very well-known and used by every textbook
treating the details of undulator radiation. Again, we will follow
notation in \cite{WIED} during our derivation. The method consists
in using the identity

\begin{equation}
e^{i a \sin(\psi)} = \sum_{p=-\infty}^{\infty} J_p(a) e^{ip\psi}~,
\label{alfeq}
\end{equation}
where $J_p$ indicates the Bessel function of the first kind of
order $p$.

After introduction of

\begin{equation}
Q = {2 K \beta_{av} \gamma \theta \cos(\phi)\over{1+
K^2/2+\gamma^2\theta^2}} \label{CQ}
\end{equation}
and

\begin{equation}
S =  {K^2 \beta_{av} \over{4(1+ K^2/2+\gamma^2\theta^2)}}
\label{S}
\end{equation}
one can express the exponential function in Eq. (\ref{undurad})
as:

\begin{equation}
e^{i\Phi_T} = \sum_{m=-\infty}^{\infty} \sum_{n=-\infty}^{\infty}
J_m(u) J_n(v) e^{i\pi n/2} e^{-i R_\omega k_w z'} \label{transf}
\end{equation}
where

\begin{equation}
u = {\omega\over{\omega_1}} S~,~~ v = {\omega\over{\omega_1}}
Q~,~~\mathrm{and}~~ R_\omega = {\omega\over{\omega_1}} - n
-2m~.\label{defi}
\end{equation}
Then, Eq. (\ref{undurad}) can be re-written in the more suggestive
form:

\begin{eqnarray}
\widetilde{\boldsymbol{E}}_\bot(z_o, \boldsymbol{r_{\bot o}},\omega)= {i \omega
e\over{c^2 z_o}} e^{i \Phi_s} \sum_{m,n=-\infty}^{\infty} J_m(u)
J_n(v) &&\cr \times e^{i\pi n/2}
\int_{-N_w\lambda_w/2}^{N_w\lambda_w/2} dz' e^{-i R_\omega k_w z'}
&&\cr \times\left[\left({K\over{\gamma}} \sin\left(k_w
z'\right)+\theta\cos\left(\phi\right)\right)\boldsymbol{\hat{x}}
+\theta\sin\left(\phi\right)\boldsymbol{\hat{y}}\right]~.&&\cr
\label{undurad2}
\end{eqnarray}
Performing the integral in Eq. (\ref{undurad2}) one gets:

\begin{eqnarray}
\widetilde{\boldsymbol{E}}_\bot= {i \omega e
 \lambda_w N_w e^{i \Phi_s} \over{c^2 z_o}}
\sum_{m,n=-\infty}^{\infty} J_m(u) J_n(v)  e^{i\pi n/2} &&\cr
\times \left\{\left[{K\over{2\gamma}}i\left({\sin (\pi N_w
(R_\omega+1)) \over{\pi N_w (R_\omega+1)}}
\right.\right.\right.&&\cr\left.\left.\left.-{\sin (\pi N_w
(R_\omega-1)) \over{\pi N_w (R_\omega-1)}}\right)
+\theta\cos\left(\phi\right) {\sin (\pi N_w R_\omega) \over{\pi
N_w R_\omega }} \right]\boldsymbol{\hat{x}}
\right.&&\cr\left.+\theta\sin\left(\phi\right) {\sin (\pi N_w
R_\omega) \over{\pi N_w R_\omega }} \boldsymbol{\hat{y}}\right\}.&&\cr
\label{undurad3}
\end{eqnarray}
As is well known, and can be seen by inspection, terms in Eq.
(\ref{undurad3}) exhibit resonant character. Maxima of different
terms are found as $R_\omega =0$, $R_\omega + 1 =0$ and $R_\omega
- 1 =0$. These correspond to particular frequencies multiples of
the fundamental $\omega_1$. Always following \cite{WIED} we
introduce the harmonic number $k$ and we set $\omega_k = k
\omega_1$. Then, after setting $\Delta \omega_k = \omega -
\omega_k$, the following expression can be obtained for the field
around frequency $\omega_k$:

\begin{eqnarray}
\widetilde{\boldsymbol{E}}_{\bot,k}= {i \omega e \lambda_w N_w e^{i
\Phi_s} \over{c^2 z_o}} {\sin (\pi N_w \Delta \omega_k/\omega_1)
\over{\pi N_w \Delta \omega_k/\omega_1}} &&\cr \times
\sum_{m=-\infty}^{\infty} \left\{\left[{K\over{2\gamma}}i\left(
J_m(u) J_{k-2m-1}(v) e^{i\pi (k-2m-1)/2}
\right.\right.\right.&&\cr\left.\left.\left.- J_m(u) J_{k-2m+1}(v)
e^{i\pi (k-2m+1)/2}\right)\right.\right.&&\cr\left. \left.
+\theta\cos\left(\phi\right)  J_m(u) J_{k-2m}(v)  e^{i\pi
(k-2m)/2} \right]\boldsymbol{\hat{x}} \right.&&\cr\left.
+\theta\sin\left(\phi\right) J_m(u) J_{k-2m}(v) e^{i\pi (k-2m)/2}
\boldsymbol{\hat{y}}\right\}.&&\cr \label{undurad3b}
\end{eqnarray}
The factor $\sin(\cdot)/(\cdot)$ represents the well-known
resonant character of the device: our electron produces field
peaked at frequencies $\omega_k$ only. $\omega_k$ is a function of
the angle $\theta$. Once an observation angle $\theta$ is fixed, a
resonant frequency $\omega_k(\theta)$ is also defined and the
bandwidth of the radiation is determined by $\Delta
\omega_k(\theta)$ through the resonant factor
$\sin(\cdot)/(\cdot)$. If we are interested in the angular width
of the peak around the observation angle $\theta$, we can
introduce an angular displacement $\Delta \theta$ with respect to
$\theta$. The resonant frequency at angle $\theta + \Delta
\theta$, that is $\omega_k(\theta+\Delta \theta)$, has a different
value with respect to $\omega_k(\theta)$. When $\Delta \theta$
becomes large enough, $\omega_k(\theta+\Delta \theta)$ gets
outside the bandwidth $\Delta \omega_k(\theta)$. This happens for
an angular displacement $\Delta \theta$ such that

\begin{equation}
N_w (\omega_k(\theta+\Delta
\theta)-\omega_k(\theta))/\omega_1(\theta) \sim 1
\label{angueqgen}
\end{equation}
which is the first zero of the $\sin(\cdot)/(\cdot)$ function in
Eq. (\ref{undurad3b}). If we are interested in the angular width
of the peak around $\theta=0$ in the case $k=1$, that is for the
fundamental harmonic, we should solve the equation

\begin{equation}
N_w (\omega_1(\Delta \theta)-\omega_1(0))/\omega_1(0) = 1
\label{angueq}
\end{equation}
with respect to $\Delta \theta$. Eq. (\ref{angueq}) should be
taken only as a rough indication of the angular width of the
radiation peak. In fact, taking the first zero of the
$\sin(\cdot)/(\cdot)$ function in Eq. (\ref{undurad3b}) is just a
convention: as we will see, only through similarity techniques we
will be able to determine the natural angle which fits with the
physical situation under study: $\theta$ can be compared in a
natural way only with that angle, and not with the solution of Eq.
(\ref{angueq}). Yet, following textbooks and tradition we will
call with $\theta_{c}$ the solution in $\Delta \theta$ of Eq.
(\ref{angueq}), and we will refer to it as a rough indication of
the angular width of the radiation for the $1st$ harmonic. The
cone with aperture $\theta_{c}$ is usually called central cone. It
can be found that (see \cite{WIED})

\begin{equation}
\theta_{c} = {1\over{\sqrt{2 N_w}\gamma_z }} \label{ccone}
\end{equation}
where $\gamma_z=\gamma/\sqrt{1+K^2/2}$. From the previous
discussion follows that Eq. (\ref{undurad3b}) can be drastically
simplified under the approximation $N_w \gg 1$ and within the
central cone $\theta \lesssim \theta_{c}$. In particular when
$k=1$, that is for the fundamental harmonic, we have

\begin{eqnarray}
\widetilde{\boldsymbol{E}}_{\bot}= -{K \omega e \lambda_w N_w e^{i \Phi_s}
\over{c^2 z_o \gamma}} {\sin (\pi N_w \Delta \omega_1/\omega_1)
\over{\pi N_w \Delta \omega_1/\omega_1}} &&\cr \times
A_{JJ}\left({\omega K^2 \over{8 k_w c
\gamma^2}}\right)\boldsymbol{\hat{x}} ~. \label{undurad4}
\end{eqnarray}
Here we neglected all terms but the one proportional to $N_w$.
Also, $A_{JJ}(u) \equiv J_0(u) - J_1(u)$, and we have used the
fact that $K\ll\gamma$.

Note that at this point we are not able to specify the accuracy of
this approximation precisely: we can only say that, when $\theta
\sim \theta_c$, our accuracy scales as $1/N_w$. This can be seen
showing that the third term in $\boldsymbol{\hat{x}}$ and the term in
$\boldsymbol{\hat{y}}$ in the series of Eq. (\ref{undurad3b}) can be
neglected with an error $\sim 1/N_w$. First we fix $k=1$, that is
the fundamental harmonic. Second we note, from Eq. (\ref{defi}),
that for any value of $K$ and $\theta$ much smaller than
$1/\gamma_z$ we have $u$ of order unity while $v$ is much smaller
than unity, so that we can take it as a small parameter: then, the
largest contributions from the sum in Eq. (\ref{undurad3b}) will
be for the smallest indexes of the Bessel functions $J(v)$ because
of the asymptotic behavior of $J_q(v) \sim v^q$. It follows that
the third term in $\boldsymbol{\hat{x}}$ and the term in $\boldsymbol{\hat{y}}$ in
the series of Eq. (\ref{undurad3b}) are of magnitude $\theta
J_0(u) J_1(v)$ or $\theta J_1(u) J_{-1}(v)$ to be compared with
the first two terms in $\boldsymbol{\hat{x}}$ scaling as $K/(2\gamma)
J_0(u)J_0(v)$ and $K/(2\gamma) J_1(u)J_0(v)$. 

This situation refers to the case when trigonometric  functions  in
$\phi$ are of order unity, otherwise extra factors $\cos^2(\phi)$
or $\sin(\phi)\cos(\phi)$ should be included as well. Taking the
trigonometric factors of order unity gives an upper limit to the
accuracy of our calculation and we will always do this. Whatever
the value of $u$, neglecting the third term in $\boldsymbol{\hat{x}}$ and
the term in $\boldsymbol{\hat{y}}$ in Eq. (\ref{undurad3b}) can be done
with an accuracy given by the ratio

\begin{equation}
R = {\theta Q/2\over{K/(2\gamma)}} = \gamma_z^2 \theta^2~.
\label{accuratio}
\end{equation}
Then, when $\theta \sim \theta_c$ we obtain $R\sim 1/N_w$. Yet, as
has already been remarked, $\theta_c$ is not the natural angle to
which $\theta$ should be compared and as a result we cannot
specify numerical factors in front of the $1/N_w$. Later we will
comment further on this point.

Eq. (\ref{undurad4}) is indeed well-known,  but usually no
attention is paid to its deeper meaning. As we have said before,
it does not make sense to calculate the intensity distribution
from Eq. (\ref{undurad}) alone, without  extra interfering terms.
Yet, we have told that, in \textit{particular situations},  the
contribution from Eq. (\ref{undurad}) is dominant with respect to
others and that in this case Eq. (\ref{undurad}), alone, is
endowed with physical meaning. On the other hand, we have seen
that Eq. (\ref{undurad}) simplify to Eq. (\ref{undurad4}) if
$k=1$, $N_w \gg 1$ and within the central cone $\theta \lesssim
\theta_{c}$. Again, when $\theta \sim \theta_{c}$ such
simplification is valid within an accuracy scaling with $1/N_w$.
By inspection of Eq. (\ref{undurad4}) we see that, due to
resonance, the only surviving term scales with $N_w$. Any extra
term added to Eq. (\ref{undurad4}) due to non resonant devices
like bending magnets or straight lines will be simply negligible
with respect the field in Eq. (\ref{undurad4}) with an accuracy
scaling with $1/N_w$. Note that Eq. (\ref{undurad4}) is valid
\textit{regardless} the value of $K$ (in particular, in the limit
for $K \ll 1$ we have an extra small parameter, then $A_{JJ}(u)
\simeq 1$ and Eq. (\ref{undurad4}) can be further simplified).
This is the power of the resonant approximation.

We can show that the same result is achievable starting with Eq.
(\ref{generalfin}) and neglecting the gradient terms in Eq.
(\ref{generalfin}) (including, therefore, the entire
$\boldsymbol{\hat{y}}$-polarization contribution) and also the constrained
particle motion in that part of the phase $\Phi_T$ which follows
from the Green's function, that can be found in the  second term
in Eq. (\ref{totph}). In fact, with these prescriptions, Eq.
(\ref{undurad}) and Eq. (\ref{phitundu}) reduce to

\begin{eqnarray}
\widetilde{\boldsymbol{E}}_\bot= {\omega e K\over{2\gamma c^2 z_o}}
\int_{-N_w\lambda_w/2}^{N_w\lambda_w/2} dz' {e^{i \Phi_T}}
\left(e^{i k_w z'} - e^{-i k_w z'}\right)\boldsymbol{\hat{x}}&&\cr
\label{unduradsimpli}
\end{eqnarray}
and

\begin{eqnarray}
\Phi_T = {\omega} \left[{z'\over{2 \gamma_z^2 c}} +
{\theta^{2}z'\over{2 c}}  - {1\over{\beta c}} {K^2\over{8\gamma^2
k_w}} \sin(2 k_w z')\right]+\Phi_s~.&&\cr\label{phitundusimpli}
\end{eqnarray}
Eq. (\ref{unduradsimpli}) can be rewritten as

\begin{eqnarray}
\widetilde{\boldsymbol{E}}_\bot= {\omega e K\over{2\gamma c^2 z_o}}
\int_{-N_w\lambda_w/2}^{N_w\lambda_w/2} dz'  \left\{\exp\left[i
\left({\omega \over{2\gamma_z^2c}}+k_w\right)z'\right]
\right.&&\cr\left. -\exp\left[i \left({\omega
\over{2\gamma_z^2c}}-k_w\right)z'\right]\right\} \exp\left[i\omega
\left({\theta^2 z'\over{2c}}-{1\over{\beta c}}
\right.\right.&&\cr\left.\left.\times {K^2\over{8\gamma^2 k_w}}
\sin(2 k_w z')\right)+\Phi_s\right] \boldsymbol{\hat{x}}&&\cr
~.\label{unduradsimpli2}
\end{eqnarray}
%
%

Introducing the detuning parameter $C$

\begin{equation}
C = {\omega \over{2\gamma_z^2c}}-k_w ~,\label{detC}
\end{equation}
Eq. (\ref{unduradsimpli2}) can be written as

\begin{eqnarray}
\widetilde{\boldsymbol{E}}_\bot= -{\omega e K\over{2\gamma c^2 z_o}}
\int_{-N_w\lambda_w/2}^{N_w\lambda_w/2} dz' &&\cr \times
\left[1-\exp(2 i k_w z')\right]\exp \left[i \left(C + {\omega
\theta^{2}\over{2 c}}\right)z'\right] &&\cr \times \exp
\left\{i\omega \left[- {1\over{\beta c}} {K^2\over{8\gamma^2 k_w}}
\sin(2 k_w z')\right]+\Phi_s \right\} \boldsymbol{\hat{x}}
\label{undusimpli2}
\end{eqnarray}
Note that $\widetilde{\boldsymbol{E}}_\bot$ has maximal magnitude when $C
+ {\omega \theta^{2}/{2 c}}=0$ because otherwise the integrand in
Eq. (\ref{undusimpli2}) starts displaying oscillatory behavior.
For $\theta=0$ this simply means $C=0$. Condition

\begin{equation}
C + {\omega \theta^{2}\over{2 c}}=0
 \label{reeescon} \end{equation}
is called the resonant condition.

With the help of Eq. (\ref{alfeq}),  Eq. (\ref{undusimpli2}) can
be transformed to

\begin{eqnarray}
\widetilde{\boldsymbol{E}}_\bot= -{\omega e K\over{2\gamma c^2 z_o}}
e^{i\Phi_s} \sum_{m=-\infty}^{\infty} J_m\left(-
{\omega\over{\beta c}} {K^2\over{8\gamma^2 k_w}} \right)&&\cr
\times \int_{-N_w\lambda_w/2}^{N_w\lambda_w/2} dz' \left[1-\exp(2i
k_wz')\right]&&\cr \times \exp \left[i \left(C + {\omega
\theta^{2}\over{2 c}}\right)z'\right] \exp \left[2 i m k_w
z'\right] \boldsymbol{\hat{x}} \label{undusimpli3a}
\end{eqnarray}
The only non-zero terms are for $m = 0$ and $m=-1$ so that

\begin{eqnarray}
\widetilde{\boldsymbol{E}}_\bot= -{\omega e K\over{2\gamma c^2 z_o}}
e^{i\Phi_s}  A_{JJ}&&\cr \times \int_{-L_w/2}^{L_w/2} dz' \exp
\left[i \left(C + {\omega \theta^{2}\over{2 c}}\right)z'\right]
\boldsymbol{\hat{x}} ~,\label{undusimpli3}
\end{eqnarray}
where the argument of $A_{JJ}$ is implied. Eq. (\ref{undusimpli3})
can be integrated leading to

\begin{eqnarray}
\widetilde{\boldsymbol{E}}_{\bot}= -{K \omega e L_w e^{i \Phi_s} \over{c^2
z_o \gamma}} {\sin [C L_w/2 + \omega L_w\theta^{2}/(4 c)] \over{C
L_w/2 + \omega L_w \theta^{2}/(4 c)}} A_{JJ}\boldsymbol{\hat{x}} ~.&&\cr
\label{undurad4bis}
\end{eqnarray}
Eq. (\ref{undurad4bis}) is equivalent to Eq. (\ref{undurad4}).
This can be seen noting that the argument in the resonant term can
be written as

\begin{eqnarray}
\pi N_w {\Delta \omega_1(\theta)\over{\omega_1(\theta)}} = \pi N_w
\left[{\omega - \omega_1(0)\over{\omega_1(0)}}+{\omega
\theta^{2}\over{2 k_w c}}\right] &&\cr = \left[{L_w
C\over{2}}+{\omega L_w \theta^{2}\over{4c}} \right]~,
\label{resoterm}
\end{eqnarray}
as it can be readily shown. The previous result can be cast in a
more compact way, suitable for further manipulations, using the
following normalized quantities

\begin{eqnarray} \hat{E}_{\bot} = -{c^2 z_o \gamma
\tilde{E}_{\bot} \over{K \omega e L_w A_{JJ}}} ~,&&\cr \hat{C} =
L_w C = 2 \pi N_w \Delta
\omega_1/\omega_1~,&&\cr\hat{\theta}=\theta \sqrt{{\omega
L_w\over{c}}}~,&&\cr \hat{z}={z\over{L_w}}~.\label{C}
\end{eqnarray}
%
As remarked before, introduction of similarity techniques involves
recognition of natural quantities which enter in the normalization
of equations in one possible way only and fit with the physical
nature of the problem. For instance, after the introduction of the
normalized angle $\hat{\theta}$ we can say that $\theta$ is to be
compared, naturally, with the angle $[c/(\omega L_w)]^{1/2}$ and
not with $\theta_c$. When $\theta\sim [c/(\omega L_w)]^{1/2}$, Eq.
(\ref{accuratio}) gives, naturally, the numerical factor in front
of $1/N_w$ that we were not able to specify before. In particular,
substitution in Eq. (\ref{accuratio}) gives the following accuracy
for our calculations:

\begin{equation}
R = {1\over {4\pi N_w}}~.\label{accufinal}
\end{equation}
Using Eq. (\ref{C}), Eq. (\ref{undusimpli3}) and Eq.
(\ref{undurad4bis}) can be written respectively as

\begin{eqnarray}
\hat{E}_\bot=  e^{i\Phi_s}  \int_{-1/2}^{1/2} d\hat{z}' \exp
\left[i \left(\hat{C} + {\hat{
\theta}^{2}\over{2}}\right)z'\right] &&\cr \label{undunorm}
\end{eqnarray}
and

\begin{eqnarray}
\hat{E}_{\bot}=  e^{i \Phi_s} {\sin [\hat{C} /2 +
\hat{\theta}^{2}/4 ] \over{\hat{C} /2 + \hat{\theta}^{2}/4 }} ~,
\label{unduradnorm}
\end{eqnarray}
which are valid in the limit for $\hat{z} \gg 1$, $4 \pi N_w \gg
1$, and $\hat{C} + \hat{\theta}^{2}/{2}\ll 4 \pi N_w$. The last
relation comes from the fact that  we are interested in the
fundamental harmonic of our device; then $\Delta\omega_1/\omega_1
\ll 1$; use of Eq. (\ref{resoterm}) gives then  $\hat{C} +
\hat{\theta}^{2}/{2}\ll 4 \pi N_w$. Note that for $\hat{C} +
\hat{\theta}^{2}/{2}\lesssim 1$, that is near resonance, the
formation length of the system becomes $L_f \sim L_w$, as one can
readily see imposing that the phase of the integrand in Eq.
(\ref{unduradnorm}) be of order unity.

As already remarked, although Eq. (\ref{undunorm}) and Eq.
(\ref{unduradnorm}) are well known in literature (at least in
their dimensional form), our discussion is far from being a mere
repetition of textbook material. In fact we put particular
attention to the assumptions used in order to obtain them, which
are completely neglected in textbooks. Our method helped to
clarify these assumptions. Now, after this step, Eq.
(\ref{undunorm}) and Eq. (\ref{unduradnorm}) constitute the
starting point for further manipulations:  the investigation of
the near-field effects treated in the next Paragraph, and the
derivation of the field in case of an electron with offset and
deflection, treated in Paragraph \ref{sec:ofdef}.

\subsection{\label{sec:near} Near-field effects}

As it has just been said, Eq. (\ref{unduradnorm}) is derived under
several assumptions: $\hat{z}_o \gg 1$, $4\pi N_w \gg 1$, and
$\hat{C} + \hat{\theta}^{2}/{2}\ll 4\pi N_w$. Here we will relax
the condition $\hat{z}_o \gg 1$ and we will treat near field
effects, showing how we can control the accuracy of calculations
using our method.

We start considering the near field region, where $\hat{z}_o
\gtrsim 1/2$. In practice, the field at the very end of the
undulator is not interesting because detectors are, in practice,
never put at $\hat{z}_o = 1/2$. Therefore in practical situation
we will be always interested in situation where $\hat{z}_o \gtrsim
1$ or $\hat{z}_o \gtrsim 2$. In this region, suitable for our
investigations, the assumption $4 \pi N_w \gg 1$, and the resonant
approximation are still retained valid, and allow substantial
simplification of the equation for the field at the fundamental
harmonic, much like Eq. (\ref{unduradnorm}) is a simplification of
Eq. (\ref{undurad3b}). Using the same line of reasoning in the
last Paragraph, but without expanding expressions $(\hat{z}_o
-\hat{z}')^{-1}$ around $\hat{z}_o$, we obtain the following
simplified expression for the field, that is

\begin{eqnarray}
\hat{E}_\bot=  \hat{z}_o  \int_{-1/2}^{1/2} d\hat{z}'
{1\over{\hat{z}_o-\hat{z}'}}&&\cr \times {\exp \left[i
\left(\hat{C} z'+
{\hat{x}_o^{2}+\hat{y}_o^2\over{2(\hat{z}_o-\hat{z}')}}\right)\right]
}  ~.\label{undundu}
\end{eqnarray}
Here we introduced normalized units $\hat{x}_o = x_o [\omega/(L_w
c)]^{1/2}$ and $\hat{y}_o = y_o [\omega/(L_w c)]^{1/2}$. These
definitions for normalized units, together with the ones given in
(\ref{C}), are naturally dictated by the system itself, through
the non-normalized equations which describe it. In particular,
$\hat{x}_o$ and $\hat{y}_o$ can be derived from the definitions of
$\hat{\theta}$ and $\hat{z}$ in (\ref{C}). Using normalized units
allows one to compare each physical quantity with its natural
measure and has the advantage of reducing the number of parameters
that the system depends on to the few fitting the
physical characteristics of the problem. Note that because of the
resonant approximation, the phase factor in Eq. (\ref{undundu}) is
much smaller than $4\pi N_w$.

Eq. (\ref{undundu}) is valid for any value of $\hat{z}_o$ with an
accuracy $1/(4\pi N_w)$. We may wish to push analytical
investigations further: in fact we have now full control over the
expansion of $(\hat{z}_o -\hat{z}')^{-1}$, meaning that we can
decide when to truncate the series

\begin{equation}
(\hat{z}_o -\hat{z}')^{-1} =
{1\over{\hat{z}_o}}\sum_{n=0}^{\infty}
\left({\hat{z}'\over{\hat{z}_o}}\right)^n \label{seriezz}
\end{equation}
both in the first factor and in the exponential factor of Eq.
(\ref{undundu}).

Neglecting terms of order higher than $n=m$ in the expansion of
the first factor in Eq. (\ref{undundu}) can be done with an
accuracy better than $({\hat{z}'/{\hat{z}_o}})^{m+1} \sim
1/(\hat{z}_o)^{m+1}$ so that, with this accuracy, we have

\begin{eqnarray}
\hat{E}_\bot=  \sum_{n=0}^{m} {1\over{\hat{z}_o^n}}
\int_{-1/2}^{1/2} d\hat{z}' \hat{z}'^{n}&&\cr \times {\exp \left[i
\left(\hat{C} z'+
{\hat{x}_o^{2}+\hat{y}_o^2\over{2(\hat{z}_o-\hat{z}')}}\right)\right]
}  \label{undutrunc1}
\end{eqnarray}
Now we should study the exponential factor. Let us keep orders up
to  $l=j$:

\begin{equation}
{\hat{x}_o^{2}+\hat{y}_o^2\over{2(\hat{z}_o-\hat{z}')}} =
{\hat{\theta}^2\hat{z}_o\over{2}}
\sum_{l=0}^{j}\left({\hat{z}'\over{\hat{z}_o}}\right)^l+
O(\hat{z}'^{j+1})~, \label{expphase}
\end{equation}
where $\hat{\theta}^2 = \hat{x}_o^2/\hat{z}_o^2 +
\hat{y}_o^2/\hat{z}_o^2$. The first term to be neglect will give a
contribution to the integrand equal to $\exp [i {\hat{\theta}^2
\hat{z}'^{j+1}/(2\hat{z}_o^{j}})]$. Now, if $(\hat{z}_o
)^{-j}\cdot \hat{\theta}^2/2< 1$, this exponential contribution
can be expanded too, and the dominant term in the expansion, after
unity,  will be just of order ${\hat{\theta}^2/(2
\hat{z}_o^{j})}$. Neglecting this term can therefore be done with
an accuracy $(\hat{z}_o )^{-j}\cdot \hat{\theta}^2/2$. If we
impose that this accuracy be of order $1/(\hat{z}_o)^{m+1}$, we
have automatically that $(\hat{z}_o )^{-j}\cdot \hat{\theta}^2/2<
1$  and we find the useful condition

\begin{equation}
{\hat{\theta}^{2}\over{2}} \sim (z_o)^{j-m-1} ~.\label{accuaccu}
\end{equation}
This can be used to retain important terms: for instance, given a
maximal observation angle of interest and an accuracy to be
reached, for a certain setup, we find $j$ such that $(\hat{z}_o
)^{-j}\cdot \hat{\theta}^2/2$ is the desired accuracy; then we
find $m$ such that Eq. (\ref{accuaccu}) is satisfied. To give a
numerical example, if we choose $\hat{z}_o \simeq 5$ and
$\hat{\theta}^2/2 \simeq 1$ and we want to get our result with an
accuracy of about $4\%$ we can put $j=2$; of course we should
compare this $4\%$ accuracy with the accuracy of the resonant
approximation  $1/(4 \pi N_w)$, which is usually about $1\%$ or
smaller. Then, solving Eq. (\ref{accuaccu}) for $m$ we find $m=1$.
Note that expansions up to $m=1$ and $j=2$ are easy to solve
analytically. One obtains the following result:

\begin{eqnarray}
\hat{E}_\bot=    e^{i\Phi_s} &&\cr \times \left\{
\int_{-1/2}^{1/2} d\hat{z}'
  \exp \left[i \left(\bar{U} z'+ \bar{W}
\hat{z}'^2\right)\right] \right.&&\cr\left. +
 {1\over{\hat{z}_o}}\int_{-1/2}^{1/2}
d\hat{z}' \hat{z}' \exp \left[i \left(\bar{U} z'+ \bar{W}
\hat{z}'^2\right)\right] \right\} \label{undundu2}
\end{eqnarray}
where

\begin{equation}
\bar{U} = \hat{C} + {\hat{\theta}^2\over{2}} ~.\label{Upar}
\end{equation}
and

\begin{equation}
\bar{W} =  {\hat{\theta}^2\over{2 \hat{z}_o}} \label{Wpar}
\end{equation}
The parameter $\bar{W}$ is closely related to the near field
parameter $W = L_w^2 \theta^2/(2 \lambda z_o)$ introduced in
\cite{WALK}. In fact, once translated in dimensional units,
$\bar{W} = 2\pi W$. We are therefore able to reproduce the results
in \cite{WALK}, but this time accounting, thanks to our approach,
for accuracy and applicability region of the approximations.

\begin{figure}
\begin{center}
\includegraphics*[width=85mm]{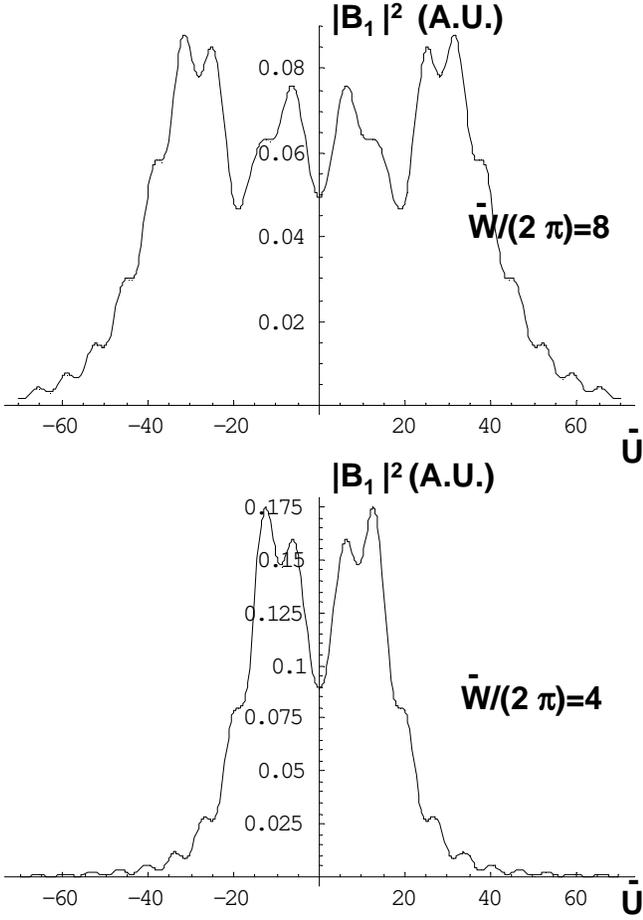}
\caption{\label{fresn1} Intensity, in arbitrary units, from the
term $B_1$, as a function of $\bar{U}$ for $\bar{W}/(2\pi) = 8$
(upper plot) and for $\bar{W}/(2\pi) = 4$ (lower plot).}
\end{center}
\end{figure}

\begin{figure}
\begin{center}
\includegraphics*[width=85mm]{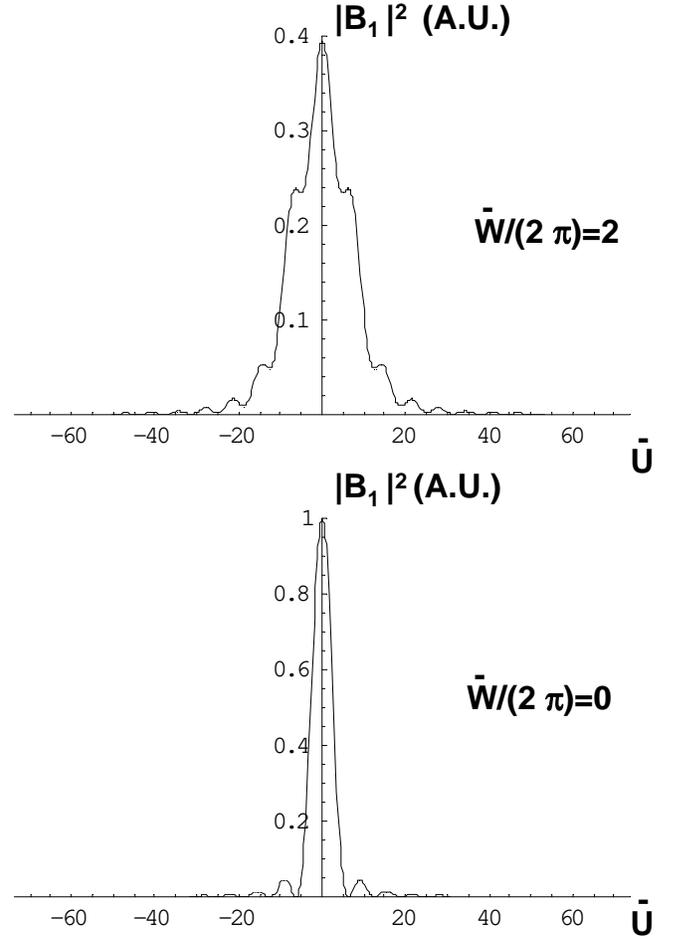}
\caption{\label{fresn2} Intensity, in arbitrary units, from the
term $B_1$, as a function of $\bar{U}$ for $\bar{W}/(2\pi) = 2$
(upper plot) and for $\bar{W}/(2\pi) = 0$ (lower plot).}
\end{center}
\end{figure}

The integrals in Eq. (\ref{undundu2}) can be calculated
analytically leading to

\begin{eqnarray}
\hat{E}_\bot=    e^{i\Phi_s} \left[ {B_1}+ {B_2\over{\hat{z}_o}}
\right] \label{undundu2biss}
\end{eqnarray}
where

\begin{eqnarray}
B_1 = {\sqrt{\pi} (1+i) \over{2\sqrt{2
\bar{W}}}}\left[\mathrm{erf}\left({e^{3i\pi/4}
(\bar{U}-\bar{W})\over{2\sqrt{\bar{W}}}}\right) \right.
&&\cr\left. -\mathrm{erf}\left({e^{3i\pi/4}
(\bar{U}+\bar{W})\over{2\sqrt{\bar{W}}}}\right)\right]
\exp\left[{i {\bar{U}^2\over{4\bar{W}}}}\right] \label{I1}
\end{eqnarray}
and

\begin{eqnarray}
B_2 = {1\over{8 \bar{W}^{3/2}}} \left\{-4i \exp\left[i {2
\bar{U}^2+\bar{W}^2\over{4\bar{W}}}\right]\sqrt{\bar{W}}
\right.&&\cr \left.\times \left[-1+\exp(i \bar{U})\right]+ (1+i)
\bar{U} \exp\left[i
\bar{U}{\bar{U}+2\bar{W}\over{4\bar{W}}}\right] \sqrt{2\pi}
\right.&&\cr\left.\times \left[-\mathrm{erf}\left({e^{3i\pi/4}
(\bar{U}-\bar{W})\over{2\sqrt{\bar{W}}}}\right)
\right.\right.&&\cr\left.\left. +\mathrm{erf}\left({e^{3i\pi/4}
(\bar{U}+\bar{W})\over{2\sqrt{\bar{W}}}}\right)\right]\right\}
\exp\left[{i
{\bar{U}\left(\bar{U}+\bar{W}\right)\over{2\bar{W}}}}\right]
\label{I2}
\end{eqnarray}
It might be remarked that the $\mathrm{erf}$ function can be
represented in terms of Fresnel functions which were used in
presentation of results in \cite{WALK}. For the sake of comparison
with \cite{WALK} it may be interesting to plot $\mid B_1 \mid^2$
as a function of $\hat{\theta}$. When terms in $B_2$ are
negligible this represents, in normalized units, the intensity in
the near field. This is the case, for instance, if we choose
$\hat{z}_o \simeq 5$, $\hat{\theta}^2/2 \simeq 1$ and we want to
get our result with an accuracy of about $20\%$: then we may set
$j=2$ and $m=0$. Results are plotted in Fig. (\ref{fresn1}) and in
Fig. (\ref{fresn2}) as intensity as a function of $\bar{U}$ for
different values of $\bar{W}$.

The understanding of the region of applicability involved in the
derivation of Eq. (\ref{undundu2biss}) allows us to make
interesting remarks. For instance, note that for any given
negative value of the detuning parameter $\hat{C}$ there is a
value of $\hat{\theta}$ such that $\bar{U} = 0$. This means that
even when $\hat{C} = -40$, for example, the value
$\hat{\theta}^2/2 = 40$ is such that $\bar{U}=0$ and the
contribution of the near field $\bar{W}$ is dominant. In this
case, even choosing $\hat{z}_o = 40$, $\bar{W} = 1$ and letting
$m=0$ and $j=2$, the term in $\bar{W}$ is the only contribution up
to an accuracy of $5\%$ (note that this is consistent with Eq.
(\ref{accuaccu}), which gives a maximal angle $\hat{\theta}^2/2
\sim 40$. This is a very particular region of parameters, where
the electric field depends on the near field parameter $\bar{W}$
even though $\hat{z}_o \gg 1$. In this particular situation we can
write

\begin{eqnarray}
\hat{E}_\bot= -   e^{i\Phi_s}
{\sqrt{\pi}(1+i)\over{\sqrt{2\bar{W}}}} \mathrm{erf}
\left[{e^{3i\pi/4}\sqrt{\bar{W}}\over{2}} \right]~.
\label{undundu3}
\end{eqnarray}

\subsection{\label{sec:ofdef} Electron motion with offset and
deflection}

The same remark given in Paragraph \ref{sub:circledef} for
electrons on a circular trajectory is valid here for particles in
an undulator: the nominal trajectory is just an approximation.
Electron beams have always some finite geometrical emittance and
they can be thought, in agreement with paraxial treatment, as a
composition of perfectly collimated beams with different
deflection angles with respect to the orbital plane of the nominal
trajectory. As remarked in Paragraph \ref{sub:circledef}, this
representation will be of great importance in the calculation of
complicated quantities like the field autocorrelation function,
which is of uttermost importance in the characterization of the
statistical properties of a light source. Let us now discuss how
to calculate $\widetilde{\boldsymbol{E}}$ from a single particle moving in
an undulator with a given angular deflection and offset with
respect to the orbital plane of a nominal electron. Once we answer
this question for a single particle, we can add up contributions
from different electrons the way we want, in perfect symmetry with
Paragraph \ref{sub:circledef}. Let us introduce electron
deflection angles $\eta_x$ and $\eta_y$ and displacements $l_x$
and $l_y$. The particle velocity will be characterized, as a
function of $z'$, as

\begin{equation}
\boldsymbol{v_\bot}(z') = \left[- {c K\over{\gamma}} \sin{\left(k_w
z'\right)}+\eta_x v_z\right] \boldsymbol{x}+\left[\eta_y v_z
\right]\boldsymbol{y}~, \label{vuz2}
\end{equation}
so that

\begin{eqnarray}
\boldsymbol{r'_\bot}(z') = \left[ {K\over{\gamma k_w}}
\left(\cos{\left(k_w z'\right)}-1\right)+\eta_x z' +l_x\right]
\boldsymbol{x} &&\cr + \left[\eta_y z' +l_y\right]\boldsymbol{y}~. \label{erz2}
\end{eqnarray}
Using

\begin{eqnarray}
s(z') = \int_{0}^{z} dz' \left[1+\left({d
x'\over{dz'}}\right)^2+\left({d
y'\over{dz'}}\right)^2\right]^{1/2} \simeq &&\cr \int_{0}^{z} dz'
\left[1+{1\over{2}}\left({d
x'\over{dz'}}\right)^2+{1\over{2}}\left({d
y'\over{dz'}}\right)^2\right] \label{curvascgen}
\end{eqnarray}
one has

\begin{eqnarray}
s(z') =
\left({{\beta}\over{\beta_{av}}}+{\eta_x^2+\eta_y^2\over{2}}\right)z'
- {K \eta_x\over{k_w \gamma}} && \cr- {K^2\over{8\gamma^2
k_w}}\sin\left(2 k_w z' \right)+ {K \eta_x\over{\gamma
k_w}}\cos\left(k_w z' \right)~.\label{zsss2}
\end{eqnarray}
We will now work under the approximations:  $\hat{z}_o \gg 1$, $4
\pi  N_w \gg 1$ and we are interested in the fundamental harmonic
of our device. We can use a procedure analogous to the one used in
Paragraph \ref{sec:res} (and Paragraph \ref{sec:near}) to get the
following simplified expression for the field in normalized units
in resonant approximation:

\begin{eqnarray}
\hat{E}_\bot=  e^{i\Phi_U} \int_{-1/2}^{1/2} d\hat{z}' \exp
\Bigg[i \hat{z}' \Bigg(\hat{C} &&\cr+
{1\over{2}}\left(\hat{{\theta}}_x-{\hat{l}_x\over{\hat{z}_o}}-
\hat{\eta}_x\right)^2
+{1\over{2}}\left(\hat{{\theta}}_y-{\hat{l}_y\over{\hat{z}_o}}-\hat{\eta}_y
\right)^2 \Bigg)\Bigg] ~,~ \label{undunormfin}
\end{eqnarray}
where
\begin{equation}
\Phi_U =
\left[\left(\hat{\theta}_x-{\hat{l}_x\over{\hat{z}_o}}\right)^2
+\left(\hat{\theta}_y-{\hat{l}_y\over{\hat{z}_o}}\right)^2\right]
{\hat{z}_o\over{2}} \label{phisnorm}
\end{equation}
The same normalization for $\theta$ holds also for $\eta_{x,y}$.
The accuracy of Eq. (\ref{undunormfin}) is given by the ratio of
$\hat{C} + {1/{2}}\left(\hat{{\theta}}_x-{\hat{l}_x/{\hat{z}_o}}-
\hat{\eta}_x\right)^2+{1/{2}} \left(\hat{{\theta}}_y-
{\hat{l}_y/{\hat{z}_o}}-\hat{\eta}_y \right)^2$ to $4 \pi N_w$.
When $\hat{C}\lesssim 1$ and $(\hat{{\theta}}_{x,y}
-{\hat{l}_{x,y}/{\hat{z}_o}}- \hat{\eta}_{x,y})^2 \lesssim 1$ the
accuracy is just $\sim 1/(4\pi N_w)$. Finally, Eq.
(\ref{undunormfin}) can be integrated giving

\begin{equation}
\hat{E}_{\bot}= e^{i \Phi_U} {\sin \zeta\over{\zeta}} ~,
\label{endangle}
\end{equation}
where $\Phi_U$ is given in Eq. (\ref{phisnorm}), while

\begin{equation}
\zeta = {\hat{C}\over{2}} +
{1\over{4}}\left(\hat{{\theta}}_x-{\hat{l}_x\over{\hat{z}_o}}-\hat{\eta}_x\right)^2
+{1\over{4}}\left(\hat{{\theta}}_y-{\hat{l}_y\over{\hat{z}_o}}-\hat{\eta}_y\right)^2~.
\label{zeta}
\end{equation}
In the limit when $\hat{l}_{x,y}/\hat{z}_o \ll 1$, one can
simplify further Eq. (\ref{endangle}) thus getting:

\begin{equation}
\hat{E}_{\bot}= e^{i \Phi_s} e^{i\Phi_o} {\sin \zeta\over{\zeta}}
~, \label{endangle22}
\end{equation}
where
\begin{equation}
\Phi_s = \left(\hat{\theta}_x^2+\hat{\theta}_y^2\right)
\hat{z}_o~, \label{phisnorm22}
\end{equation}
\begin{equation}
\Phi_o = -\hat{\theta}_x \hat{l}_x - \hat{\theta}_y \hat{l}_y~,
\label{phionorm22}
\end{equation}
and

\begin{equation}
\zeta = {\hat{C}\over{2}} +
{1\over{4}}\left(\hat{{\theta}}_x-\hat{\eta}_x\right)^2
+{1\over{4}}\left(\hat{{\theta}}_y-\hat{\eta}_y\right)^2~.
\label{zeta22}
\end{equation}
The same remarks given for Eq. (\ref{srtwoang2}) hold for Eq.
(\ref{endangle}), and we will repeat them here, since we consider
them of great importance. Eq. (\ref{endangle}) is an extremely
useful tool, because it describes the radiation from an electron
with offset and deflection as in an electron beam with finite
emittance, including the correct phase factor for the field. Eq.
(\ref{srtwoang2}) was derived here from first principles. An
alternative derivation based on the intuitive picture that the
undulator radiation can be approximated as a sum of spherical
waves emitted at the entrance of each pole is given in
\cite{TAK2}. Accounting for correct phase means that contributions
from different electrons can be simply added up to give the field
from a beam with given emittance at any observer position.
Starting from Eq. (\ref{endangle}) then, it is possible to
calculate the field correlation function, and to provide a study
of transverse coherence properties of the radiation from a given
electron beam by means of analytical techniques. Again, in perfect
analogy with what has been remarked in Paragraph
\ref{sub:circledef}, a numerical code can always be developed,
either starting from Eq. (\ref{generalfin}) or just from the
Lienard-Wiechert fields, which calculates the field correlation
function in a generic case, but such a code would not help in
physical understanding of the situation. On the contrary, Eq.
(\ref{endangle}) includes all relevant information about an
electron in a realistic beam (i.e. with offset and deflection)
and, being an analytically manageable expression, constitutes a
first step towards the more ambitious goal of characterization of
transverse coherence properties from undulator radiation, which we
leave for future work.

\subsection{\label{sec:ofdef2} The effects of electron beam emittance
on the basic characteristics of undulator radiation}

In the next Section \ref{sec:appl}, we will make use of Eq.
(\ref{endangle}) to calculate  radiation from a complex system: we
will demonstrate in this way the power and the practical
convenience of our computational method. In this Paragraph
instead, we will study in detail how to apply Eq. (\ref{endangle})
to calculate angular distribution and frequency spectrum of
radiation produced by an electron beam in a standard undulator
when electron beam emittance is present. We restrict our attention
to the asymptotic case of a large horizontal and a small vertical
emittance. This limiting situation is, in fact, of great practical
interest for today's third generation light sources and it will be
discussed here in order to illustrate the effectiveness of
similarity techniques.

First we assume that we can use Eq. (\ref{endangle22}) in place of
Eq. (\ref{endangle}). Indicating with $\beta_{o x,y}$ the minimal
values of the betatron function in the $x$ and $y$ directions and
with $h_{x,y}(\eta_{x,y})$ the angular distributions of the
particles, again with respect to horizontal and vertical
directions we have

\begin{equation}
h_{x,y}(\eta_{x,y}) = {{N_p}\over{\sqrt{2\pi} \sigma_{x',y'}}}
\exp\left( -{\eta_{x,y}^2\over{2\sigma_{x',y'}^2}} \right)~,
\label{angledi}
\end{equation}
where $\sigma_{x',y'}^2 = \varepsilon_{x,y}/\beta_{o x,y}$. Upon
introduction of

\begin{eqnarray}
\hat{\beta}_o = L_w^{-1} \beta_o~,&&\cr \hat{\varepsilon}=
({\omega/{c}}) \varepsilon~,&&\cr \label{adqua}
\end{eqnarray}
where we neglected all indexes in $x, y, x'$ and $y'$ for
notational simplicity, Eq. (\ref{angledi}) can be rewritten as a
function of normalized quantities as

\begin{equation}
h(\hat{\eta}) = {1\over{\sqrt{2\pi
\hat{\varepsilon}/\hat{\beta}_o}}} \exp\left( -{\hat{\eta}^2
\hat{\beta_o} \over{2\hat{\varepsilon}}} \right)~,
\label{anglediadi}
\end{equation}
We will treat the limiting case $\hat{\varepsilon}_y /
\hat{\beta}_{o y} \ll 1$. When $\hat{\beta}_{o y} \sim 1$ as it
usually is the case, this is simply a condition on the normalized
emittance $\hat{\varepsilon}_y \ll 1$.

In this particular case, the particle density distribution in the
vertical phase space behaves like a $\delta$-distribution. Then,
the beam intensity is simply given by

\begin{equation}
{I}_{b} = {N_p c\over{4 \pi \sqrt{2\pi} \sigma_{x'}}}
\int_{-\infty}^{\infty} d{\eta}_x |\tilde{E}_\bot|^2 \exp \left(
-{{\eta}_x^2 \over{2 \sigma_{x'}^2}}\right)~. \label{Ilimitnonorm}
\end{equation}
or, in normalized units

\begin{equation}
\hat{I}_b = {1\over{\sqrt{2\pi \hat{\varepsilon}_x/\hat{\beta}_{o
x}}}} \int_{-\infty}^{\infty} d\hat{\eta}_x
{\sin^2\zeta\over{\zeta^2}} \exp \left( -{\hat{\eta}_x^2
\hat{\beta}_{o x} \over{2 \hat{\varepsilon}_x}}\right)~,
\label{Ilimit}
\end{equation}
where

\begin{equation}
\hat{I}_b = {4 \pi \over {N_p c}}\left({c^2 z_o \gamma \over{K
\omega e L_w A_{JJ} }}\right)^2I_b\label{blab}
\end{equation}

Eq. (\ref{Ilimit}) simplifies further in the limit
$\hat{\varepsilon}_x/\hat{\beta}_{o x} \gg 1$. Also in this case,
when $\hat{\beta}_{o x} \sim 1$ as it is usually verified, this is
a condition on the normalized emittance $\hat{\varepsilon}_x \gg
1$. In this limiting situation, the function $
{\sin^2\zeta/{\zeta^2}}$ in  Eq. (\ref{Ilimit}) behaves as the
$\delta$-distribution $\delta(\hat{\theta}_x-\hat{\eta}_x)$ when
compared with the exponential function, so that Eq. (\ref{Ilimit})
can be simplified as

\begin{equation}
\hat{I}_b = {  \exp \left[ -{\hat{\theta}_x^2 \hat{\beta}_{o x}
/({2 \hat{\varepsilon}_x})}\right] \over{\sqrt{2\pi
\hat{\varepsilon}_x/\hat{\beta}_{o x}}}} \int_{-\infty}^{\infty}
d\hat{\eta}_x {\sin^2\zeta\over{\zeta^2}} ~ \label{Ilimitb}
\end{equation}

\begin{figure}
\begin{center}
\includegraphics*[width=90mm]{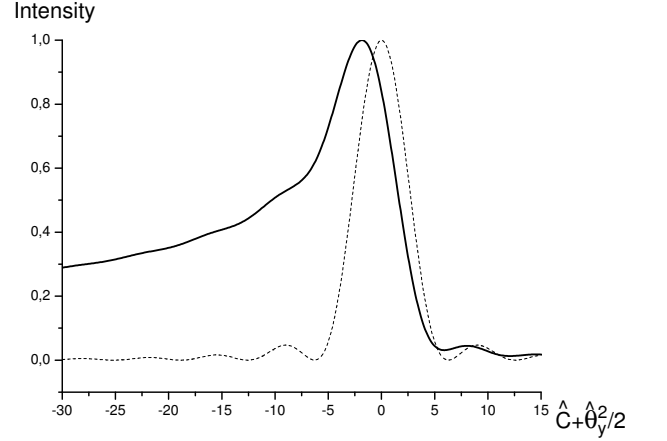}
\caption{Solid line : relative intensity $\hat{\bar{I}}$ from a
beam in the limit $\hat{\varepsilon}_x \rightarrow \infty$,
$\hat{\varepsilon}_y \rightarrow 0$. The intensity is plotted as a
function of $\hat{C}+\hat{\theta}_y^2/2$. Dashed line : single
particle intensity $\hat{I}$ as a function of
$\hat{C}+\hat{\theta}_y^2/2$ for $\theta_x=0$\label{cth} }
\end{center}
\end{figure}
\begin{figure}
\begin{center}
\includegraphics*[width=90mm]{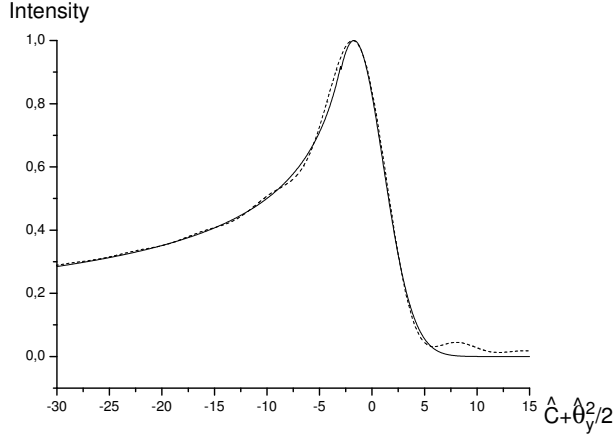}
\caption{Solid line : approximation of $\hat{\bar{I}}$ using Eq.
(\ref{HTbunch}). Dashed line : relative intensity $\hat{\bar{I}}$
from a beam in the limit $\hat{\varepsilon}_x \rightarrow \infty$,
$\hat{\varepsilon}_y \rightarrow 0$. The intensity is plotted as a
function of $\hat{C}+\hat{\theta}_y^2/2$, as the solid line in
Fig. \ref{cth}. \label{cthfit} }
\end{center}
\end{figure}
\begin{figure}
\begin{center}
\includegraphics*[width=90mm]{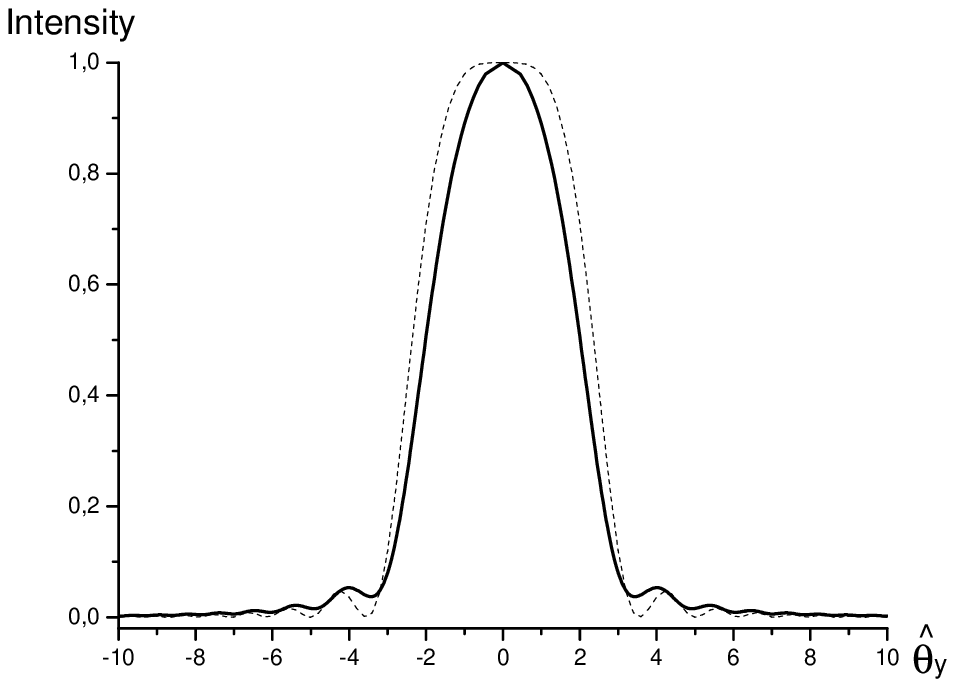}
\caption{Solid line : relative intensity $\hat{\bar{I}}$ from a
beam in the limit $\hat{\varepsilon}_x \rightarrow \infty$,
$\hat{\varepsilon}_y \rightarrow 0$. The intensity is plotted as a
function of $\hat{\theta}_y$ for $\hat{C}=0$. Dashed line : single
particle intensity $\hat{I}$ as a function of $\hat{\theta}_y^2$
for $\theta_x=0$ and $\hat{C}=0$\label{thyzero} }
\end{center}
\end{figure}
\begin{figure}
\begin{center}
\includegraphics*[width=90mm]{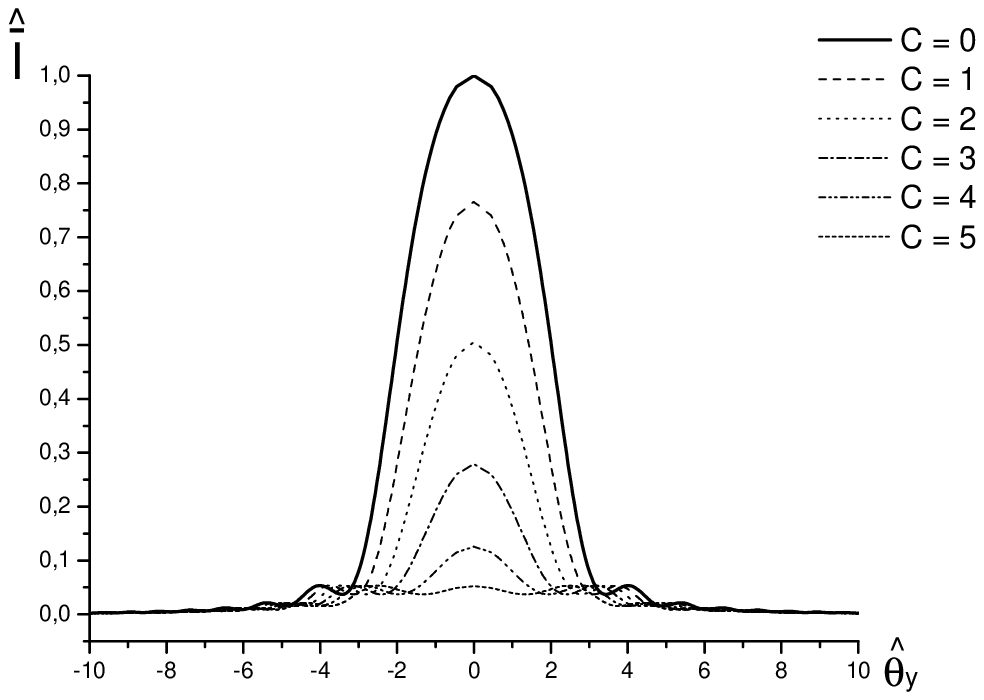}
\caption{Relative intensity $\hat{\bar{I}}$ from a beam in the
limit $\hat{\varepsilon}_x \rightarrow \infty$,
$\hat{\varepsilon}_y \rightarrow 0$. The intensity is plotted as a
function of $\hat{\theta}_y$ for different values of $\hat{C}$.
The symbol $C$ in the figure legend should be read as $\hat{C}
\geq 0$. \label{thyplus} }
\end{center}
\end{figure}
\begin{figure}
\begin{center}
\includegraphics*[width=90mm]{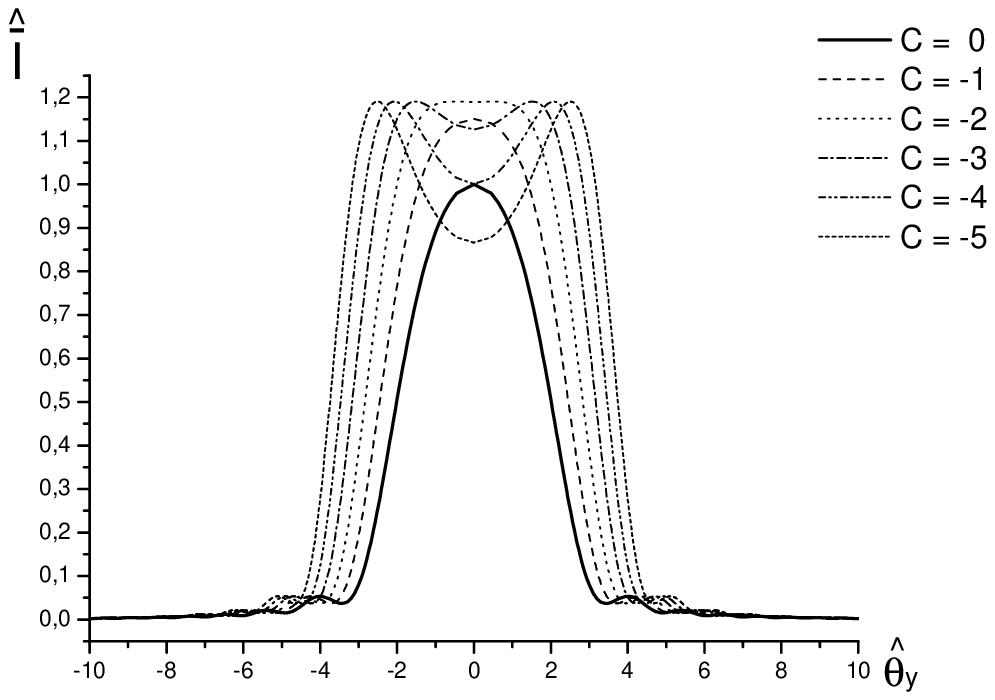}
\caption{Relative intensity $\hat{\bar{I}}$ from a beam in the
limit $\hat{\varepsilon}_x \rightarrow \infty$,
$\hat{\varepsilon}_y \rightarrow 0$. The intensity is plotted as a
function of $\hat{\theta}_y$ for different values of $\hat{C} \leq
0$. The symbol $C$ in the figure legend should be read as
$\hat{C}$. \label{thyminus} }
\end{center}
\end{figure}
Now we can introduce the relative intensity

\begin{equation}
\hat{\bar{I}} = {3\sqrt{2
\hat{\varepsilon}_x}\over{8\sqrt{\hat{\beta}_{o x}}}} \left[\exp
\left( -{\hat{\theta}_x^2 \hat{\beta}_{o x} \over{2
\hat{\varepsilon}_x}}\right)\right]^{-1} \hat{I}_b \label{fine}
\end{equation}
so that

\begin{equation}
\hat{\bar{I}} = {3\over{8 \sqrt{\pi}}} \int_{-\infty}^{\infty}
d\hat{\eta}_x {\sin^2\zeta\over{\zeta^2}} ~, \label{Ilimit2}
\end{equation}
and according to our approximations

\begin{equation}
\zeta = {\hat{C}\over{2}} +
{1\over{4}}\left(\hat{{\theta}}_x-\hat{\eta}_x\right)^2
+{\hat{{\theta}}_y^2\over{4}}~. \label{zetacaselim}
\end{equation}
With an obvious change of integration variable $\hat{\eta}_x
\rightarrow \hat{{\theta}}_x-\hat{\eta}_x$ we finally obtain

\begin{equation}
\hat{\bar{I}} = {3\over{8 \sqrt{\pi}}} \int_{-\infty}^{\infty}
d\hat{\eta}_x {\sin^2 \left[ \left({\hat{C}/{2}}
+{\hat{{\theta}}_y^2/{4}}\right)+ {\hat{\eta}_x^2/{4}}\right]
\over{\left[ \left({\hat{C}/{2}} +{\hat{{\theta}}_y^2/{4}}\right)+
{\hat{\eta}_x^2/{4}}\right]^2}} ~, \label{Ilimit3}
\end{equation}
It is now evident that similarity techniques we used are not only
a theoretical tool: in fact $\hat{\bar{I}}$ is a universal
function of $\hat{C}+\hat{\theta}_y^2/2$ and  Eq. (\ref{Ilimit3})
can be integrated with the help of simple numerical techniques to
get the universal plot in Fig. \ref{cth} (solid line). This
universal plot is compared, always in Fig. \ref{cth} (dashed
line), with the relative intensity in the case when also
$\hat{\varepsilon}_x / \hat{\beta}_{o x} \ll 1$. In this case we
simply have the single particle intensity

\begin{equation}
\hat{I} = {\sin^2 \left[{\hat{C}/{2}} +{\hat{{\theta}}_y^2/{4}}+
{\hat{\theta}_x^2/{4}}\right] \over{\left[{\hat{C}/{2}}
+{\hat{{\theta}}_y^2/{4}}+ {\hat{\theta}_x^2/{4}}\right]^2}} ~.
\label{Ilimit3bis}
\end{equation}
The dashed line in Fig. \ref{cth} refers to the case
$\hat{\theta}_x = 0$; note the different normalization factor
between $\hat{\bar{I}}$ and $\hat{I}$. It may be useful to provide
a fit of the universal plot in Fig. \ref{cth} using the following
function:

\begin{equation}
H(\xi) =  \left\{
\begin{array}{c}
~~A
\exp\left[-{(\xi-\xi_1)^2\over{2\xi_2^2}}\right]~~~~~~~\xi>\xi_0~,~~\xi_0<0
\\
\\
~~B
{\exp\left[\xi/\xi_3\right]\over{\sqrt{-\xi/\xi_3}}}~~~~~~~~~~~~~~~~~\xi<\xi_0<0
\end{array}~.\right.\label{HTbunch}
\end{equation}
Fitting for $A$, $B$, $\xi_1$, $\xi_2$ and $\xi_3$ and using, for
instance, a reasonable value for $\xi_0 = -3.00$ we find $A \simeq
5.7$, $B \simeq 0.24$, $\xi_1 \simeq 1.7$, $\xi_2 \simeq 2.8$ and
$\xi_3 \simeq 1.4 \cdot 10^3$ which result in the solid curve in
Fig. \ref{cthfit}, the dashed curve being a comparison with the
real universal function plotted in Fig. \ref{cth}. It should be
noted that $H(\xi)$ has no particular physical sense: it simply
gives an analytical approximation for the universal plot in Fig.
\ref{cth} with graphical accuracy shown in Fig. \ref{cthfit}.

After Fig. \ref{cth} is tabulated, one can plot the intensity as a
function of $\hat{\theta}_y$ for given values of $\hat{C}$ simply
solving $\hat{C}+\hat{\theta}_y^2/2=X$ for $\hat{\theta}_y$, $X$
being any of the tabulated abscissas with constraint
$X-\hat{C}>0$. By fixing $\hat{C}=0$ we obtain the universal plot
in Fig. \ref{thyzero} (solid line). This universal plot is
compared, always in Fig. \ref{thyzero} (dashed line), with the
normalized intensity as a function of $\hat{\theta}_y$ in the case
when also $\hat{\varepsilon}_x / \hat{\beta}_{o x} \ll 1$. The
dashed line in Fig. \ref{thyzero} refers to the case
$\hat{\theta}_x = 0$; again, note the different normalization
factor between $\hat{\bar{I}}$ and $\hat{I}$. Note that, for this
case, Eq. (\ref{Ilimit3bis}) calculated for $\hat{C}=0$  already
constitutes a relatively good approximation for the plot in Fig.
\ref{thyzero}.

Letting $\hat{C}$ vary we obtain universal plots of the intensity
parameterized with respect to $\hat{C}$; we show some of them in
Fig. \ref{thyplus} and Fig. \ref{thyminus} with respect to
$\hat{C}\ge 0$ and $\hat{C} \le 0$.

\begin{figure}
\begin{center}
\includegraphics*[width=90mm]{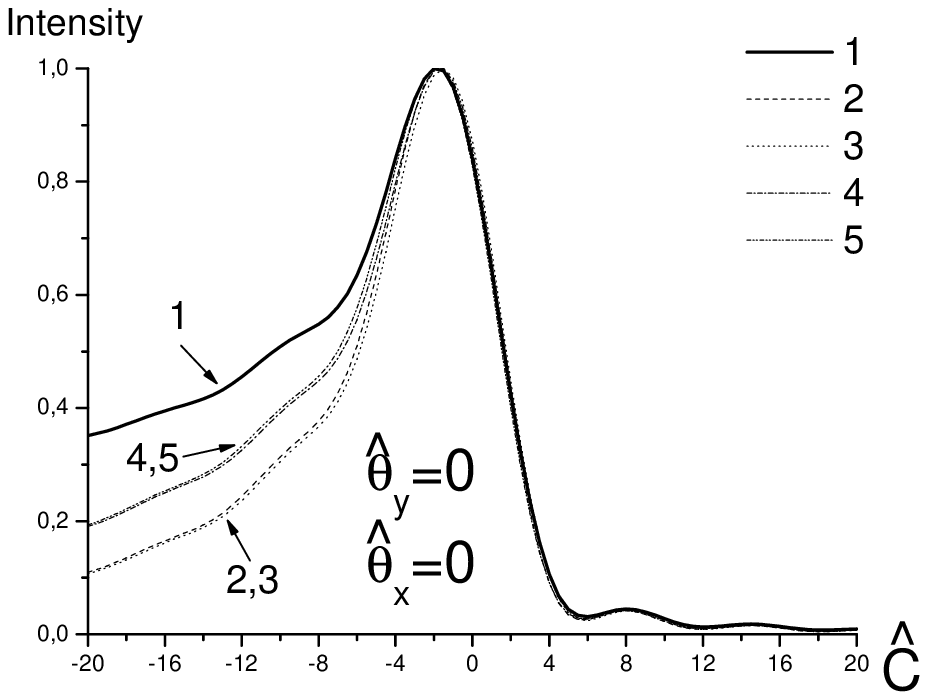}
\caption{Plots of relative intensities $\hat{\bar{I}}$ as a
function of $\hat{C}$ for $\hat{\theta}_x=0$ and
$\hat{\theta}_y=0$. Here $\hat{\beta}_{ox} = \hat{\beta}_{oy} =
1.0$. Curve 1 : relative intensity $\hat{\bar{I}}$ from a beam in
the limit $\hat{\varepsilon}_x \rightarrow \infty$,
$\hat{\varepsilon}_y \rightarrow 0$, as in Fig. \ref{cth}.  Curve
2 : relative intensity from a beam with $\hat{\varepsilon}_x=40$,
$\hat{\varepsilon}_y =1$. Curve 3 : relative intensity from a beam
with $\hat{\varepsilon}_x=40$, $\hat{\varepsilon}_y \rightarrow
0$. Curve 4 : relative intensity from a beam with
$\hat{\varepsilon}_x=80$, $\hat{\varepsilon}_y \rightarrow 0$.
Curve 5 : relative intensity from a beam with
$\hat{\varepsilon}_x=80$, $\hat{\varepsilon}_y =1$. \label{accuc}
}
\end{center}
\end{figure}

\begin{figure}
\begin{center}
\includegraphics[width=90mm]{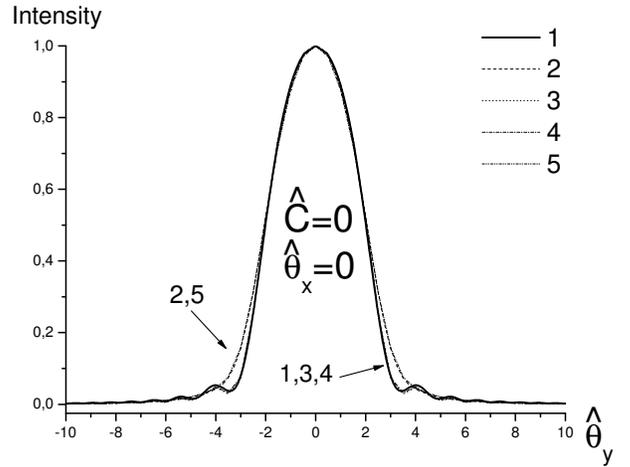}
\caption{Plots of relative intensities $\hat{\bar{I}}$ as a
function of $\hat{\theta}_y$ for $\hat{\theta}_x=0$ and
$\hat{C}=0$. Here $\hat{\beta}_{ox} = \hat{\beta}_{oy} = 1.0$.
Curve 1 : relative intensity $\hat{\bar{I}}$ from a beam in the
limit $\hat{\varepsilon}_x \rightarrow \infty$,
$\hat{\varepsilon}_y \rightarrow 0$, as in Fig. \ref{thyzero}.
Curve 2 : relative intensity from a beam with
$\hat{\varepsilon}_x=40$, $\hat{\varepsilon}_y =1$. Curve 3 :
relative intensity from a beam with $\hat{\varepsilon}_x=40$,
$\hat{\varepsilon}_y \rightarrow 0$. Curve 4 : relative intensity
from a beam with $\hat{\varepsilon}_x=80$, $\hat{\varepsilon}_y
\rightarrow 0$. Curve 5 : relative intensity from a beam with
$\hat{\varepsilon}_x=80$, $\hat{\varepsilon}_y =1$.
\label{accuthy} }
\end{center}
\end{figure}

Finally, with the help of a Monte Carlo simulation described in
more detail in the following Section \ref{sec:appl}, we compared
our results with the case of a beam with different values of
$\hat{\varepsilon}_x$ and $\hat{\varepsilon}_y$, always assuming
$\hat{\beta}_{ox}= \hat{\beta}_{oy} = 1.0$ and no energy spread is
present. Note that in general, the spectrum (at
$\hat{\theta}_{x,y} = 0$) or the angular intensity (at
$\hat{C}=0$, $\theta_x=0$) is a function of two parameters:
$\hat{\epsilon}_x/\hat{\beta}_{ox}$ and
$\hat{\epsilon}_y/\hat{\beta}_{oy}$. Results are shown in Fig.
\ref{accuc} and Fig. \ref{accuthy}, where relative intensities are
plotted as a function of $\hat{C}$ for $\hat{\theta}_x=0$ and
$\hat{\theta}_y=0$ and as a function of $\hat{\theta}_y$ for
$\hat{C}=0$ and $\hat{\theta}_x=0$, respectively. Fig. \ref{accuc}
and Fig. \ref{accuthy} illustrate the accuracy of the asymptotic
limit Eq. (\ref{Ilimit3}).

%


\section{\label{sec:appl} Set of two undulators with a focusing triplet in between}

\begin{figure}
\begin{center}
\includegraphics*[width=90mm]{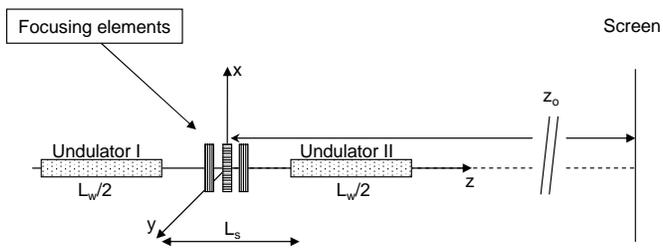}
\caption{\label{applgeo} Undulator system geometry: two undulator
segments are separated by strong focusing quadrupoles.}
\end{center}
\end{figure}

In this Section we use our understanding of approximations and
region of applicability discussed before to deal with a difficult
setup constituted by two or more undulator segments separated by
strong focusing quadrupoles. In this Section, and in Fig.
\ref{applgeo}, we consider the particular case of two undulator
segments separated by a focusing triplet.

This scheme is known as segmented undulator scheme.

Insertion devices at PETRA III are due to work according to an
\textit{on top} injection scheme and must therefore be compatible
with injection of new particles during operation. In general,
injection operations require a larger beam acceptance with respect
to an already circulating beam and this poses a strict limit to a
safe choice of the length of the insertion devices.

Since no focusing element is present inside the undulator the
$\beta_T$ function increases quadratically with the distance from
the minimal value in the middle of the undulator. In practice,
then, the maximal length of the insertion device is also fixed and
one has the rule of thumb $L_w \sim \beta_T$. Here we can safely
talk about the vertical direction only, because the vertical
$\beta_T$ function is also the shortest.

In principle, given a certain acceptance $\varepsilon_{acc}$, it
is possible to increase the $\beta_T$-function by increasing the
undulator gap according to $\beta_T \sim \sigma^2_{gap}
\varepsilon_{acc}$. However, the particle energy and the radiation
wavelength are nearly fixed at PETRA III. Moreover restrictions
related with the number of photons required and with operation of
third harmonic scheme apply, which allow only values of $K$ around
unity. Then, $\lambda_w$ is fixed as well as the magnetic field on
axis and, as a result, the undulator gap too. This means that an
upper limit on the $\beta_T$ function and therefore to the
insertion device length $L_w$ is also fixed.

In practical situation, when enough experience with the machine,
safety margins might be bent. For instance a single $25$ m-long
\textit{in vacuum} undulator operates at SPring-8 in \textit{on
top} injection mode with a vertical $\beta_T \simeq 10$ m.
However, safety limits indicate, for a first design, much more
stringent constraints.

For a first conservative design, the segmented undulator scheme
constitutes a reliable alternative for filling the first straight
section at the new arc of PETRA III, which can accommodate a $20$
m long insertion device. Breaking a long insertion device in
several parts allows one to "keep the average vertical
$\beta_T$-function rather small which in turn is essential to
allow for a small geometrical vertical aperture and magnetic gap"
(cited from \cite{PETR}, Section 4.1.8). "The maximal vertical
$\beta$-function along this undulator will be about $10$ m.
Therefore a vertical inner size of $9.5$ mm is necessary for the
corresponding chambers" (cited from \cite{PETR}, Section 3.2.1).

With this motivation in mind we start our investigation. We are
interested in computing the radiation intensity seen by an
observer at angles $\hat{\theta}_x$ and $\hat{\theta}_y$. We will
be interested in both angular and spectral intensity distribution.
Although it is possible to take advantage of the analysis
performed in Paragraph \ref{sec:near} and treat near field effect,
in this paper we will restrict our attention to the far field
limit. Then we will compare our results with the case of an
uniform undulator.

Of course radiation intensity can be obtained, numerically, from
the Lienard-Wiechert expressions for the electromagnetic field
without approximations of any kind. The previous statement though,
sounds like declaring that solution to all electromagnetic
problems can be found solving Maxwell's equation: it is
undoubtedly true, but quite generic. The most general expressions
for the Lienard-Wiechert field can be used, and have been used
(again, see \cite{CHUB} and \cite{TANA} for instance) as a basis
for numerical codes, but understanding of correct approximations
and their region of applicability can simplify many tasks a lot,
including practical and non-trivial ones like the one we are going
to study here. Instead of relying on computational power we
propose a simple technique for dealing with the segmented
undulator scheme which minimizes numerical difficulties and
computation time and can be implemented straightforwardly also by
non-expert programmers. In order to reach this result we will use
our understanding of approximations in undulator radiation theory,
thus simplifying equations as much as possible before using, in
the final step, some numerical integration technique. It is
worthwhile to stress the fact that we will have full control of
our approximations at every step, meaning that we will be able to
get the radiation intensity \textit{and} the accuracy with which
this is computed as well.

We will consider the undulators tuned at the fundamental
frequency. Therefore we start reminding the resonant approximation
seen in Section \ref{sec:undu}. In particular, using our method we
were able to demonstrate that under the resonant approximation the
field contribution from non-resonant structures can be neglected
with respect to the undulator field with an accuracy of $1/(4\pi
N_w)$. Here we will restrict our attention to a region of
parameters where such an approximation is valid. Having indicated,
as in Fig. \ref{applgeo}, with $L_s$ the straight section length,
we will therefore be interested in the fundamental harmonic of the
system within the assumptions: $\hat{z}_o \gg 1+L_s/L_w$, $4\pi
N_w \gg 1$, and $\hat{C} + \hat{\theta}^{2}/{2}\ll 4\pi N_w$.

Also, for the same reason and with the same accuracy we can
neglect the constrained motion in the Green's function phase and
consider the particle moving on a straight line when calculating
this phase term.

In Paragraph \ref{sec:ofdef} we treated the case of a single
electron with offset and deflection. If one is interested in the
far field limit case, one can use Eq. (\ref{endangle22}) to obtain
the contributions from each segment but special attention must be
taken in summing up the fields with the correct relative phase
factor.

With the help of Eq. (\ref{endangle22}) and the coordinate system
depicted in Fig. \ref{applgeo} we start accounting for the field
contributions from the first and the second undulator in the case
there is no straight section between the two segments that is
$\hat{L}_s$, defined as $L_s/L_w$, is zero. In this case we have:

\begin{equation}
\hat{E}_1 = {1\over{2}} e^{i(\Phi_s+\Phi_o)}e^{-i \zeta_1/2 }
{\sin{\zeta_1/2}\over{\zeta_1/2}}~, \label{primound}
\end{equation}
while
\begin{equation}
\hat{E}_2 = {1\over{2}} e^{i(\Phi_s+\Phi_o)}e^{i \zeta_2/2 }
{\sin{\zeta_2/2}\over{\zeta_2/2}} \label{secound}
\end{equation}
where

\begin{equation}
\Phi_o = - \hat{\theta}_x\hat{l}_{cx}
-\hat{\theta}_y\hat{l}_{cy}~, \label{phase2}
\end{equation}
\begin{eqnarray}
\zeta_1 = {\hat{C}\over{2}} + {1\over{4}}\left(\hat{{\theta}}_x-
\hat{\eta}_{x1}\right)^2 +{1\over{4}}\left(\hat{{\theta}}_{y}-
\hat{\eta}_{y1}\right)^2~ \label{zeta1s}
\end{eqnarray}
and
\begin{eqnarray}
\zeta_2 = {\hat{C}\over{2}} + {1\over{4}}\left(\hat{{\theta}}_x-
\hat{\eta}_{x2}\right)^2 +{1\over{4}}\left(\hat{{\theta}}_{y}-
\hat{\eta}_{y2}\right)^2~. \label{zeta2s}
\end{eqnarray}
Here $\Phi_s$ is defined as usual by Eq. (\ref{phisnorm}); offsets
$\hat{l}_{c(x,y)}$ refer to the middle point of the system (i.e.
the point with zero value of $z$ in Fig. \ref{applgeo}) and
deflections $\hat{\eta}_{(x,y)1}$ and $\hat{\eta}_{(x,y)2}$ refer
to the first and the second undulator respectively, and the
relation between the two is determined by the focusing elements
only: therefore it makes sense to talk about deflections for the
first and the second undulator without specifying the point.

The next step is to account for the phase shift in the case
$\hat{L}_s \neq 0$. On the one hand, shifting the first segment of
a quantity $-{L}_s/2$ gives a phase contribution:

\begin{equation}
\Delta \Phi_1 = -{\omega L_s\over{2}}\left[{1\over{2c\gamma^2}}+
{1\over{2}}\left({{\theta}}_x- {\eta}_{x1}\right)^2
+{1\over{2}}\left({{\theta}}_{y}- {\eta}_{y1}\right)^2\right]~.
\label{df1}
\end{equation}
On the other hand, shifting the second segment of a quantity
${L}_s/2$ leads to an extra phase term:

\begin{equation}
\Delta \Phi_2 = {\omega L_s\over{2}}\left[{1\over{2c\gamma^2}}+
{1\over{2}}\left({{\theta}}_x- {\eta}_{x2}\right)^2
+{1\over{2}}\left({{\theta}}_{y}- {\eta}_{y2}\right)^2\right]~.
\label{df2}
\end{equation}
To complicate the situation further, we should account for the
fact that, in order to practically control the relative phase
between the two contributions phase shifters will be installed.
These devices are also useful in FEL technology (see \cite{PFLU});
in the case of PETRA III, they will be usually designed and tuned
in order to provide matching condition for particles at nominal
energy moving at zero angle with respect to the undulator axis,
and with zero detuning. This means that the phase shifter will
contribute for a relative phase $\phi_{sh}$ such that

\begin{equation}
\phi_{sh} + {\omega_1 L_s\over{2c\gamma^2}} = 2 \pi n
\label{shifter}
\end{equation}
where $n$ is an integer number. Taking out the unessential $2 \pi
n$ contribution (or setting $n=0$) and accounting for $\phi_{sh}$
one gets

\begin{equation}
\Delta \Phi_1 = -{\omega_1
L_s\over{2}}\left[{\omega-\omega_1\over{2c\gamma^2 \omega_1}}+
{1\over{2}}\left({{\theta}}_x- {\eta}_{x1}\right)^2
+{1\over{2}}\left({{\theta}}_{y}- {\eta}_{y1}\right)^2\right]~
\label{df1b}
\end{equation}
and

\begin{equation}
\Delta \Phi_2 = {\omega_1
L_s\over{2}}\left[{\omega-\omega_1\over{2c\gamma^2\omega_1}}+
{1\over{2}}\left({{\theta}}_x- {\eta}_{x2}\right)^2
+{1\over{2}}\left({{\theta}}_{y}- {\eta}_{y2}\right)^2\right]~.
\label{df2b}
\end{equation}
Finally, we make use of the relation $C = \Delta
\omega/(2\gamma_z^2 c)$ and we express $\Delta \Phi_1$ and $\Delta
\Phi_2$ in normalized units thus writing

\begin{equation}
\Delta \Phi_1 = -{\hat{C}\over{2}} \hat{L}'_s - \left[
{1\over{4}}\left(\hat{{\theta}}_x- \hat{\eta}_{x1}\right)^2
+{1\over{4}}\left(\hat{{\theta}}_{y}-
\hat{\eta}_{y1}\right)^2\right]\hat{L}_s~ \label{df1c}
\end{equation}
and

\begin{equation}
\Delta \Phi_2 = {\hat{C}\over{2}} \hat{L}'_s + \left[
{1\over{4}}\left(\hat{{\theta}}_x- \hat{\eta}_{x2}\right)^2
+{1\over{4}}\left(\hat{{\theta}}_{y}-
\hat{\eta}_{y2}\right)^2\right]\hat{L}_s~ \label{df2c}
\end{equation}
where

\begin{equation}
\hat{L}'_s = {\hat{L}_s \over{1+{K^2/{2}}}} ~.\label{LSprime}
\end{equation}
We can account for a given energy deviation $\Delta\gamma/\gamma$
expanding $\hat{C}$ around the nominal energy. Trivial
calculations show that, for $\omega = \omega_1$, $\omega_1$ being
now calculated at $\theta=0$ and at nominal energy we obtain a
shift $\hat{\Delta}_E$ in the detuning parameter given by

\begin{equation}
\hat{\Delta}_E = 4\pi N_w {\Delta \gamma\over{\gamma}}~.
\label{delthatgamma}
\end{equation}
Including $\hat{\Delta}_E$ in our equations we obtain the final
results:

\begin{equation}
\hat{E}_1 = {1\over{2}} e^{i(\Phi_s+\Phi_o)}e^{-i \zeta_1/2
}e^{i\Delta \Phi_1} {\sin{\zeta_1/2}\over{\zeta_1/2}}~
\label{primoundf}
\end{equation}
and
\begin{equation}
\hat{E}_2 = {1\over{2}} e^{i(\Phi_s+\Phi_o)}e^{i \zeta_2/2 }
e^{i\Delta \Phi_2}{\sin{\zeta_2/2}\over{\zeta_2/2}}
\label{secoundf}
\end{equation}
with

\begin{eqnarray}
\zeta_1 = {\hat{C}+\hat{\Delta}_E\over{2}} +
{1\over{4}}\left(\hat{{\theta}}_x- \hat{\eta}_{x1}\right)^2
+{1\over{4}}\left(\hat{{\theta}}_{y}- \hat{\eta}_{y1}\right)^2~,
\label{zeta1sf}
\end{eqnarray}
\begin{eqnarray}
\zeta_2 = {\hat{C}+\hat{\Delta}_E\over{2}} +
{1\over{4}}\left(\hat{{\theta}}_x- \hat{\eta}_{x2}\right)^2
+{1\over{4}}\left(\hat{{\theta}}_{y}- \hat{\eta}_{y2}\right)^2~,
\label{zeta2sf}
\end{eqnarray}
\begin{equation}
\Delta \Phi_1 = -{\hat{C}+\hat{\Delta}_E\over{2}} \hat{L}'_s -
\left[ {1\over{4}}\left(\hat{{\theta}}_x- \hat{\eta}_{x1}\right)^2
+{1\over{4}}\left(\hat{{\theta}}_{y}-
\hat{\eta}_{y1}\right)^2\right]\hat{L}_s~, \label{df1cf}
\end{equation}
\begin{equation}
\Delta \Phi_2 = {\hat{C}+\hat{\Delta}_E\over{2}} \hat{L}'_s +
\left[ {1\over{4}}\left(\hat{{\theta}}_x- \hat{\eta}_{x2}\right)^2
+{1\over{4}}\left(\hat{{\theta}}_{y}-
\hat{\eta}_{y2}\right)^2\right]\hat{L}_s~, \label{df2cf}
\end{equation}
where $\hat{C}$ is now the detuning parameter for a particle with
nominal energy and other quantities are as defined before.

As we have said, focusing elements simply change the electron
trajectory, but they do not contribute directly to the total field
at the observer position. We will treat them like a thin lens with
focal length $f_x$ and $f_y$ in the horizontal and vertical planes
respectively. We can, therefore, introduce normalized focal
lengths $\hat{f}_x = f_x/L_w$ and $\hat{f}_y = f_y/L_w$.

From now on, when possible, we will neglect indexes $x$ or $y$ for
notational simplicity. If we have offset and deflection
$\hat{l}_1$ and $\hat{\eta}_1$ at the entrance of the system
(where $\hat{z} = -\hat{L}_2/2 -1/2$, see Fig. \ref{applgeo}), we
can easily calculate

\begin{equation}
\hat{\eta}_2 =  -{\hat{l}_1\over{\hat{f}}}+\left(1-
{1+\hat{L}_s\over{2\hat{f}}} \right)\hat{\eta}_1 ~.\label{eta2}
\end{equation}
It is worth to note here that if  we take the limit for $\hat{f}
\rightarrow\infty$ and $\hat{L}_s \rightarrow 0$, the sum
$\hat{E}_1+\hat{E}_2$ gives back Eq. (\ref{endangle}), as it must.

Since we have many particles here we can consider a collective
description of the beam, for instance in terms of the Twiss
parameters $\alpha_{T}$, $\beta_{T}$ and $\gamma_{T}$ for the
horizontal and vertical planes at the entrance of the system,
where $\gamma_{T}=(1+\alpha_{T}^2)/\beta_{T}$. Knowing the Twiss
parameters and the emittance $\varepsilon$ at the entrance of the
first undulator it is easy to write down the number of particles
with offset between $l_1$ and $l_1+\delta l_1$ and with deflection
angles between $\eta_1$ and $\eta_1+\delta \eta_1$, always at the
entrance of the system; in fact, one can simply write it as
$g(l_1,\eta_1) \delta l_1 \delta \eta_1$, where $g$ is the
following density distribution \cite{ROSE}:

\begin{eqnarray}
g(l_1,\eta_1)= {N_p\over{2\pi \varepsilon}} \exp{\left[-{\gamma_T
l_1^2 + 2\alpha_T l_1\eta_1+\beta_T \eta_1^2\over{2
\varepsilon}}\right]} ~.&&\cr \label{distri}
\end{eqnarray}
Here $N_p$ is the number of particles in the beam and again, for
notational simplicity, we are taking indexes $x$ and $y$ as
implicit. Upon introduction of

\begin{eqnarray}
\hat{\alpha}_T = \alpha_T~,&&\cr \hat{\beta}_T = L_w^{-1}
\beta_T~,&&\cr \hat{\gamma}_T = L_w \gamma_T~,&&\cr
\hat{\varepsilon}= (\omega/c) \varepsilon~&&\cr \label{adquaV}
\end{eqnarray}
we can write Eq. (\ref{distri}) as a function of dimensionless
quantities as

\begin{eqnarray}
{g}(\hat{l}_1,\hat{\eta}_1)= {1\over{2\pi \hat{\varepsilon}}}
\exp{\left[-{\hat{\gamma}_T \hat{l}_1^2 + 2\hat{\alpha}_T
\hat{l}_1\hat{\eta}_1+\hat{\beta}_T \hat{\eta}_1^2\over{2
\hat{\varepsilon}}}\right]}~. &&\cr \label{distriad}
\end{eqnarray}
\begin{table}
\caption{Set of parameters for the segmented undulator scheme
\label{tavola1}}
\begin{ruledtabular}
\begin{tabular}{ccc}
&& \\
Quantity                & Unit & Value \\
&& \\ \tableline && \\
${L}_w             $  &   m        &  $ 20.0$ \\
$N_w               $  &   -        &  $ 690$ \\
${L}_s             $  &   m        &  $ 6.5$ \\
$\gamma            $  &   -        &  $ 11742.0 $\\
$\lambda             $  &   m        &  $ 2.5 \cdot 10^{-10}$ \\
${\alpha}_{Tx}       $  &   -        &  $ 0.218$ \\
${\beta}_{Tx}        $  &   m        &  $ 24.005$ \\
${\gamma}_{Tx}       $  &   ~~~m$^{-1}$ &  $ 0.044$ \\
${\varepsilon}_{x}   $  &   m        &  $ 1.0 \cdot 10^{-9}$ \\
${f}_{x}             $  &   m        &  $ 5.0$ \\
${\alpha}_{Ty}       $  &   -        &  $ 1.291$ \\
${\beta}_{Ty}        $  &   m        &  $ 10.272$ \\
${\gamma}_{Ty}       $  &   ~~~m$^{-1}$ &  $ $ 0.260\\
${\varepsilon}_{y}   $  &   m        &  $ 1.0 \cdot 10^{-11}$ \\
${f}_{y}             $  &   m        &  $ 5.8$ \\
$\sigma_E            $  &   -        &  $ 1.1 \cdot 10^{-3}$ \\
$K                   $  &   -        &  $ 1.66 $\\
&&\\
\end{tabular}
\end{ruledtabular}
\end{table}
\begin{table}
\caption{Set of normalized parameters for the segmented undulator
scheme \label{tavola2}}
\begin{ruledtabular}
\begin{tabular}{cc}
&\\
$\hat{\alpha}_{Tx}    $  &  $ 0.218$ \\
 $\hat{\beta}_{Tx}     $  &  $ 1.20$ \\
$\hat{\gamma}_{Tx}    $  &  $ 0.873$ \\
$\hat{\varepsilon}_{x}$  &  $25.1$ \\
$\hat{f}_{x}          $  &  $ 0.250$ \\
 $\hat{\alpha}_{Ty}    $  &  $ 1.29$ \\
 $\hat{\beta}_{Ty}     $  &  $ 0.514$ \\
 $\hat{\gamma}_{Ty}    $  &  $ 5.19$ \\
$\hat{\varepsilon}_{y}$  &  $ 0.251$ \\
$\hat{f}_{y}          $  &  $ 0.290$ \\
$\hat{L}_s            $  &  $ 0.325$ \\
$\hat{\sigma}_E       $  &  $ 9.54   $ \\
$\hat{L}'_s                 $  &  $ 0.137   $ \\
&\\
\end{tabular}
\end{ruledtabular}
\end{table}
The local energy spread of the beam will be assumed to be a
gaussian function with standard deviation $\hat{\sigma}_E$:

\begin{equation}
F(\hat{\Delta}_E) = {1\over{\sqrt{2\pi \hat{\sigma}^2_E}}}
e^{-\hat{\Delta}^2_E/(2\hat{\sigma}^2_E)}~,\label{gaussenspr}
\end{equation}
where $\hat{\sigma}_E = 4 \pi N_w \sigma_E$ and

\begin{equation}
\sigma_E^2 = \left\langle \left({\Delta \gamma
\over{\gamma}}\right)^2\right\rangle ~,\label{defsigmae}
\end{equation}
brackets indicating the ensamble average over the particle
distribution.

A set of realistic normalized parameters for the segmented
undulator is given in Table \ref{tavola1} and Table \ref{tavola2}
for dimensional and dimensionless quantities, respectively. From
these tables we can already foresee  a large effect of the
normalized energy spread parameter $\hat{\sigma}_E$ which is, in
our case, comparable with the normalized horizontal emittance.

We can now average the normalized intensity over the particle
distribution function and the energy spread of the beam at the
entrance of the system thus finding the final expression for the
total intensity $\hat{I}_b$

\begin{eqnarray}
\hat{I}_b = |\hat{E}_1+\hat{E}_2|^2=
\int_{-\infty}^{\infty}d\hat{\Delta}_E F(\hat{\Delta}_E)
\int_{-\infty}^{\infty}d\hat{l}_{x1} &&\cr \times
\int_{-\infty}^{\infty}d\hat{\eta}_{x1}
\int_{-\infty}^{\infty}d\hat{l}_{y1}
\int_{-\infty}^{\infty}d\hat{\eta}_{y1}
~{g}_x(\hat{l}_{x1},\hat{\eta}_{x1})
{g}_y(\hat{l}_{y1},\hat{\eta}_{y1}) &&\cr \times \left| e^{-i
\zeta_1/2 } e^{i\Delta \Phi_1} {\sin{\zeta_1/2}\over{\zeta_1/2}} +
e^{i\zeta_2/2 } e^{i\Delta \Phi_2}
{\sin{\zeta_2/2}\over{\zeta_2/2}}\right|^2~.
 \label{intfinal}
\end{eqnarray}
As in Eq. (\ref{blab}), the normalized intensity $\hat{I}_b$ is
related to the non-normalized intensity $I_b$ by

\begin{equation}
\hat{I}_b = {4\pi\over{N_p c}}{\left( c^2 z_o \gamma  \over{K
\omega e L_w A_{JJ}}\right)^{2}}  I_b~. \label{hatin}
\end{equation}
Moreover $\hat{\eta}_{2}$ is expressed in terms of $\hat{l}_{1}$
and $\hat{\eta}_{1}$ in Eq. (\ref{eta2}).

We now reached an expression, Eq. (\ref{intfinal}), which can be
integrated with the help of numerical techniques. Results of
numerical integration are the intensity as a function of the
detuning parameter $\hat{C}$ for fixed observation angles or the
intensity as a function of angles for a fixed detuning parameter.

We developed a short code which uses a Monte Carlo method to
compute Eq. (\ref{intfinal}). A quasi-random generator is used to
fill the horizontal and vertical transverse phase space with a
given number of particles. The actual simulation makes use of
Sobol sequences to reduce the formation of cluster of particles, a
well known phenomenon described, for instance, in \cite{REICH} and
references therein. Then, the intensity for each particle is
calculated and all contributions are summed up to give an
estimation of Eq. (\ref{intfinal}). The critical parameter in this
kind of codes is the number of particles used. In general the
higher the number of particles is, the smaller the fluctuation of
the results from run to run and the higher the accuracy of the
integration. On the other hand numerical algorithm designers warn
that quasi-random sequences start to display unwanted correlations
after several tens of millions calls to the generating routine. We
observed that a choice of several million particles is more than
sufficient to ensure that results converge fast enough so that no
relevant fluctuation is present in the outcome (with graphical
accuracy of the results which will be described in the next few
figures).

First, we compared results from the Monte Carlo approach in the
limiting case $\hat{L}_s \rightarrow 0$, $\hat{f}_{x,y}\rightarrow
\infty$, $\hat{\varepsilon}_x \rightarrow \infty$,
$\hat{\varepsilon}_y \rightarrow  0$ and $\hat{\sigma}_E
\rightarrow 0$. Results were in perfect agreement with the ones
shown in Fig. \ref{cth}, Fig. \ref{thyzero}, Fig. \ref{thyplus}
and Fig. \ref{thyminus} as it should be. Then, as has already been
said in Paragraph \ref{sec:ofdef} we compared this asymptotic
situation with a more practical case of a uniform undulator with
$\hat{\varepsilon}_x \simeq 40$ and $\hat{\varepsilon}_y \simeq
1$. Results have been already described, and shown in Fig.
\ref{accuc} and Fig. \ref{accuthy}.

\begin{figure}
\begin{center}
\includegraphics*[width=90mm]{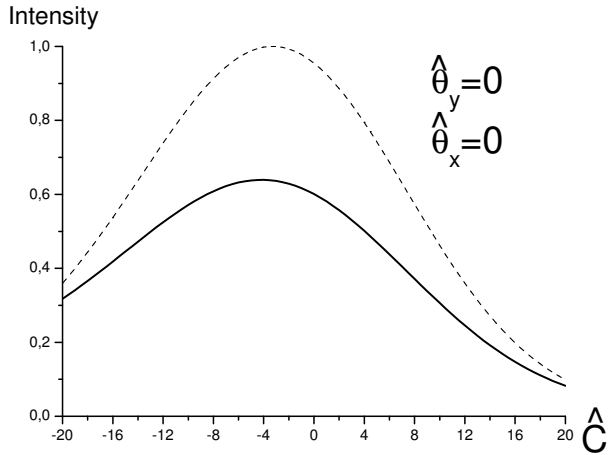}
\caption{Solid line : relative intensity  from a beam with
parameters listed in Table \ref{tavola2}. The intensity is plotted
as a function of $\hat{C}$ for $\hat{\theta}_x=0$ and
$\hat{\theta}_y=0$. It is normalized to the maximal intensity from
a beam in a uniform undulator. Dashed line : relative intensity
from a beam in a uniform undulator ($\hat{L}_s \rightarrow 0$ m
and $\hat{f} \rightarrow \infty$). We set the minimal betatron
function values $\hat{\beta}_{ox} = 1$ in the horizontal direction
and $\hat{\beta}_{oy} = 0.5$ m in the vertical one.
$\hat{\varepsilon}_{x,y}$ and $\hat{\sigma}_E$ are left unvaried
as in Table \ref{tavola2}. \label{intc} }
\end{center}
\end{figure}
\begin{figure}
\begin{center}
\includegraphics*[width=90mm]{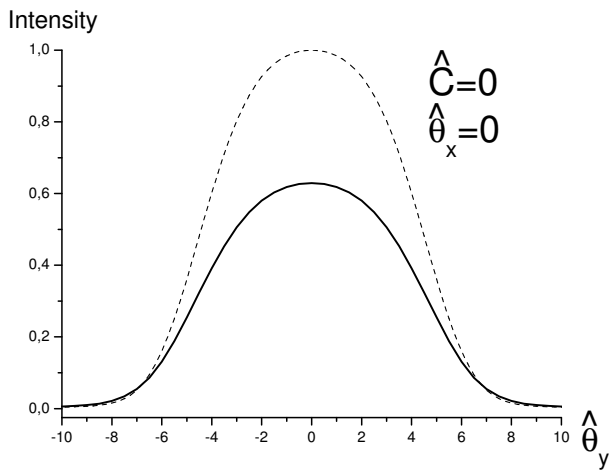}
\caption{Solid line : relative intensity from a beam with
parameters listed in Table \ref{tavola2}. The intensity is plotted
as a function of $\hat{\theta}_y$ for $\hat{\theta}_x=0$ and
$\hat{C}=0$. It is normalized to the maximal intensity from a beam
in a uniform undulator. Dashed line : comparison with the relative
intensity from a beam in a uniform undulator ($\hat{L}_s
\rightarrow 0$ m and $\hat{f} \rightarrow \infty$). We set the
minimal betatron function values $\hat{\beta}_{ox} = 1$ in the
horizontal direction and $\hat{\beta}_{oy} = 0.5$ m in the
vertical one. $\hat{\varepsilon}_{x,y}$ and $\hat{\sigma}_E$ are
left unvaried as in Table \ref{tavola2}. \label{inthy} }
\end{center}
\end{figure}
Finally we used all the features of our program in order to deal
with the segmented undulator case. We set up parameters as in
Table \ref{tavola1}, which correspond to a design set for the
PETRA III case. The solid line in Fig. \ref{intc} represents the
intensity as a function of the detuning parameter $\hat{C}$ at
$\hat{\theta}_x = 0$ and $\hat{\theta}_y=0$. The dashed line in
Fig. \ref{intc} is the intensity for the uniform undulator case,
corresponding to the limit  $\hat{L}_s \rightarrow 0$ m and
$\hat{f} \rightarrow \infty$, while $L_w = 20$ m. In this case we
set the minimal betatron function values to $\beta_{ox} = 20$ m in
the horizontal direction and $\beta_{oy} = 10$ m in the vertical
one. Beam energy, wavelength, undulator deflection parameter,
relative energy spread and electron beam emittances are left
unvaried as in Table \ref{tavola1}. Assuming the minimal $\beta_T$
function in the center of the undulator we can calculate the Twiss
parameters at the system entrance using for both horizontal and
vertical plane (see, for instance, \cite{ROSE}) :

\begin{equation}
\beta_T(z) = \beta_{o}
\left[1+\left({z-z_w\over{\beta_{o}}}\right)^2\right]~,
\label{betwiss} \end{equation}
\begin{equation}
\gamma_T(z) = \gamma_o = {1\over{\beta_o}}~, \label{gammatwiss}
\end{equation}
\begin{equation}
\alpha_T(z) = -\gamma_o(z-z_w) \label{alphatwiss}~,
\end{equation}
where $z_w$ indicates the $z$-position at the beam waist, while
subscripts "o", as usual, denote minimal values of the Twiss
parameters. Taking $z_w = 0$ m at the center of the system and $z
= -10$ m we obtain the initial Twiss parameters to enter in our
program: $\alpha_{Tx} = 0.5$, $\beta_{Tx} = 25.0$ m and
$\gamma_{Tx} = 0.05$ m$^{-1}$, $\alpha_{Ty} = 1.0$, $\beta_{Ty} =
20.0$ m and $\gamma_{Ty} = 0.1$ m$^{-1}$.

It is important to remark that we normalized the solid line in
Fig. \ref{intc}, that is the result for the segmented undulator,
to the maximal intensity found for the uniform undulator while the
dashed line represents the relative intensity for the uniform
undulator. This allows direct comparison between uniform undulator
and segmented undulator schemes.

To conclude, in Fig. \ref{inthy} (solid line) we plot again the
intensity as a function of $\hat{\theta}_y$ at $\hat{C}=0$ and
$\hat{\theta}_x=0$. Similarly as for Fig. \ref{intc}, the dashed
line in Fig \ref{intc} represents a comparison with the uniform
undulator case $\hat{L}_s = 0$ m and $\hat{f} \rightarrow \infty$,
where the other parameters are left unvaried as in Table
\ref{tavola1}. Normalization of the curves are as in Fig.
\ref{intc}. The final results for the two schemes do not appear to
be dramatically different. 

As a comment for both Fig.
\ref{intc} and Fig. \ref{inthy} it is interesting to note the
smoothing action of finite beam emittance combined with finite
energy spread.

\section{\label{sec:conc} Conclusions}

In this paper we have shown that even well-known and long studied
subjects, like Synchrotron Radiation, continue to retain interest
from a theoretical viewpoint. In particular we have proposed a
technique to compute harmonic contributions of the electric field
from a moving charge particularly suitable for analytical
investigations. We showed that our general expression, Eq.
(\ref{generalfin}), can be obtained from already known expression
Eq. (\ref{rev2}) with the use of a paraxial approximation, but
also much more straightforwardly from Maxwell's equation in
paraxial form, with the help of a parabolic Green's function
method. The main advantage of our approach with respect to
standard techniques is that we separated from very beginning
approximations independent from the system under study (paraxial
approximation) and assumptions that, from time to time, arise in
the study of a particular magnetic system. This gave us the
possibility to specify region of applicability and accuracy of
well known results, which we were, of course, able to recover
using our method, like the field from a charge moving on a circle,
in a short bending magnet or in an undulator and also edge
radiation effects from a particular setup. Again, our particular
approach is logically much less involved with respect to the
others, since it relies on direct integration of the paraxial
Maxwell's equations for a given harmonic. As a result, our
expression gives a clearer and completely fresh insight in old
results. Besides being able to discuss their region of
applicability and their accuracy, we pointed out that phase
corrections with respect to the spherical wavefront for an
electron on a circle arise naturally from analytical treatment. We
have seen that short magnet radiation is a complete separate
phenomenon with respect to edge radiation.  We stressed the
importance of the knowledge of the entire electron trajectory in
order to obtain a correct calculation of the field: from this
viewpoint we have seen how edge radiation from two dipoles
separated by a straight section arises, in our method, from the
contribution  due to the straight section alone, and that the
field term from the magnets is completely negligible. We
discussed, under resonance approximation, accuracy and
applicability region of the far field undulator radiation and we
dealt with near field effects as well: in particular we were able
to reproduce already known results studying their region of
applicability in detail, and we found also other regions of
interest, where the near field parameter $\bar{W}$ is of
fundamental importance.

Besides this, our method proved to be a reliable basis for
calculation of more complicated quantities like field
autocorrelation functions, either from bending magnets or
undulators, which are of fundamental importance in the
understanding of spatial coherence properties of Synchrotron
Radiation. In fact, by direct application of our general formula
Eq. (\ref{generalfin}) we were able to provide analytically
manageable expressions for bending magnet or undulator radiation
in the case of an electron with a transverse offset and
deflection.  These expressions can be integrated over  realistic
beam particle distributions and can be used in very practical
calculations. Their strength stems from the fact that they
correctly account for the field phase and that they can be simply
summed up to give results for apparently complicated situations.
Complete treatment of spatial coherence would probably double the
size of this paper. For this reason, this first work focuses on
the basis of Synchrotron Radiation theory only.

A few applications which exploit the power of our approach have
been selected for the final part of the paper. At the end of
Section \ref{sec:undu} we addressed basic characteristics of
undulator radiation in the presence of electron beam emittance. We
discussed the asymptotic situation for a large horizontal and a
small vertical emittance.

Finally, in Section \ref{sec:appl}, we considered a novel
undulator configuration (setup of two undulators separated by a
focusing element) which is planned for installation at PETRA III.
Computation of radiation characteristics  from new setups  by
means of numerical techniques alone requires, almost always,
modifications of existing simulation codes which can be better
done by the code authors themselves. Up to date, the particular
case we chose to study has not been included in existing
simulation codes. For this reason we selected this particular
example: our goal was to demonstrate that, in general, the
application of our method allows solution of apparently difficult
problems relying on simple computer algorithm which can be
developed by non-expert programmers, with obvious advantages in
saving time and better physical understanding.
%

We expect that our technique will be used in the future as a basis
for developments of simulation codes and for physical
understanding of complex situation as well.

\section{\label{sec:graz} Acknowledgements}

The authors are grateful to Klaus Balewski, Werner Brefeld,
Winfried Decking, Martin Dohlus, Hermann Franz, Petr Ilinski and
Edgar Weckert for many useful discussions and to Oliver Grimm,
Jochen Schneider and Ivan Vartanyants for their interest in this
work.

%

\end{document}